\documentclass{article} 
\usepackage{times}
\usepackage[letterpaper, margin=1in]{geometry}

\usepackage{fancyhdr}
\fancyheadoffset{0pt}
\PassOptionsToPackage{hyphens}{url}
\usepackage{hyperref}
\usepackage{url}
\usepackage{xurl} 
\usepackage{amsmath}

\usepackage{amssymb}
\usepackage{bm}
\usepackage{graphicx}
\usepackage{adjustbox}
\usepackage{tabularx}
\usepackage{multirow}
\usepackage{rotating}
\usepackage{placeins}
\usepackage{subcaption}
\usepackage{microtype}

\usepackage{caption}
\usepackage{algorithm}
\usepackage{algpseudocode}
\usepackage{booktabs} 
\usepackage{breqn}
\usepackage{tikz}
\usepackage{pifont}
\newcommand{\rebuttal}[1]{\textcolor{blue}{#1}}
\renewcommand{\rebuttal}[1]{#1}

\def\scititle{
	\textsc{Flowr.root} -- A flow matching based foundation model for joint multi-purpose structure-aware 3D ligand generation and affinity prediction
}
\title{\bfseries \boldmath \scititle}

\author{
    Julian Cremer$^{1,\ast}$,
    Tuan Le$^{1}$,
    Mohammad M. Ghahremanpour$^{2}$,
    Emilia Sługocka$^{3,4}$,\\
    Filipe Menezes$^{5,6,\ast}$,
    Djork-Arné Clevert$^{1}$\\
    \small $^{1}$Machine Learning \& Computational Sciences, Pfizer Worldwide R\&D, Berlin, Germany\\
    \small $^{2}$Computational Chemistry, Medicine Design, Pfizer Worldwide R\&D, Cambridge, USA\\
    \small $^{3}$Doctoral School of Medical and Health Sciences, Jagiellonian University Medical College, Cracow, Poland\\
    \small $^{4}$Department of Physicochemical Drug Analysis, Faculty of Pharmacy, Jagiellonian University Medical College, Cracow, Poland\\
    \small $^{5}$Institute of Structural Biology, Molecular Targets and Therapeutics Center, Helmholtz Munich, Neuherberg, Germany\\
    \small $^{6}$TUM School of Natural Sciences, Department of Bioscience, Bayerisches NMR Zentrum,\\ \small Technical University of Munich, Garching, Germany\\
    \small $^{\ast}$Corresponding authors. Email: julian.cremer@pfizer.com, filipe.menezes@helmholtz-munich.de\\
}

\begin{document}

\maketitle

\begin{abstract}
We present \textsc{Flowr.root}, an $SE(3)$-equivariant flow-matching foundation model that unifies pocket-aware 3D ligand generation with multi-endpoint binding affinity prediction ($\mathrm{pIC}_{50}$, $\mathrm{p}K_{\mathrm{i}}$, $\mathrm{p}K_{\mathrm{d}}$, $\mathrm{pEC}_{50}$) and pLDDT-based confidence estimation in a single backbone. One trained model supports \textit{de novo} pocket-conditional generation, interaction- and pharmacophore-conditional sampling, scaffold hopping and elaboration, and fragment growing or replacement, enabled by a mixed isotropic--anisotropic prior placement strategy. Training proceeds in three stages: large-scale pre-training on billions of ligand conformations and millions of mixed-fidelity protein--ligand complexes, refinement on curated co-crystal data, and project-specific adaptation via parameter-efficient LoRA finetuning. Joint structure--affinity modelling enables inference-time importance-sampling guidance for single- and multi-objective design without external scoring functions. Case studies on kinase selectivity (CK2$\alpha$/CLK3) and scaffold elaboration on TYK2, ER$\alpha$, and BACE1 illustrate utility from hit identification through lead optimization.
\end{abstract}

\section*{Introduction}

\rebuttal{
Diffusion- and flow-based generative models have become central tools for learning complex data distributions and have recently enabled rapid progress in generative chemistry and structure-based drug design (SBDD)~\cite{ho2020ddpm, song2020sde, rombach2022ldm, lipman2022flowmatching, liu2022rectified, albergo2022building, albergo2023stochinterp, model:edm, model:midi, model:eqgatdiff, model:diffsbdd, model:targetdiff, model:pilot, folding:alphafold3, model:multiflow, model:semlaflow, model:flowr}.
One key subfield is pocket-conditional ligand generation, where $SE(3)$-equivariant models generate ligands directly in the protein binding site~\cite{model:diffsbdd, model:targetdiff}. Subsequent work has improved chemical validity and pose quality and explored pre-training and inference-time guidance to steer sampling toward desired objectives~\cite{model:eqgatdiff, model:molcraft, model:pilot}. Complementary to \textit{de novo} design, fragment-based approaches support practical lead optimization; however, combining pocket conditioning with flexible fragment elaboration and  editing (for example, scaffold replacement or fragment growing under interaction constraints) within a single generative backbone remains challenging~\cite{imrie2020delinker, voloboev2024fragment2d, zhang2024fraggen, lee2025fragfm, guo2023linkinvent, model:difflinker, model:flowr}.}

\rebuttal{
Beyond generation, the utility of proposed ligands depends on reliable potency and binding affinity prediction to prioritize candidates during iterative design. Classical scoring functions (for example, docking) are computationally efficient but can be insufficiently accurate or are not representing affinities, whereas more elaborate physics-based methods such as free energy perturbation (FEP) and absolute binding free energies (ABFE) are typically more reliable but too costly to apply at the scale of modern generative campaigns~\cite{trott2010vina, friesner2004glide, jones1997gold, wang2015fepplus,mey2020bestpractices,mobley2012alchemical,alibay2022abfe,feng2022abfe,Ries2024_abfe}. Machine learning scoring functions improve throughput but may suffer from dataset bias and limited generalization~\cite{jimenez2018kdeep, volkov2022frustration, model:aev-plig}. Co-folding approaches such as Boltz-2 can reach high accuracy on selected targets, but require structure prediction per query and, due to decoupled training, may limit co-adaptation between pocket geometry and affinity~\cite{boltz2_preprint}.}

\rebuttal{
These considerations motivate models that jointly learn (conditional) structure generation and affinity prediction in a protein pocket context. Joint modeling supports in-distribution ranking of generated structures and enables inference-time steering (for example, via importance sampling) without relying on external scoring functions~\cite{model:pilot}. At the same time, drug discovery campaigns are characterized by assay- and series-specific structure--activity relationships (SARs) that often differ from public training data. Consequently, strong benchmark performance does not necessarily translate to a new project without adaptation; efficient finetuning and, where appropriate, preference-based alignment~\cite{drugflow}, can help align the model to the relevant local SAR landscape.}

\rebuttal{
Training such models is further constrained by the scarcity of high-quality protein--ligand complexes with reliable affinity annotations relative to ligand-only resources. This motivates multi-stage training strategies that exploit large-scale, lower-fidelity data to learn broad chemical and structural priors, followed by refinement on curated experimental datasets to improve structural accuracy and affinity prediction~\cite{dataset:zinc,dataset:pubchem,dataset:pdbbind,dataset:hiqbind,dataset:plinder,dataset:bindingnetv2,dataset:kinodata,dataset:sair}.}

\rebuttal{
In this work, we present \textsc{Flowr.root}, a foundation model for structure-based drug design that spans large-scale pre-training to project-specific adaptation. Within a single architecture, \textsc{Flowr.root} unifies \textit{de novo}, interaction- and pharmacophore-conditional, and fragment-conditioned ligand generation with multi-endpoint affinity prediction and confidence estimation. The model employs a three-stage training strategy that leverages datasets of varying fidelity: (1) large-scale pre-training on billions of ligands and millions of protein–ligand complexes with diverse affinity labels, (2) refinement on curated, higher-quality structural data, and (3) project-specific adaptation via parameter-efficient Low-Rank Adaptation (LoRA)~\cite{hu2022lora} and multi-objective guidance through importance sampling~\cite{model:pilot}. This joint training paradigm allows affinity prediction to directly guide the generative process without external scoring functions, while supporting rapid adaptation to project-specific data. In addition, we introduce flexible fragment-conditional generation modes—including local fragment growing and replacement for spatially targeted modifications enabled by a mixed isotropic-anisotropic prior placement strategy. Combining joint structure and affinity prediction with fragment-based elaboration, affinity guidance enables effective sampling and ranking of local chemical spaces. On established benchmarks, \textsc{Flowr.root} achieves $0.97$ PoseBusters-validity for pocket-conditional ligand generation and Pearson correlations of up to $0.86$ for binding affinity prediction on the FEP+/OpenFE benchmark. Together, these capabilities position \textsc{Flowr.root} as a versatile tool for early-stage drug discovery, from hit identification through lead optimization.}

\FloatBarrier
\section*{Results}

\subsection*{Overview of \textsc{Flowr.root}}
\label{sec:flowr_root_overview}

\begin{figure}[!htbp]
    \centering
    \includegraphics[width=1.0\textwidth]{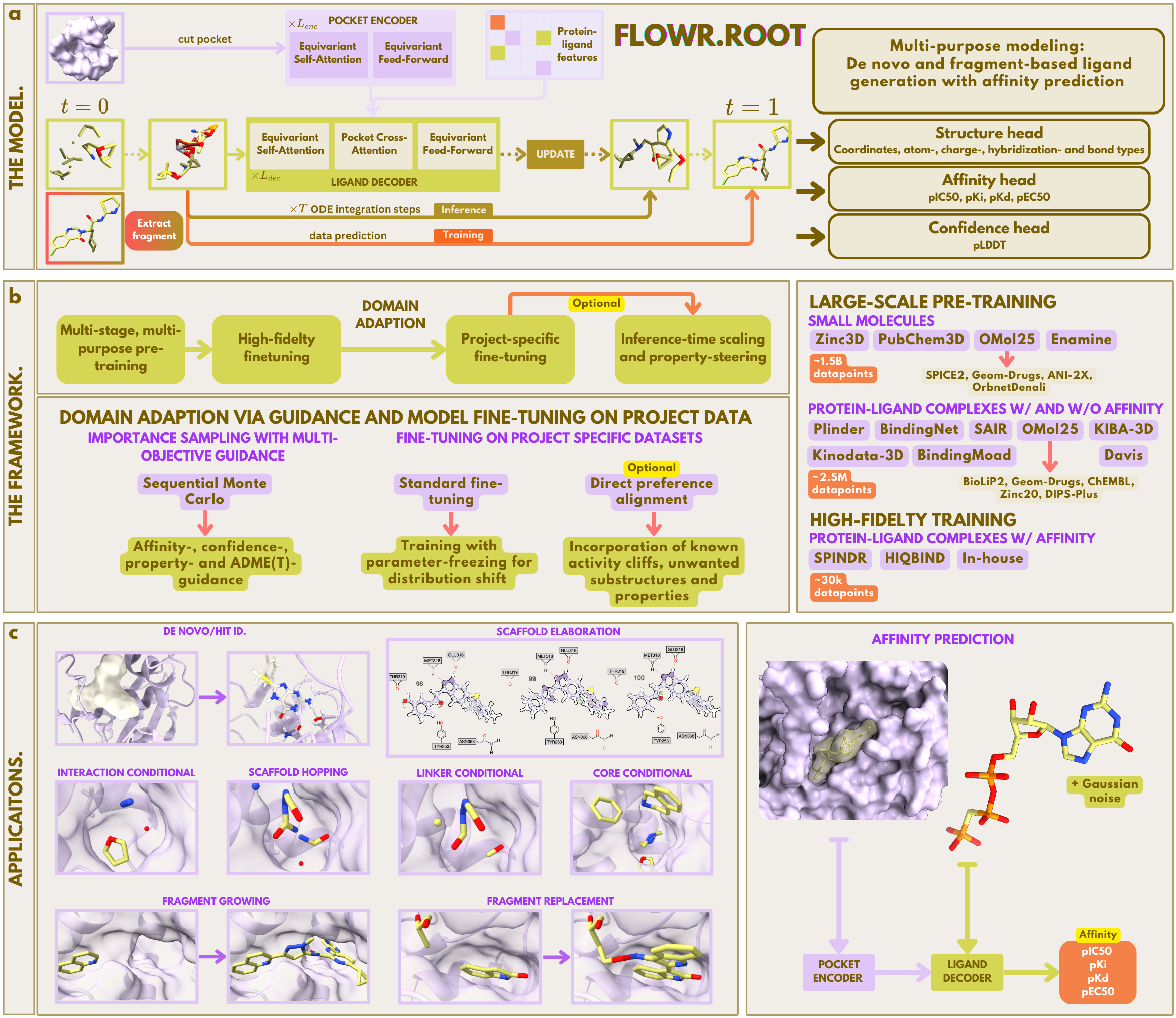}
    \caption{\textbf{Overview of the~\textsc{Flowr.root} framework for joint structure-aware ligand generation and affinity prediction.}
    \textbf{a} The \textsc{Flowr.root} model is built on an $SE(3)$-equivariant flow matching backbone that learns a transport map transforming noise (and optionally fragment anchors) into ligand structures within protein pockets. The architecture comprises an equivariant pocket encoder processing protein pocket features, and a ligand decoder processing ligands, while ligand-pocket interactions are integrated via equivariant cross-attention. A structure head predicts atomic coordinates, atom types, bond orders, charges, and hybridization states, a multi-affinity head predicts affinity with separate predictors for $\mathrm{pIC}_{50}$, $\mathrm{p}K_{\mathrm{i}}$, $\mathrm{p}K_{\mathrm{d}}$, and $\mathrm{pEC}_{50}$, and a confidence head provides pLDDT-based uncertainty estimation. At inference, samples are generated by numerically integrating the learned flow ODE with an Euler solver.
    \textbf{b} The \textsc{Flowr.root} framework consists of three stages combining multi-stage training with finetuning and domain adaptation. Training follows a three-stage paradigm exploiting data of varying fidelity. Large-scale pre-training on $\sim$1.5B small molecule conformations is followed by training on mixed-fidelity $\sim$2.5M protein-ligand complexes with and without affinity labels. Then, the model is finetuned on curated high-fidelity co-crystal datasets to sharpen structural accuracy and affinity prediction, followed by project-specific domain adaptation through parameter-efficient finetuning, or direct preference alignment, and inference-time scaling via importance sampling with multi-objective guidance.
    \textbf{c} \textsc{Flowr.root} supports multiple ligand generation modes within a single backbone: ligand-only and pocket-conditional \textit{de novo} generation, interaction/pharmacophore-conditional generation for preserving specific protein-ligand contacts, scaffold hopping and elaboration, and local fragment growing/replacement. Given a protein pocket and a bound ligand, \textsc{Flowr.root} can also be used to predict potency and binding affinity.
    }
    \label{fig:flowr.root}
\end{figure}

\textsc{Flowr.root} is a unified framework for structure-based drug design that jointly trains pocket-aware 3D ligand generation with multi-endpoint affinity prediction and confidence estimation (Fig.~\ref{fig:flowr.root}).
The framework addresses key challenges in computational drug discovery: generating chemically valid, geometrically realistic ligands within protein binding sites; accurately predicting potency and binding affinity across diverse assay types; supporting flexible generation modes from \textit{de novo} design to targeted fragment modifications; and enabling efficient adaptation to project-specific structure-activity relationships.
A comprehensive overview is given in Methods.

The model is built on an $SE(3)$-equivariant flow matching backbone that learns a mixed continuous/discrete transport map, transforming a prior distribution (noise or fragment anchors for coordinates, uniform for categorical features) to the target ligand distribution within a given protein pocket. The architecture comprises two main components: a pocket encoder that processes full-atom protein features through equivariant self-attention layers, generating invariant and equivariant representations; and a ligand decoder that employs equivariant self-attention for intra-ligand dependencies followed by cross-attention to integrate pocket context. Protein pockets are extracted using a 7{\AA} cutoff radius around binding sites, balancing computational tractability with interaction coverage.
The ligand decoder features three output heads that serve complementary purposes. The structure head predicts atomic coordinates, atom types, bond orders, partial charges, and hybridization states. The multi-affinity head comprises four separate predictors for $\mathrm{pIC}_{50}$, $\mathrm{p}K_{\mathrm{i}}$, $\mathrm{p}K_{\mathrm{d}}$, and $\mathrm{pEC}_{50}$---explicitly avoiding treatment of heterogeneous affinity labels as interchangeable. The confidence head provides pLDDT-based uncertainty estimation for generated structures.
\rebuttal{Training the structure and affinity heads jointly on a shared backbone is what enables the downstream modes used throughout this work: importance-sampling-based multi-objective optimisation, multi-target selectivity from a single model, coherent LoRA domain adaptation of structure and affinity, and affinity-guided fragment elaboration. A modular generator-plus-external-scorer pipeline cannot provide these capabilities without re-introducing the generator--scorer distribution mismatch that the joint formulation removes. A controlled structure-only vs.\ joint comparison on \textsc{HiQBind} confirms that this comes at no cost to structural quality (Supplementary Information).}

Training follows a three-stage paradigm that exploits data of varying fidelity to address the scarcity of high-quality experimental complexes:
Stage 1 builds broad generative priors across chemical and structural distributions. The model trains on $\sim$1.5B small molecule conformations (ZINC3D, PubChem3D, Enamine REAL, OMol25) for chemical coverage, and $\sim$2.5M mixed-fidelity protein-ligand complexes with and without affinity labels from both computational sources (BindingNet, SAIR, KIBA-3D, Davis-3D, Kinodata-3D) and experimental sources (Plinder, BindingMOAD).
Stage 2 adapts pre-trained weights to curated, drug-like datasets. We use SPINDR and HiQBind for multi-target finetuning, teaching the model fine details of protein-ligand interactions and binding affinity from high-quality co-crystal structures.
Stage 3 bridges the gap between public data and project-specific structure-activity landscapes through multiple strategies: inference-time scaling via importance sampling guides generation toward desired properties (affinity, ADME/T, synthetic accessibility); parameter-efficient finetuning enables rapid adaptation while avoiding catastrophic forgetting.
This multi-stage paradigm addresses a fundamental limitation: strong benchmark performance does not guarantee generalization to unseen structure-activity landscapes. Real-world drug discovery requires understanding nuanced SAR across related molecules within specific binding sites---richness uniquely captured by dense bioactivity data that differs substantially from sparse public datasets. Rather than expecting universal generalization, \textsc{Flowr.root} is designed as an adaptive companion that refines through sustained interaction with project-specific data.

The model is trained using mean squared error (MSE) loss for coordinates and categorical cross-entropy (CE) losses for atom types, charges, hybridization, and bond types. In addition, we introduce explicit geometric supervision through bond length and bond angle losses. The bond length loss penalizes deviations in predicted inter-atomic distances for bonded pairs.
The bond angle loss uses Huber loss over valid angle triplets to preserve local molecular geometry while remaining robust to outliers. These auxiliary losses substantially reduce strain energies in generated structures.
For training the affinity head, invariant and equivariant features from ligand, pocket, and their interactions via the ligand decoder's latent representations are extracted, and gated equivariant blocks combine invariant and equivariant tensors, followed by multi-layer perceptrons (MLPs) that produce aggregated ligand, pocket, and interaction embeddings. Task-specific heads then predict each affinity type separately, with Huber loss for robust training. This design explicitly models the distinct experimental setups underlying $\mathrm{IC}_{50}$, $K_{\mathrm{i}}$, $K_{\mathrm{d}}$, and $\mathrm{EC}_{50}$ measurements rather than treating them as interchangeable.

\textsc{Flowr.root} supports multiple generation modes within a single backbone, enabling flexible application across hit identification and lead optimization: (1) \textit{de novo} pocket-conditional generation for exploring diverse chemical matter; (2) interaction/pharmacophore-conditional generation for preserving or enforcing specific protein-ligand contacts; (3) scaffold hopping and elaboration for core and R-group replacement; and (4) fragment growing and local fragment replacement for targeted structural modifications. For the latter mode, we use a flexible prior placement strategy that shifts the generative prior from the zero center-of-mass to the local replacement site, enforcing locality in partial modifications and enabling precise control over which atoms to replace while preserving the remainder of the molecule. In addition, we propose a mixed isotropic and anisotropic Gaussian prior placement during training to enable a more targeted ligand generation strategy at inference. We provide more details in Methods.
\textsc{Flowr.root} also supports inference-time steering via importance sampling to sample from conditional distributions allowing for single- and multi-objective optimization without retraining, for example, steering toward higher predicted affinities while maintaining sample diversity.

To train \textsc{Flowr.root}, we leveraged a diverse collection of datasets. A detailed description of all datasets used in this work, including preprocessing and curation pipelines, dataset statistics, and chemical space analyses, is provided in Methods.
In short, all protein-ligand complexes undergo rigorous preprocessing using Schrödinger's LigPrep and PrepWizard, including protonation state determination at physiological pH, protein preparation with side chain completion, and constrained minimization. Protein pockets are extracted using a 7{\AA} cutoff radius around reference ligands.
Crucially, if not stated otherwise, throughout all protein-ligand datasets we kept a consistent dataset split following the provided Plinder~\cite{dataset:plinder} train, validation and test set splits to avoid data leakage. Plinder employs comprehensive similarity metrics (protein sequence, pocket-level Jaccard, interaction-level PLIP, and ligand-level Tanimoto) to ensure minimal leakage and high-quality, non-redundant testing. To our knowledge, this represents the most rigorous publicly available splitting strategy for structure-based modeling.

\subsection*{Unconditional and Pocket-Conditional Generation}
We first establish the improvements conferred by the architectural changes incorporated into \textsc{Flowr.root}. To this end, we demonstrate the expressiveness of the ligand decoder backbone for unconditional 3D molecule generation on the well-established \textsc{Geom-Drugs} dataset, comparing against recent state-of-the-art models including \textsc{Eqgat-diff}~\cite{model:eqgatdiff}, \textsc{SemlaFlow}~\cite{model:semlaflow}, \textsc{ADiT}~\cite{model:adit}, \textsc{Megalodon}~\cite{model:megalodon}, and \textsc{FlowMol3}~\cite{model:flowmol3}. Here, our non-pretrained base model, \textsc{Flowr.root}$^{\text{base}}$, achieves 0.94$\pm0.02$ PoseBusters-validity (mean), surpassing all baselines including the recently released \textsc{FlowMol3} model (0.92$\pm0.07$), with a median relaxation energy of 3.65$\pm0.2$\,kcal/mol and a median relaxation RMSD of only 0.07$\pm0.02$\,\AA---demonstrating high geometric precision. Complete benchmark results and comparisons are provided in Supplementary Table 1.

For pocket-conditional ligand generation, we benchmark on \textsc{CrossDocked2020}, where \textsc{Flowr.root}$^{\text{base}}$ achieves a mean PoseBusters-validity of $0.97\pm0.22$ and a mean strain energy of $67.13\pm53.05$\,kcal/mol, both of which are substantially lower than all competing models, including \textsc{Flowr} (mean PoseBusters-validity of $0.92\pm0.22$ and mean strain energy of $87.83\pm74.30$\,kcal/mol) as well as the next best model \textsc{Pilot} (mean PoseBusters-validity of $0.83\pm0.33$ and strain energy of $110.48\pm87.47$\,kcal/mol), and the highest mean AutoDock-Vina score of $-7.76\pm0.55$\,kcal/mol (vs. $-6.29\pm1.56$ and $-5.73\pm1.72$\,kcal/mol). Complete benchmark results and comparisons are provided in Supplementary Table 2.

\subsection*{Pocket-Conditional Ligand Generation: \textsc{Spindr}}

\begin{table}[!htbp]
\caption{\textbf{Evaluation and comparison of~\textsc{Flowr.root} with ~\textsc{Pilot} and~\textsc{Flowr} on~\textsc{Spindr}.} Benchmark comparison of the non-pretrained~\textsc{Flowr.root}$^{\text{base}}$ model against~\textsc{Flowr} and the diffusion-based~\textsc{Pilot} model on the~\textsc{Spindr} test dataset. We sample n = 100 ligands per target across n = 225 \textsc{Spindr} test-set targets. The evaluation includes PoseBusters-validity (PB-valid), strain energy calculated using GenBench3D, and AutoDock-Vina scores (kcal/mol). Additionally, we report the Wasserstein distance of the generated ligands' bond angles (BondA.W1), bond lengths (BondL.W1) and dihedral angles (DihedralW1) distributions relative to those in the~\textsc{Spindr} test set. Ligand sizes for all models are sampled uniformly with a -10\%/+10\% margin around the respective reference ligand size. Values denote the mean across ligands and targets; subscripts denote the standard deviation across targets.}
\label{tab:spindr_results}
\centering
\begin{adjustbox}{width=1.0\textwidth,center}
\begin{sc}
\begin{tabular}{l|c|c|c|c|c|c|c}
\toprule
Model & PB-valid $\uparrow$& Strain energy $\downarrow$ & Vina score $\downarrow$ & Vina score$^{\text{min}}$ $\downarrow$ & BondA.W1 $\downarrow$ & BondL.W1 [$10^{-2}$] $\downarrow$ & DihedralW1 $\downarrow$ \\
\toprule
\textsc{Pilot} & 0.79 {\tiny$\pm$.21} & 120.10 {\tiny$\pm$71.61} & -6.30 {\tiny$\pm$.96} & -6.68 {\tiny$\pm$1.07} & 1.82 & 0.42 & 5.52 \\
\midrule
\textsc{Flowr} & 0.93 {\tiny$\pm$.22} & 90.05 {\tiny$\pm$52.18} & -6.93 {\tiny$\pm$.92} & -7.22 {\tiny$\pm$.92} & 1.08 & 0.35 & 3.88 \\
\textsc{Flowr.root}$^{\text{base}}$ & 0.97 {\tiny$\pm$.10} & 50.36 {\tiny$\pm$34.59} & -7.52 {\tiny$\pm$.84} & -7.71 {\tiny$\pm$.85} & 0.60 & 0.43 & 3.62 \\
\toprule
Test set & 0.99 {\tiny$\pm$.04} & 43.27 {\tiny$\pm$41.85} & -7.69 {\tiny$\pm$2.00} & -7.88 {\tiny$\pm$2.00} & - & - & -\\
\bottomrule
\end{tabular}
\end{sc}
\end{adjustbox}
\end{table}

Next, we evaluate \textsc{Flowr.root}'s ability to generate valid, physically plausible ligands, but on the more challenging test set of the \textsc{Spindr} dataset. We compare the non-pretrained \textsc{Flowr.root}$^{\text{base}}$ model with \textsc{Flowr} and the diffusion-based \textsc{Pilot} model. Table~\ref{tab:spindr_results} reveals that \textsc{Flowr.root}$^{\text{base}}$ substantially outperforms both \textsc{Flowr} and \textsc{Pilot} across all metrics.
\textsc{Flowr.root}$^{\text{base}}$ achieves a mean PoseBusters-validity of $0.97\pm0.10$, nearing the test set reference ($0.99\pm0.04$). Additionally, the strain energy statistics substantially decreases to $50.36\pm34.59$ kcal/mol, further demonstrating the model's ability to generate chemically valid and energetically favorable ligands. \textsc{Flowr.root}$^{\text{base}}$ also performs better in docking accuracy, with a mean Vina score of $-7.52\pm0.84$ kcal/mol.

In addition, we analyzed a wide range of complementary metrics (Supplementary Table 3-6). \textsc{Flowr.root}$^{\text{base}}$ achieves 100\% novelty ($1.00\pm0.00$) with high diversity ($0.83\pm0.10$) and uniqueness ($0.89\pm0.18$) rates, comparable to \textsc{Flowr}, while demonstrating consistent and substantial performance improvements across fragment-conditional generation modes, including scaffold hopping and pharmacophore-conditional generation.

\subsection*{Affinity prediction: \textsc{HiQBind}}
\begin{figure}[!htbp]
    \centering
    \includegraphics[width=1.0\textwidth]{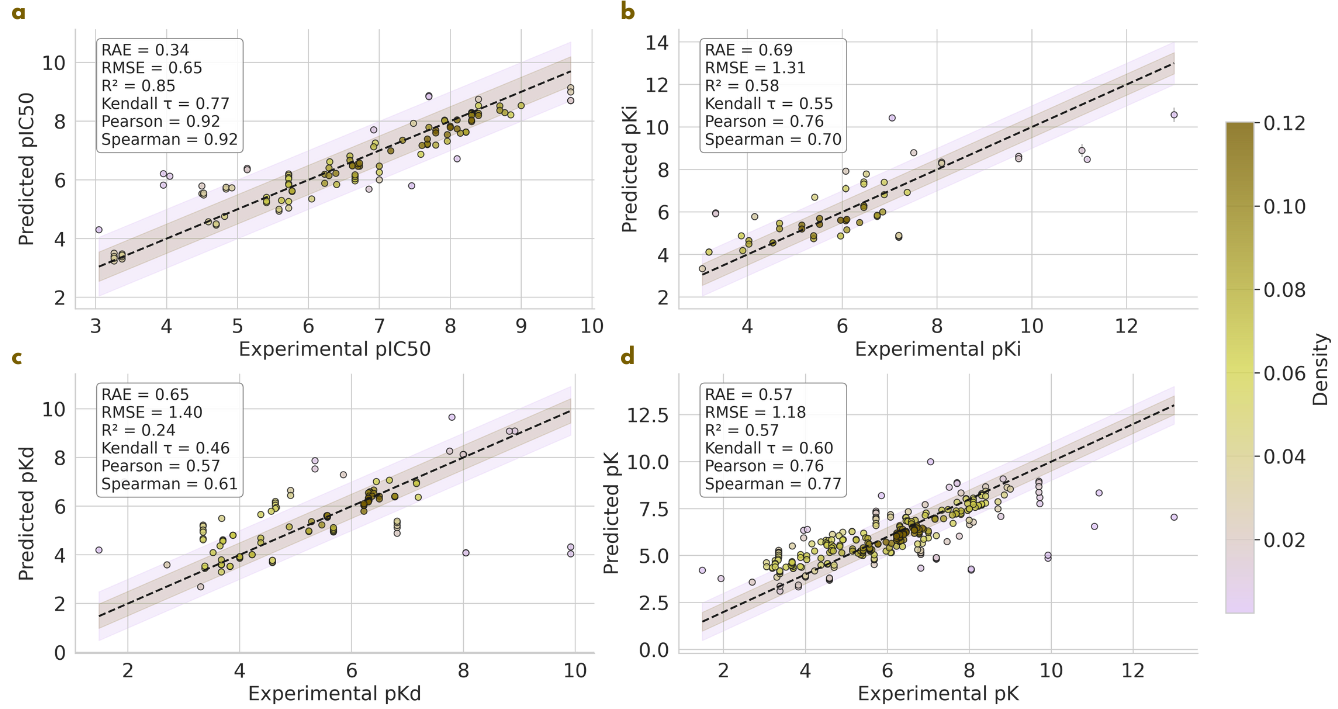}
    \caption{
    \textbf{Affinity prediction performance of \textsc{Flowr.root} on the \textsc{HiQBind} test dataset}
    Correlation plot of~\textsc{Flowr.root}-predicted $\mathrm{pIC}_{50}$ vs. experimental $\mathrm{pIC}_{50}$ binding affinities across protein-ligand complexes on the~\textsc{HiQBind} test set (based on Plinder splits; n = 278). \textbf{a} Correlation with experimental $\mathrm{pIC}_{50}$ affinities (n = 116). \textbf{b} Correlation with experimental $\mathrm{p}K_{\mathrm{i}}$ affinities (n = 54). \textbf{c} Correlation with experimental $\mathrm{p}K_{\mathrm{d}}$ affinities (n = 108), and \textbf{d} shows the correlation results if the median of all predicted affinities is used (n = 278). The numeric metrics (RAE, RMSE, R$^2$, Kendall~$\tau$, Pearson, Spearman) reported in each panel inset are means across n = 4 random-seed runs; standard deviations across seeds are depicted as vertical error bars per point. Shaded bands depict fixed reference envelopes at $\pm 0.5$ and $\pm 1$\,kcal/mol from the diagonal (not statistical error bands); colour encodes the local density of predictions (darker = denser). Each plotted point (mean value across seed runs for panels \textbf{a--c} and median value for panel \textbf{d}) corresponds to a single protein--ligand complex. Source data are provided as a Source Data file.}
    \label{fig:flowr.root_spindr}
\end{figure}

\begin{figure}[!htbp]
    \centering
    \includegraphics[width=1.0\textwidth]{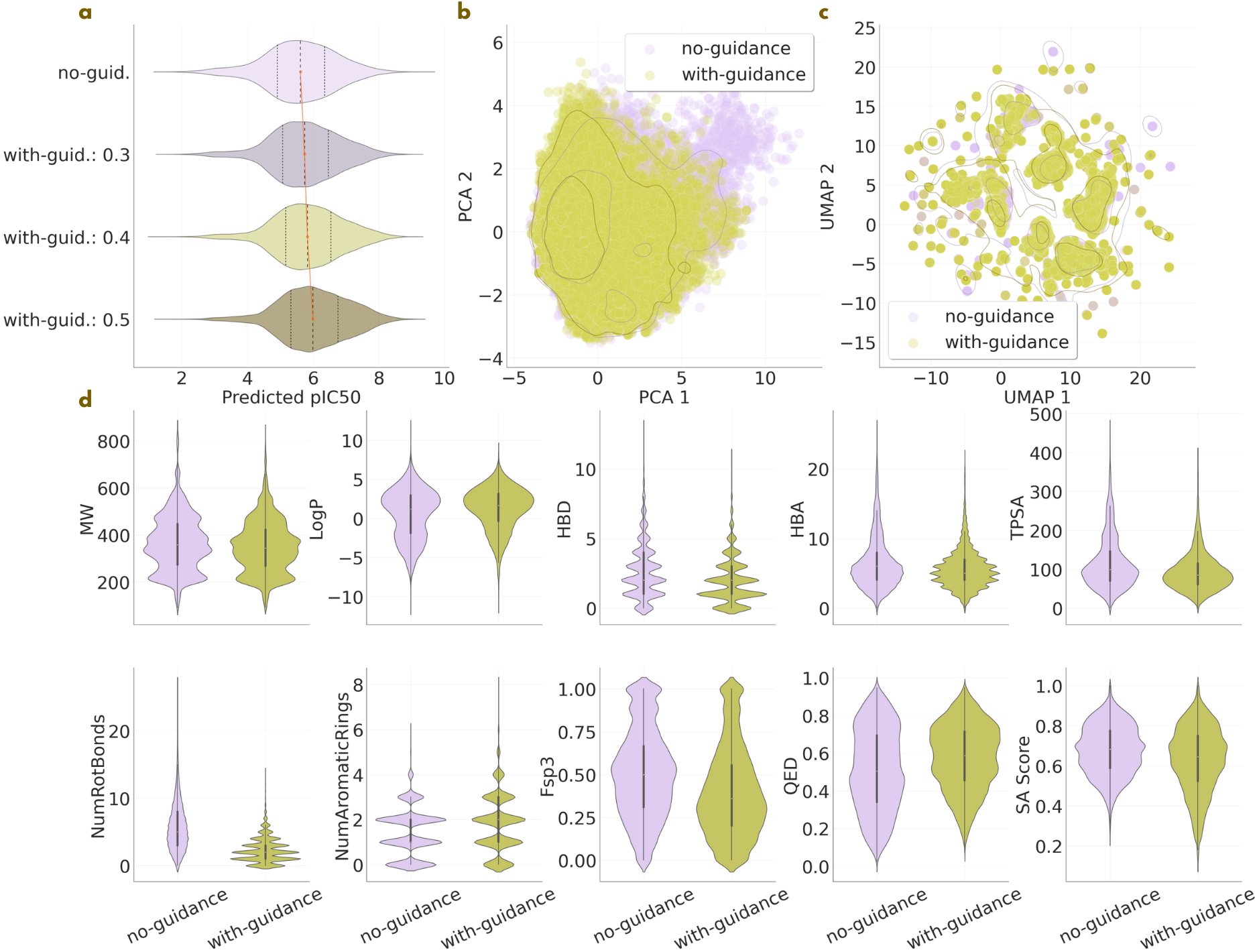}
    \caption{\textbf{Evaluation of \textsc{Flowr.root} on inference-time steering performance using affinity-guided importance sampling on the \textsc{HiQBind} test dataset.} \textbf{a} Comparing the distribution of $\mathrm{pIC}_{50}$ predictions of generated ligands across test set targets between un-guided, and mild to strongly guided steering (0.3-0.5 guidance strength). \textbf{b} Depiction of chemical space comparison between un-guided and guided samples (0.5) via PCA analysis showing first two principal components. \textbf{c} Depiction of chemical space comparison between un-guided and guided samples (0.5) via 2D UMAP analysis. \textbf{d} Distribution comparison between un-guided and guided samples (0.5) regarding different chemical properties, namely molecular weight (MW), logP, number of hydrogen donors (HBD) and acceptors (HBA), topological surface area (TPSA), number of rotatable bonds (NumRotBonds) and aromatic rings (NumAromaticRings), fraction of $sp^3$ carbons (Fsp3), druglikeness (QED) and synthesizability (SA Score). All reported distributions are descriptive. Violin plots in panels~\textbf{a} and~\textbf{d} aggregate n = 100 ligands sampled per target across n = 278 \textsc{HiQBind} test-set targets. Within each violin the white dot marks the median, the thick central bar the interquartile range (25th--75th percentile), the thin line extends to $1.5\times$ IQR (whiskers), and the outer envelope spans the full data range. Source data are provided as a Source Data file.}
    \label{fig:spindr_guidance}
\end{figure}

A central part of \textsc{Flowr.root} is its joint structure–affinity modeling. In Fig.~\ref{fig:flowr.root_spindr} we provide a detailed analysis of affinity prediction performance on the~\textsc{HiQBind} test set. The affinity head of \textsc{Flowr.root} provides accurate predictions across various affinity types, including $\mathrm{pIC}_{50}$, $\mathrm{p}K_{\mathrm{i}}$, and $\mathrm{p}K_{\mathrm{d}}$. For $\mathrm{pIC}_{50}$, the model achieves a Pearson correlation of $0.92\pm0.03$ and an R$^2$ of $0.85\pm0.06$, while for $\mathrm{p}K_{\mathrm{i}}$ and $\mathrm{p}K_{\mathrm{d}}$ we get Pearson correlations of $0.76\pm0.13$ and $0.57\pm0.20$, respectively. When aggregating predictions across all affinity types (median), the model achieves RAE $0.58\pm0.06$, RMSE $1.18\pm0.18$, R$^2$ $0.57\pm0.10$, Kendall $\tau$ $0.60\pm0.06$, and Pearson $0.76\pm0.07$. Note, reported values are mean values across four seed runs $\pm$ the mean 95\% confidence interval (CI) across seeds. These results confirm that the affinity head, trained jointly with structure generation, provides accurate and reliable potency estimates with robust performance across different experimental affinity labels.
Probing the \textsc{Flowr.root}$^{\text{base}}$ model trained solely on \textsc{HiQBind} without pre-training, we observe substantially worse performance across all metrics with RAE $0.70\pm0.13$, RMSE $1.95\pm0.24$, R$^2$ $0.39\pm0.13$, Kendall $\tau$ $0.49\pm0.09$, and Pearson $0.63\pm0.10$ underlining the importance of the proposed large-scale pre-training pipeline.
\rebuttal{Note, a structure-only ablation trained under identical conditions matches the joint model on HiQBind structural metrics within seed-level noise: mean PoseBusters-validity 0.94 vs. 0.93 with a seed-level standard deviation of each 0.02 across three seeds. We find no evidence that the joint structure–affinity objective significantly degrades structural quality.}
Note, in both cases we strictly follow the Plinder dataset split minimizing information leakage between train and test also when pre-training.

Beyond predictive accuracy, \textsc{Flowr.root} enables property-guided generation via inference-time steering. Fig.~\ref{fig:spindr_guidance} illustrates the effect of importance sampling-based steering on the predicted affinity distribution of generated ligands on the~\textsc{HiQBind} test set. Without guidance, the mean predicted $\mathrm{pIC}_{50}$ is $5.60\pm1.11$. As the steering duration increases (from $0.3$ to $0.5$ of the trajectory), the mean predicted $\mathrm{pIC}_{50}$ shifts progressively to higher values ($5.75$, $5.84$, $6.02$), while the standard deviation remains stable. This demonstrates that the model can be effectively biased toward predicted higher-affinity ligands during generation.

In Fig.~\ref{fig:spindr_guidance}, we show the distribution of important chemical properties comparing unguided with guided ($0.5$ steering duration) samples across targets on the ~\textsc{HiQBind} test set. While the distribution of SA scores shows a slight decrease towards lower scores for guided samples, we find that the overall druglikeness increases substantially with a mean Lipinski score of 4.30$\pm1.04$ for the unguided to 4.68$\pm0.65$ for the guided samples. Interestingly, we find that guidance consolidates the sample space, leading to mostly more compact distributions. This can also be seen in the chemical space comparison in Fig.~\ref{fig:spindr_guidance}\textbf{b--c}, where we overlay the chemical spaces of the unguided and guided samples using both PCA and UMAP on molecular fingerprints.

Lastly, we also analyze geometric validity and sample diversity as steering intensity increases. For unguided versus guided samples, we observe a mean PoseBusters-validity of 0.93$\pm0.20$ versus 0.91$\pm0.22$, mean strain energy of 48.46$\pm32.29$ versus 54.42$\pm36.17$\,kcal/mol, mean Vina score of -7.41$\pm0.86$ vs. -7.50$\pm0.85$, and mean per-target diversity of 0.79$\pm0.09$ versus 0.79$\pm0.10$. While we observe a slight decrease in PoseBusters-validity and increase in strain energies, the diversity of the samples remains the same, so we can rule out a potential mode collapse when using guidance. In addition, we see a slight decrease in Vina scores on guided samples, demonstrating that we can maintain physically realistic structures while optimizing for affinity.

\subsection*{Affinity prediction: \textsc{Schrodinger FEP+ and OpenFE}}

\begin{figure}[!htbp]
    \centering
    \includegraphics[width=1.0\textwidth]{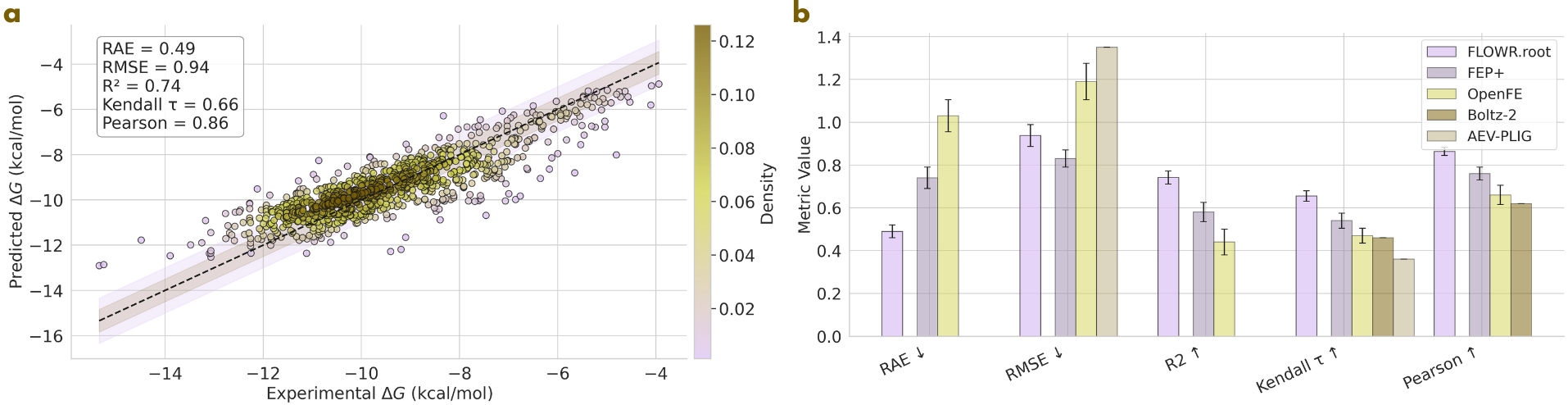}
    \caption{\textbf{Affinity prediction performance of \textsc{Flowr.root} on the FEP+/OpenFE IndustryBenchmark dataset.} \textbf{a} Correlation plot of~\textsc{Flowr.root}-predicted binding affinities as median over $\mathrm{p}K_{\mathrm{d}}$, $\mathrm{p}K_{\mathrm{i}}$ and $\mathrm{pIC}_{50}$ in kcal/mol vs. experimental binding affinities across protein-ligand complexes (n = 881). The numeric metrics (RAE, RMSE, R$^2$, Kendall~$\tau$, Pearson) reported as inset are means across n = 4 random-seed runs; standard deviations across seeds are depicted as vertical error bars per point. Shaded bands depict fixed reference envelopes at $\pm 0.5$ and $\pm 1$\,kcal/mol from the diagonal (not statistical error bands); colour encodes the local density of predictions (darker = denser). \textbf{b} Mean values of different correlation metrics with error bars indicating the 95\% confidence interval comparing \textsc{Flowr.root} with FEP+, OpenFE, AEV-PLIG and Boltz-2; \textsc{Flowr.root} values are means across n = 4 random-seed runs and error bars denote the mean 95\% confidence interval across seeds; FEP+ and OpenFE values and 95\% confidence intervals are reproduced from~\cite{openfe_results}; Boltz-2 and AEV-PLIG provide single point estimates without replicate runs; for FEP+, OpenFE, AEV-PLIG and Boltz-2 the bars represent single point estimates as reported in the cited literature. Source data are provided as a Source Data file.
    }
    \label{fig:flowr_schrodinger}
\end{figure}

Next, we evaluate the performance of the affinity prediction module of \textsc{Flowr.root} using the Schrodinger FEP+/OpenFE IndustryBenchmark datasets~\cite{openfe}, comparing \textsc{Flowr.root} to state-of-the-art methods. These benchmarks offer a robust test of the model’s prediction capabilities with experimentally determined binding affinities across various targets and affinity types for protein-ligand complexes.

In Fig.~\ref{fig:flowr_schrodinger}\textbf{a}, we present the scatter plot for \textsc{Flowr.root}-predicted combined affinity predictions (median over $\mathrm{p}K_{\mathrm{d}}$, $\mathrm{p}K_{\mathrm{i}}$, and $\mathrm{pIC}_{50}$) versus experimental binding affinities. The plot indicates strong correlation with minimal deviation, suggesting that \textsc{Flowr.root} can successfully predict binding affinities across a broad range of protein-ligand complexes. In Fig.~\ref{fig:flowr_schrodinger}\textbf{b}, we show the performance of \textsc{Flowr.root} using various error and correlation metrics in comparison to OpenFE, FEP+, AEV-PLIG and Boltz-2.
\textsc{Flowr.root} achieves an RMSE of $0.93\pm0.05$ kcal/mol, with a Pearson correlation of $0.86\pm0.02$ and a Kendall \(\tau\) of $0.65\pm0.03$. These results surpass those of FEP+ (RMSE $= 0.83\pm0.04$, Pearson $= 0.76\pm0.03$) and OpenFE (RMSE $= 1.19\pm0.08$, Pearson $= 0.66\pm0.04$), as well as those of Boltz-2 (Kendall \(\tau\) $= 0.46$, Pearson $= 0.62$) and AEV-PLIG (RMSE $= 1.35$, Kendall \(\tau\) $= 0.36$). Note, reported values for \textsc{Flowr.root} are mean values across four seed runs $\pm$ the mean 95\% CI. For FEP+ and OpenFE, the mean $\pm$ 95\% CI is shown~\cite{openfe_results}, while Boltz-2 and AEV-PLIG do not provide confidence intervals.

\textsc{Flowr.root} outperforms physics-based models, namely OpenFE and FEP+, on almost all error and correlation metrics and substantially outperforms recent state-of-the-art ML-based models AEV-PLIG and Boltz-2 across all metrics. Notably, \textsc{Flowr.root} achieves this while being 3x faster than AEV-PLIG, 200x faster than Boltz-2, and over 10000x faster than OpenFE and FEP+.
However, we also want to emphasize that the Schrodinger FEP+/OpenFE dataset is likely substantially overlapping with the training data in terms of ligand and target space. While this applies to AEV-PLIG and Boltz-2 as well allowing for a somewhat fair comparison, we do not expect these results to indicate substantial generalization capabilities.

\subsection*{Domain Adaptation via Finetuning: In-House Data}
\begin{figure}[htp!]
    \centering
    \includegraphics[width=1.0\columnwidth]{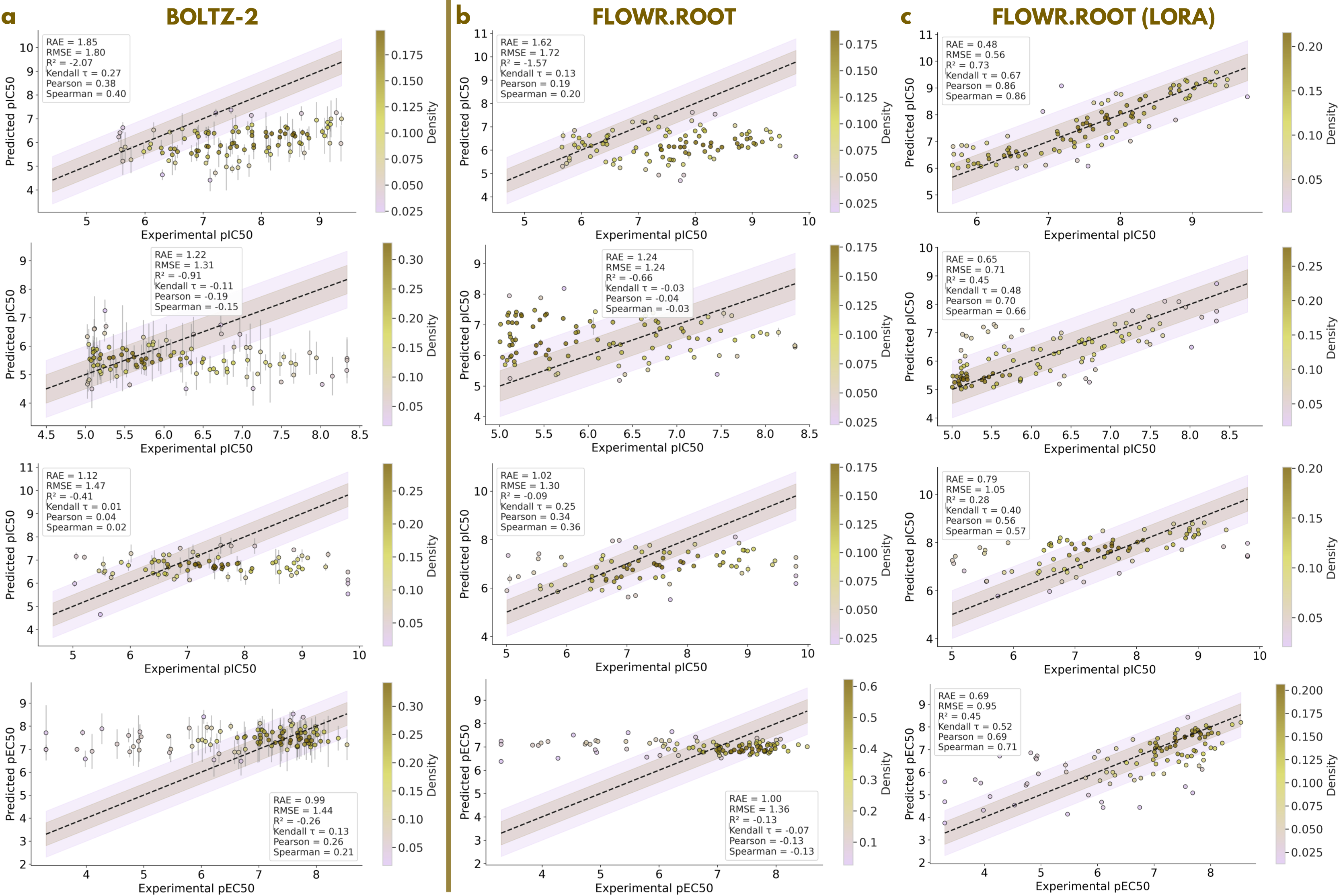}
    \caption{\textbf{Evaluation of the affinity prediction performance of \textsc{Flowr.root} on a diverse set of unseen in-house structure---activity datasets with and without LoRA-finetuning.}  across protein-ligand complexes per project. Correlation plots of~\textsc{Boltz-2}- and \textsc{Flowr.root}-predicted affinities vs. experimental measurements across protein-ligand complexes on four different projects. Each dot denotes the mean across n = 2 (Boltz-2) or n = 4 (\textsc{Flowr.root}) random-seed runs; error bars denote the standard deviation across seeds. We evaluate and compare Boltz-2, \textsc{Flowr.root} and LoRA-finetuned \textsc{Flowr.root} on root absolute error (RAE), root mean squared error (RMSE), R-squared (R$^{2}$), Kendall $\tau$, Pearson and Spearman correlation. Each row corresponds to a project (Projects 1--4), spanning distinct protein targets, therapeutic areas, ligand spaces (from small molecules to degraders), and structure-activity relationship (SAR) landscapes. Evaluation performed on held-out test sets: n = 100 complexes for Projects 1, 2, 4 and n = 80 complexes for Project 3. Shaded bands depict fixed reference envelopes at $\pm 0.5$ and $\pm 1$\,kcal/mol from the diagonal (not statistical error bands); colour encodes the local density of predictions (darker = denser). \textbf{a} Boltz-2 predictions. \textbf{b} Zero-shot \textsc{Flowr.root} predictions. \textbf{c} LoRA-finetuned \textsc{Flowr.root} predictions. Source data are provided as a Source Data file.}
    \label{fig:in_house}
\end{figure}

Next, we investigate the performance of \textsc{Flowr.root} on proprietary project-specific data. We evaluate four diverse pharmaceutical projects spanning inflammatory and neurodegenerative disease, to oncology, each encompassing distinct protein targets and ligand chemical spaces ranging from conventional small molecules to molecular degraders. Dataset sizes vary across projects: Project 1 comprises 1,073 compounds, Project 2 contains 730 compounds, Project 3 includes 229 compounds, and Project 4 contains 618 compounds. Projects 1--3 are annotated with $\mathrm{pIC}_{50}$ values, while Project 4 uses $\mathrm{pEC}_{50}$ measurements.

Pharmaceutical projects typically generate substantially more bioactivity data than co-crystal structures. To leverage existing assay data, we developed a structure generation workflow using OpenEye's docking tools~\cite{openeye:tools}. Measured compounds were first matched to existing co-crystal ligands via maximum common substructure (MCS) alignment and assigned to protein structures with the highest MCS similarity. Depending on the TanimotoCombo similarity between the query molecule and reference ligand (threshold TC $\geq 1.5$, computed via ROCS overlays~\cite{openeye:rocs} on up to 800 Omega-generated conformers~\cite{openeye:omega}), we employed either Hybrid or ShapeFit docking. Docked poses were subsequently minimized using OpenEye's Szybki with the ff14sb force field~\cite{maier2015ff14sb,openeye:szybki}, and poses with positive interaction energies were discarded. The top pose was selected based on RMSD to the reference ligand.

We evaluate both Boltz-2 and \textsc{Flowr.root} in zero-shot settings on held-out test sets comprising 100 randomly selected complexes per project (80 for Project 3 due to its smaller size), with the remaining samples used for training and validation. As shown in Fig.~\ref{fig:in_house}\textbf{a--b}, both models fail to generalize to these unseen bioactivity spaces, yielding negligible or negative correlations between predicted and experimental potencies. Specifically, Boltz-2 yields strongly negative $R^2$ across all projects; for \textsc{Flowr.root}, $R^2$ values are $-1.58\pm0.64$ (Project 1), $-0.67\pm0.45$ (Project 2), $-0.10\pm0.26$ (Project 3), and $-0.13\pm0.12$ (Project 4). These results highlight a fundamental limitation: zero-shot generalization across arbitrary structure--activity landscapes is unrealistic given the inherent complexity of structure--activity relationships.

However, \textsc{Flowr.root} distinguishes itself through its capacity for straightforward yet effective finetuning, enabling adaptation to the underlying data distribution and improved modeling of the structure--activity landscape.
We finetuned the model using Low-Rank Adaptation (LoRA)~\cite{hu2022lora}, where weight updates are parameterized as $\Delta W = BA$ with $B \in \mathbb{R}^{d \times r}$ and $A \in \mathbb{R}^{r \times k}$, using rank $r=16$ and scaling factor $\alpha=32$. This configuration yields approximately 3.3M trainable parameters, representing approximately 9\% of the full model weights. Models were trained for 200 epochs with validation on 50 held-out samples. Since our proprietary datasets comprise $\mathrm{pIC}_{50}$- and $\mathrm{pEC}_{50}$-annotated data points, only the respective affinity head and the backbone were finetuned.

As shown in Fig.~\ref{fig:in_house}\textbf{c}, LoRA-finetuned \textsc{Flowr.root} achieves strong performance across all four projects: Project 1 (R$^2$ = $0.73\pm0.11$, Pearson $r$ = $0.86\pm0.05$), Project 2 (R$^2$ = $0.45\pm0.21$, Pearson $r$ = $0.70\pm0.12$), Project 3 (R$^2$ = $0.28\pm0.16$, Pearson $r$ = $0.56\pm0.13$), and Project 4 (R$^2$ = $0.45\pm0.20$, Pearson $r$ = $0.69\pm0.11$). All reported values denote mean values across four seed runs $\pm$ the mean 95\% CI.

These findings underscore a critical observation: while achieving broad coverage of chemical space through ligand generation presents manageable challenges (see Supplementary Fig.~7), generalizing across structure--activity landscapes constitutes a fundamentally more complex problem. The consistent success of LoRA-finetuned \textsc{Flowr.root} across four therapeutically and chemically distinct projects, and both $\mathrm{IC}_{50}$ and $\mathrm{EC}_{50}$ endpoints—combined with the failure of both Boltz-2 and \textsc{Flowr.root} to generalize despite strong public benchmark performance, supports our position that SBDD models should function as adaptive tools requiring calibration to project-specific SAR patterns.
Thus, rather than expecting models to generalize universally from a single training phase, continuous refinement should be regarded as an integral component of the discovery process. Model utility grows through sustained interaction with project-specific data, positioning generative models as adaptive companions rather than standalone tools.

\subsection*{Domain Adaptation via Finetuning: PDE10A Benchmark}

\begin{table}[!htbp]
\caption{\textbf{Evaluation of the affinity prediction performance of \textsc{Flowr.root} with and without LoRA-finetuning on PDE10A.} Benchmark comparison on the PDE10A dataset~\cite{tosstorff2022pde10a} comprising 1,162 inhibitors with experimentally determined $\mathrm{pIC}_{50}$ values. We adapt the four split strategies: random split and three temporal splits (2011, 2012, 2013) that assess prospective prediction capability. Results compare the 2D3D hybrid method from~Tosstorff et~al.~\cite{tosstorff2022pde10a}, Boltz-2, zero-shot \textsc{Flowr.root}, and LoRA-finetuned \textsc{Flowr.root}. We report root mean squared error (RMSE) as well as Spearman correlation (Spear.) metrics following~\cite{tosstorff2022pde10a}. Subscripts denote 95\% bootstrap confidence intervals. Bold values denote the best result per metric column.}
\label{tab:pde10a_results}
\centering
\resizebox{\columnwidth}{!}{%
\begin{sc}
\begin{tabular}{l|cc|cc|cc|cc}
\toprule
& \multicolumn{2}{c|}{Random} & \multicolumn{2}{c|}{Temp. 2011} & \multicolumn{2}{c|}{Temp. 2012} & \multicolumn{2}{c}{Temp. 2013} \\
Model & RMSE & Spear. & RMSE & Spear. & RMSE & Spear. & RMSE & Spear. \\
\midrule
2D3D hybrid & 0.85{\tiny$\pm$.11} & 0.72{\tiny$\pm$.07} & 0.81{\tiny$\pm$.07} & 0.57{\tiny$\pm$.09} & 1.25{\tiny$\pm$.14} & 0.56{\tiny$\pm$.12} & 0.95{\tiny$\pm$.15} & 0.61{\tiny$\pm$.18} \\
Boltz-2 & 0.85{\tiny$\pm$.09} & 0.75{\tiny$\pm$.07} & 0.83{\tiny$\pm$.07} & 0.59{\tiny$\pm$.08} & 0.73{\tiny$\pm$.09} & 0.83{\tiny$\pm$.06} & 0.95{\tiny$\pm$.22} & 0.74{\tiny$\pm$.15} \\
\textsc{Flowr.root} & 1.14{\tiny$\pm$.11} & 0.67{\tiny$\pm$.08} & 0.94{\tiny$\pm$.10} & 0.64{\tiny$\pm$.09} & 1.03{\tiny$\pm$.12} & 0.68{\tiny$\pm$.11} & 1.03{\tiny$\pm$.22} & 0.70{\tiny$\pm$.16} \\
\textsc{Flowr.root}$^{\text{LoRA}}$ & \textbf{0.32}{\tiny$\pm$.06} & \textbf{0.97}{\tiny$\pm$.01} & \textbf{0.33}{\tiny$\pm$.06} & \textbf{0.93}{\tiny$\pm$.03} & \textbf{0.43}{\tiny$\pm$.10} & \textbf{0.93}{\tiny$\pm$.04} & \textbf{0.44}{\tiny$\pm$.16} & \textbf{0.95}{\tiny$\pm$.04} \\
\bottomrule
\end{tabular}
\end{sc}
}
\end{table}

Next, to complement our in-house validation, we evaluate on the publicly available PDE10A benchmark dataset introduced by~Tosstorff et~al.~\cite{tosstorff2022pde10a}. This high-quality dataset comprises 1,162 Phosphodiesterase-10A (PDE10A) inhibitors with experimentally determined $\mathrm{IC}_{50}$ values, curated from a former Roche PDE10A project with consistent assay conditions. The benchmark includes multiple evaluation strategies: a random split and three temporal splits (2011, 2012, 2013) that partition compounds by date, thereby simulating realistic prospective prediction scenarios encountered in drug discovery campaigns.

Table~\ref{tab:pde10a_results} presents results comparing the 2D3D hybrid QSAR method from~Tosstorff et~al.~\cite{tosstorff2022pde10a}, Boltz-2, zero-shot \textsc{Flowr.root}, and LoRA-finetuned \textsc{Flowr.root}. LoRA-finetuned \textsc{Flowr.root} consistently achieves the best performance across all splits and metrics. On the random split, LoRA finetuning yields RMSE = $0.31\pm0.06$ and Spearman = $0.96\pm0.01$, substantially outperforming Boltz-2 and the 2D3D hybrid method. Importantly, this strong performance is maintained on the more challenging temporal splits that assess prospective prediction capability: for temporal 2013, LoRA-finetuned \textsc{Flowr.root} achieves RMSE = $0.44\pm0.17$ and Spearman = $0.95\pm0.04$

Note, the \textsc{2d3d hybrid} approach proposed in~Tosstorff et~al.~\cite{tosstorff2022pde10a} is a trained model on the respective split. Further, PDE10A is a well-studied target, thus is present in public databases and likely overlaps with Boltz-2's comprehensive training data, making this an unreliable evaluation of true generalization. Nevertheless, these results further demonstrate and underline the effectiveness of parameter-efficient domain adaptation via LoRA finetuning within the \textsc{Flowr.root} framework, consistently transforming unreliable zero-shot predictions into accurate affinity estimates. In Supplementary Fig.~8 and Supplementary Fig.~9, we demonstrate the fragment growing mode of LoRA-finetuned \textsc{Flowr.root} applied to a co-crystal structure (PDB \texttt{5SF4}) from the PDE10A dataset. The generated ligands are annotated with affinity predictions to enable subsequent ranking and selection, illustrating the potential utility of our proposed framework in structure-based drug design workflows.

\subsection*{Case studies}
To demonstrate the practical utility and versatility of \textsc{Flowr.root} for structure-based drug design, we present three case studies that systematically evaluate different aspects of the framework's capabilities. First, we investigate multi-objective optimization through a kinase selectivity study, where we simultaneously maximize binding affinity for the on-target kinase CK2$\alpha$ while minimizing off-target activity against CLK3. This case study demonstrates how \textsc{Flowr.root} can address one of the most challenging aspects of kinase drug discovery—achieving selectivity among structurally homologous ATP-binding sites.

Second, we benchmark the framework's conditional ligand generation performance against quantum mechanical binding energy calculations using TYK2 kinase, ER$\alpha$ and BACE1 as model systems. These validation studies assess the correlation between \textsc{Flowr.root}-predicted binding affinities and computationally demanding QM binding energies, while providing mechanistic insights into the structural features that govern binding affinity in the generated ligands.

\subsection*{Ligand Selectivity: \textsc{CK2$\alpha$} and \textsc{CLK3}}
\begin{figure}[htp!]
    \centering
    \includegraphics[width=0.8\columnwidth]{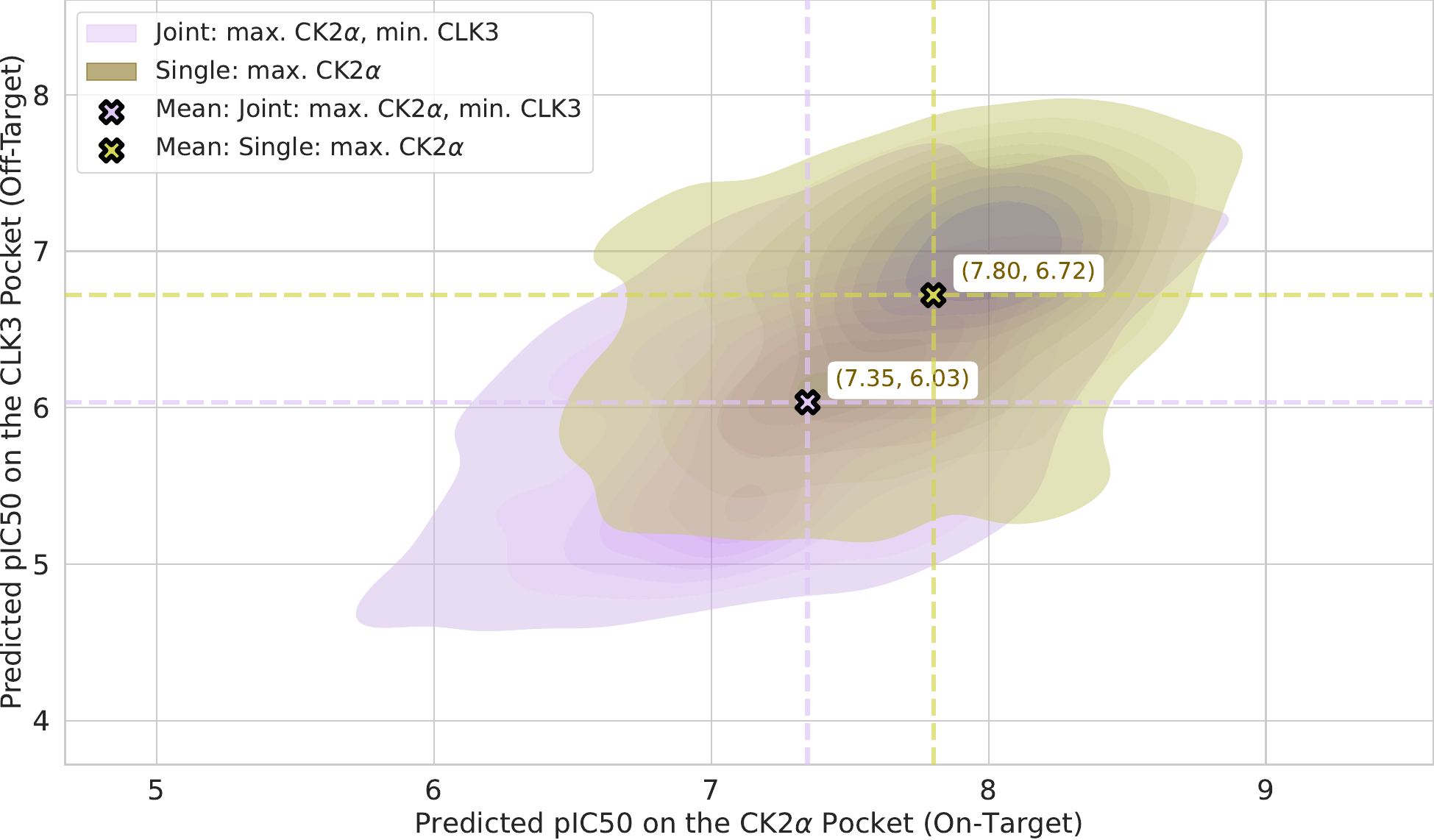}
    \caption{\textbf{Evaluation of \textsc{Flowr.root} on affinity-guided selectivity optimization of CK2$\alpha$ and CLK3}. Kernel density estimation plot comparing joint optimization (purple, maximizing CK2$\alpha$ while minimizing CLK3) versus single optimization (green, maximizing CK2$\alpha$ only) on predicted $\mathrm{pIC}_{50}$ values. Black X markers show mean values with dashed reference lines; the kernel density estimate visualises the per-strategy distribution. n = 1{,}000 ligands per optimization strategy per pocket (PDB \texttt{3PE1} for CK2$\alpha$; PDB \texttt{6KHF} for CLK3). Joint optimization achieves improved selectivity with lower off-target CLK3 activity (mean $\mathrm{pIC}_{50}$ = $6.03$ vs. $6.72$) while maintaining comparable on-target CK2$\alpha$ potency. Source data are provided as a Source Data file.}
    \label{fig:clk3-study}
\end{figure}

The human kinome consists of over 500 kinases~\cite{doi:10.1126/science.1075762} that share conserved catalytic domain structures, particularly in their ATP-binding sites, where most competitive inhibitors bind. This structural homology, while functionally important for cellular signaling pathways, presents a major challenge for kinase selectivity—the ability of small molecule inhibitors to achieve potent inhibition of specific kinase targets while avoiding off-target interactions with structurally similar kinases.
To demonstrate the capability of our framework for multi-objective optimization, we perform a selectivity study targeting selectivity between CK2$\alpha$ (on-target, PDB \texttt{3PE1}) and CLK3 (off-target, PDB \texttt{6KHF}).

We compare two optimization strategies using importance sampling: joint optimization that simultaneously maximizes the predicted binding affinity on CK2$\alpha$ while minimizing CLK3 affinity formulated as dual-objective, versus the single optimization which only maximizes CK2$\alpha$. We generated $1,000$ ligands for each set using the protein pockets from PDB \texttt{3PE1}~\cite{Battistutta2011-dh} and \texttt{6KHF}~\cite{Lee2019-fd} after rotational alignment.

Our results show that joint optimization successfully achieves improved selectivity profiles compared to single-target optimization. The joint approach generated ligands with substantially lower off-target CLK3 activity (mean $\mathrm{pIC}_{50}$ = $6.03$) compared to single optimization (mean $\mathrm{pIC}_{50}$ = $6.72$), while maintaining comparable on-target CK2$\alpha$ potency as shown in the distribution plots in Fig.~\ref{fig:clk3-study} with mean $\mathrm{pIC}_{50}$ of $7.35$ for the joint set on CK2$\alpha$.

\begin{figure}[htp!]
    \centering
    \includegraphics[width=1.0\columnwidth]{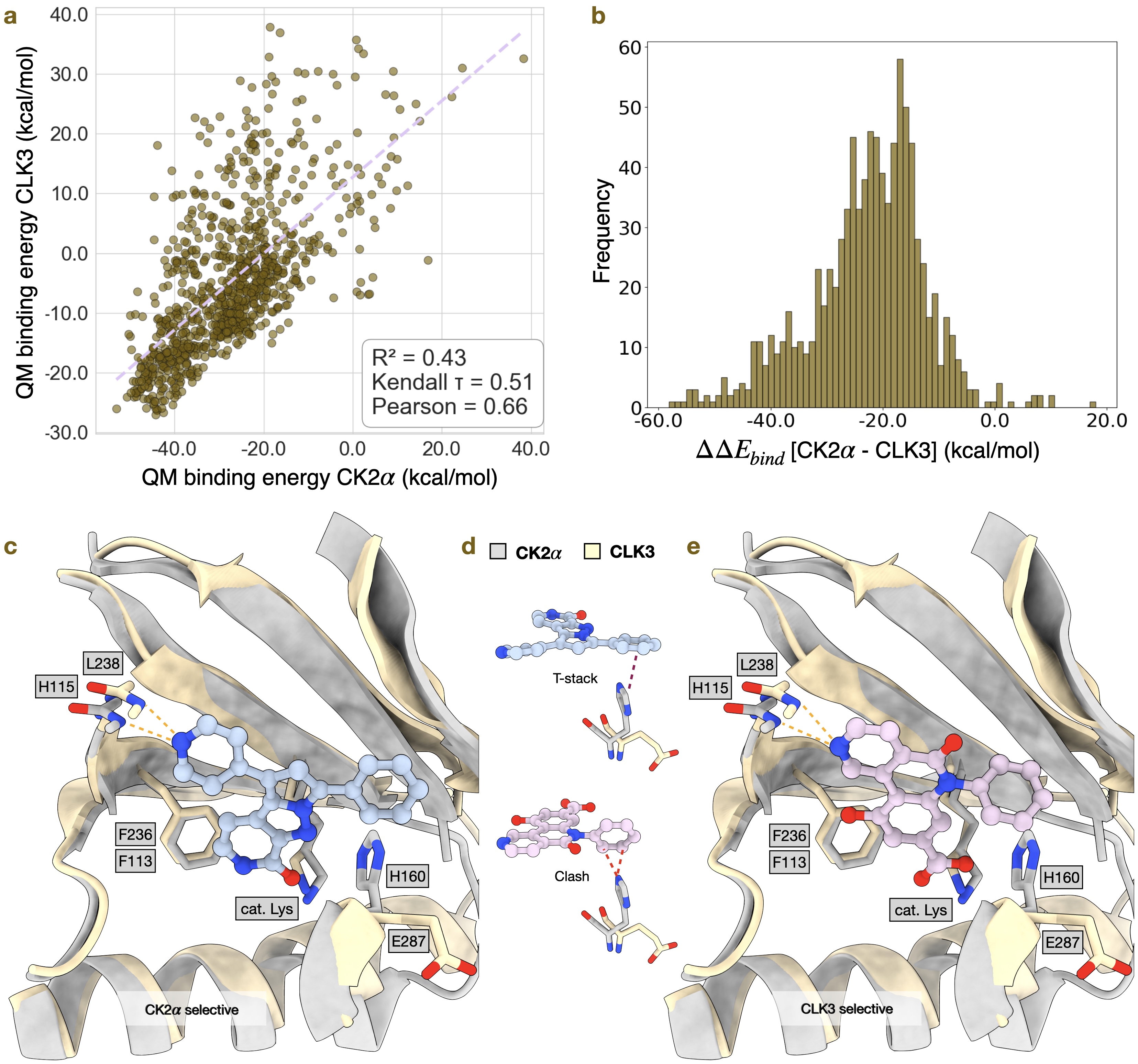}
    \caption{\textbf{Quantum mechanical analysis of \textsc{Flowr.root} on selective ligand design on CK2$\alpha$ and CLK3.} {\textbf{a}} Quantum-mechanical-calculated binding energy for CLK3 vs. CK2$\alpha$. Ligands for which the CLK3 binding energy was above $40.0$~kcal/mol were excluded ($\approx 5.6$\%, structural clashes). The plot shows systematically lower binding energies for CK2$\alpha$, indicating selectivity. \textbf{b} Distribution of the difference in binding energies for all retained ligands; the skewed distribution peaks at about $-17.0$~kcal/mol. \textbf{c} Example of a CK2$\alpha$-selective ligand and its interaction patterns with the kinases' pockets. \textbf{d} Interactions of the two ligands with the residue H160. \textbf{e} Example of a CLK3-selective ligand and its interaction patterns with the kinases' pockets. Starting from n = 1{,}000 jointly-optimised \textsc{Flowr.root}-generated ligands, n = 944 ligands enter the panel \textbf{a}/\textbf{b} statistics after the $40.0$~kcal/mol exclusion. Each ligand contributes a single GFN2-xTB binding-energy evaluation per pocket. Source data are provided as a Source Data file.}
    \label{fig:selectivity-QM-study}
\end{figure}

We further benchmark our approach using an alternative protocol. To this end, we took the $1,000$ \textsc{Flowr.root}-generated ligands from the joint optimization and performed QM binding energy calculations. We excluded all complexes for which CLK3 binding energies were above $40.0$~kcal/mol, as these resulted from structural clashes (\textit{ca.} $5.6$ \% of all complexes). Fig.~\ref{fig:selectivity-QM-study}{\bf{a}} shows a scatter plot of CK2$\alpha$ against CLK3 binding energies. The scatter shows a linearized correlation shifted towards lower CK2$\alpha$ binding energies, corroborating the success of the approach. Though other statistical metrics were calculated, we stress the calculated RMSE of 25.06 kcal/mol for this correlation. As RMSE measures deviation from the identity line, a large value implies selectivity towards one kinase, which from the data we conclude is CK2$\alpha$. Further, we record several outliers from linearity, which result from the fact that \textsc{Flowr.root} tries to further penalize ligands in the CLK3 pocket. In this task, however, the most relevant metric is the relative binding energy of a given ligand towards the pockets of each kinase. Fig.~\ref{fig:selectivity-QM-study}{\bf{b}} shows the distribution of relative binding energies in these two kinases. The distribution, peaked at about $-17.0$~kcal/mol, is skewed, with a quick decay towards higher relative binding energies (less selective) and a slower decay on lower relative binding energies (more selective) further validating our approach.

It is particularly interesting to analyze ligands that minimize and maximize the relative binding energies between the two kinases. Fig.~\ref{fig:selectivity-QM-study}{\bf{c-e}} compare a CK2$\alpha$-selective ($\Delta \Delta {E}_{bind}=-58.18$~kcal/mol) against a CLK3 binder ($\Delta \Delta {E}_{bind}=+18.00$~kcal/mol), indicating the mechanisms, with which \textsc{Flowr.root} tried to accomplish selectivity. Two major protein-ligand interactions were explored. On one hand, the interaction with the kinases' hinge. In both cases, the hydrogen bond between the generated ligand and L238 of CLK3 shows a suboptimal angular orientation. Conversely, the CK2$\alpha$-selective binder shows a quasi-optimal hydrogen bond orientation with CK2$\alpha$'s H115, with the angle $H-{\hat{N}}_{ar}-{C}_{ar}$ reaching a value of $114.2 ^{\circ}$. In the case of the CLK3-selective binder, the equivalent angle with CK2$\alpha$'s main chain H of H115 goes up to $137.9 ^{\circ}$. Additional weakening of the hydrogen bond to CK2$\alpha$'s hinge is achieved by increasing the $H-{\hat{N}}$ distance, which goes from $2.77$ {\AA} in the CK2$\alpha$-selective binder to $2.97$ {\AA} in the CLK3-selective one.

The second mechanism is related to interactions with H160. In previous work~\cite{menezes:silmitasertib}, H160 was identified as a critical residue conferring sub-nM affinity of the inhibitor silmitasertib to CK2$\alpha$. Here too, we observe that \textsc{Flowr.root} explores interactions with this residue to achieve selectivity. On the CK2$\alpha$-selective binder, \textsc{Flowr.root} orients the ligand's phenyl substituent to optimize a T-stack interaction. On the CLK3-selective binder, a similar substituent is rotated to yield a short, repulsive contact. This is possible because the side chain of E287, the residue in a similar position in CLK3, points in another direction.

We conclude that, in order to achieve selectivity between two similar kinases, \textsc{Flowr.root} exploited two previously identified elements of selectivity. This concerns the hinge region of kinases, where better binders show shorter hydrogen bond distances and more adequate angles. Simultaneously, \textsc{Flowr.root} also explored lipophilic interactions with the two kinases, optimizing a T-stack with a residue critical for the high potency of drugs currently in clinical trials. We stress that none of these interactions could have been passed to the model during training or in guiding selectivity.

\subsection*{\textsc{TYK2}, \textsc{ER$\alpha$}, and \textsc{BACE1}}

\begin{figure}[htp!]
    \centering
    \includegraphics[width=1.0\columnwidth]{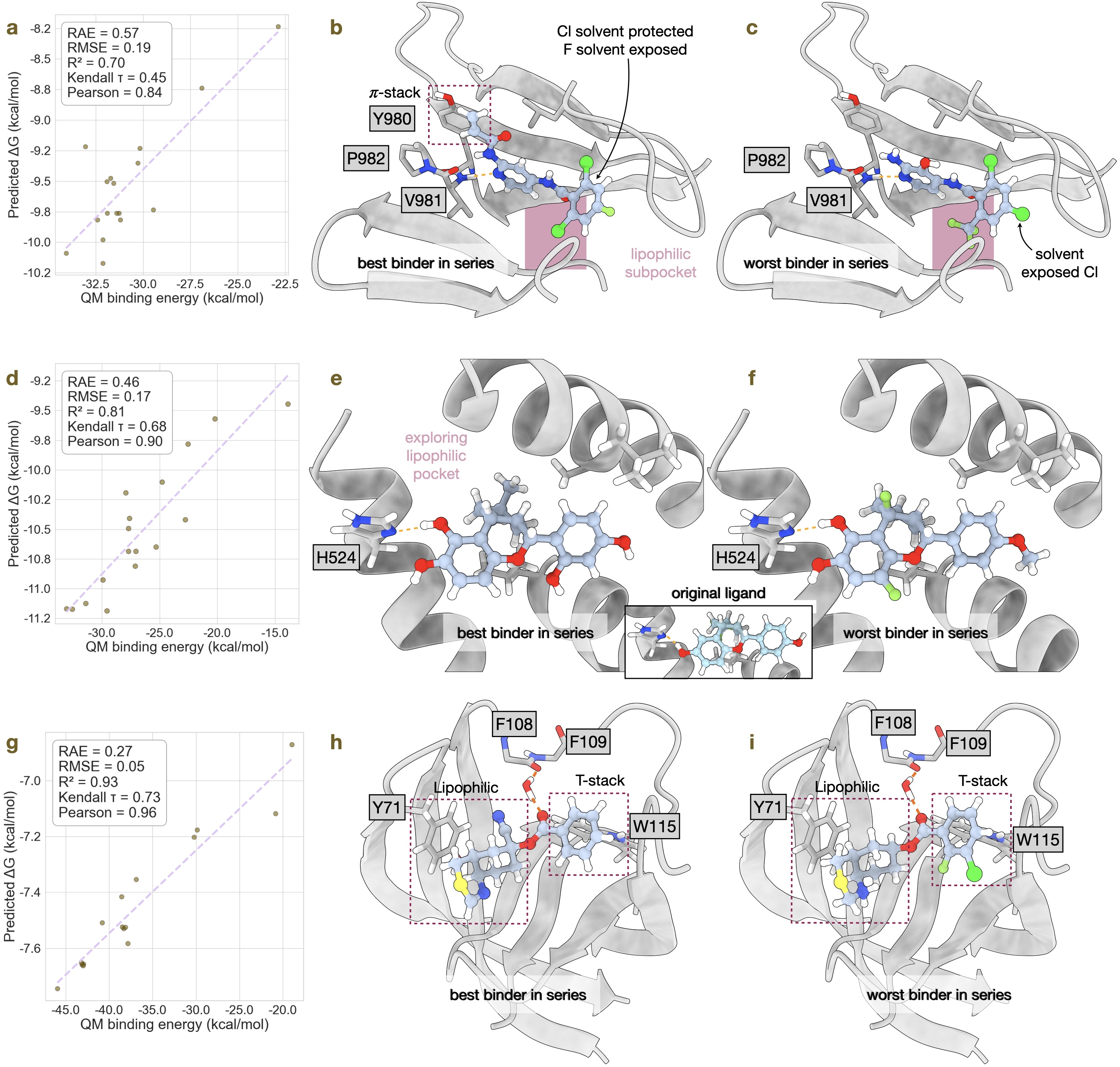}
    \caption{\textbf{\textsc{Flowr.root} binding affinity validation using quantum mechanical calculations on TYK2, ER$\alpha$ and BACE1.} Benchmark cases of the TYK2 kinase ({\bf{a-c}}), ER$\alpha$ ({\bf{d-f}}), and \textsc{BACE1} ({\bf{g-i}}). {\bf{a}} correlation plot between \textsc{Flowr.root} predicted affinities and the quantum mechanical binding energies for TYK2. {\bf{b}} example of the best binder in the series and its interactions with the pocket. {\bf{c}} example of the worst binder in the series and its interactions with the pocket.
    {\bf{d}} correlation plot between \textsc{Flowr.root} predicted affinities and the quantum mechanical binding energies for ER$\alpha$. {\bf{e}} example of the best binder in the series and its interactions with the pocket. {\bf{f}} example of the worst binder in the series and its interactions with the pocket. {\bf{g}} correlation plot between \textsc{Flowr.root} predicted affinities and the respective quantum mechanical binding energies. {\bf{h}} example of the best binder in the series and its interactions with the pocket. {\bf{i}} example of the worst binder in the series and its interactions with the pocket. Each correlation panel ({\bf a},{\bf d},{\bf g}) uses n = 17 \textsc{Flowr.root}-generated ligands per target. Each ligand contributes one GFN2-xTB binding-energy evaluation and one \textsc{Flowr.root} affinity prediction; reported R$^{2}$, Pearson, Kendall $\tau$, RMSE and RAE are point estimates over the full per-target ligand set. Reference complexes: BACE1 PDB \texttt{4ZSP}, TYK2 PDB \texttt{4GIH}, ER$\alpha$ PDB \texttt{2Q70}. Source data are provided as a Source Data file.
    }
    \label{fig:tyk2-era-study}
\end{figure}

As shown in previous sections, \textsc{Flowr.root} generates diverse chemical matter conditioned on a protein pocket. The generated molecules are chemically reasonable and are not heavily strained, unlike previous generative models. Additionally, the predicted affinity data reliably reproduces experimental distributions. In this section, we benchmark how the whole \textsc{Flowr.root} workflow behaves on the task of conditional ligand generation. To this end, we chose the \textsc{TYK2} kinase, \textsc{ER$\alpha$}, and \textsc{BACE1} as targets from the Schrodinger FEP+ dataset.

Testing whether the \textsc{Flowr.root}-generated ligands are good binders requires effective metrics. In this case, we chose to benchmark the \textsc{Flowr.root} poses using QM binding energy analysis. QM calculations on protein-ligand complexes offer scoring functions; therefore we aimed at finding a correlation between QM binding energies and the median of predicted binding affinities.

We first look at the example of the \textsc{TYK2} kinase (Fig.~\ref{fig:tyk2-era-study}{\bf{a-c}}). We find a good correlation between the QM-calculated binding energies and the \textsc{Flowr.root}-predicted median affinities, with an R$^2$ of $0.70$, a Pearson correlation coefficient of $0.84$, a Kendall $\tau$ of $0.45$, RMSE of $0.19$, and RAE of $0.57$. From a chemical viewpoint, it is more interesting to compare the structures of the best and worst \textsc{Flowr.root}-predicted ligands. We see that, despite some common features, the model generalizes effectively the hinge-binding motif, and the solvent-exposed residue of the ligand. Importantly, all ligands retained the hydrogen bond acceptor to V981, a critical hinge residue already in the original ligand structure. Binding to a kinase's hinge is very important in kinase inhibition, kept by \textsc{Flowr.root}-generated chemical matter. Interestingly, no ligand improved the hydrogen-bond geometry with P982, despite some variations in the chemotypes exhibited by the hinge-binding heads. The major differences take place in how specific interactions are exploited. Particularly critical for the best binder in this series is the $\pi$-stack interaction established with Y980. But also the fact that this ligand protects the least lipophilic halogens from solvent exposure. We also calculated ligand deformation energies, which for this series lie between $19.32$~kcal/mol and $44.29$~kcal/mol.

A similar correlation was obtained for \textsc{ER$\alpha$} (Fig.~\ref{fig:tyk2-era-study}{\bf{d-f}}), however, in this case, all the statistical metrics improved: The R$^2$ is $0.81$, the Pearson correlation coefficient is $0.90$, the Kendall $\tau$ is $0.68$, RMSE is $0.17$ and RAE $0.46$. Some variability is expected between systems, and in the case of \textsc{ER$\alpha$} we observe less outliers than in the \textsc{TYK2} example. In the case of \textsc{ER$\alpha$}, we observed that \textsc{Flowr.root} retained most of the ligand's structure, improving mostly the interactions with the residue H524. In the structure of the original ligand there is a poorly-oriented hydrogen bond with H524, which is effectively corrected even in the case of the worst binder: the \textsc{Flowr.root}-generated ligands exhibited better geometries for this hydrogen bond. Similar to the case of the \textsc{TYK2} kinase, we calculated ligand deformation energies, which were between $24.36$~kcal/mol and $38.30$~kcal/mol. 

Finally, we also analysed a benchmark on \textsc{BACE1} (Fig.~\ref{fig:tyk2-era-study}{\bf{g-i}}), where we obtained the best correlation of this series between \textsc{Flowr.root}-predicted affinities and the QM calculated ones. This is seen in the lowest RAE and RMSE coupled with the highest R$^2$, Pearson correlation coefficient, and Kendall $\tau$ of this series of benchmarks. We see that in all cases the ligands retain a water-mediated hydrogen bond with F108, which results in high affinities overall. The major differences are in the nature of the nitrogen-containing group near Y71, which seems to favour more lipophilic groups, but also the substituents in the aromatic ring T-stacking the residue W115. We note that the T-stack interaction is favoured by electron-rich phenyl rings, which is not the case of the worst binder in the series. This seems then to be one of the major effects controlling affinity to \textsc{BACE1}.

\section*{Discussion}
The integration of pocket-aware generation, multi-endpoint affinity prediction, and efficient domain adaptation within a single architecture addresses a central challenge in computational drug design: bridging the gap between generating geometrically realistic ligands and reliably estimating their binding properties, while remaining adaptable to project-specific structure--activity landscapes. \textsc{Flowr.root} realizes this integration through an SE(3)-equivariant flow matching framework that couples structural prediction with dedicated affinity and confidence heads, enabling property-guided molecule design through inference-time steering. By supporting \textit{de novo} design, interaction-guided generation, and fragment-based elaboration within a unified backbone, the framework spans applications from early-stage hit identification to late-stage lead optimization.

Across established benchmarks, the model achieves leading performance in both pocket-conditional ligand generation and affinity prediction, consistently outperforming recent diffusion- and flow-based approaches while producing geometrically realistic, low-strain structures. The joint affinity head matches or exceeds the accuracy of computationally expensive physics-based methods and clearly surpasses ML-based scoring functions, at orders of magnitude greater throughput. Importantly, targeted parameter-efficient finetuning enhances prediction reliability on project-specific datasets substantially, demonstrating practical utility in real-world medicinal chemistry workflows.

Quantum mechanical validation studies on TYK2 kinase, ER$\alpha$, and BACE1 provide evidence that \textsc{Flowr.root} generates chemically meaningful ligands that systematically exploit key structural motifs---hinge-binding interactions, hydrogen bond geometries, and aromatic stacking---indicating that the model has internalized fundamental principles of protein-ligand recognition. The kinase selectivity study between CK2$\alpha$ and CLK3 further illustrates that affinity-guided generation can surface selectivity mechanisms. Together, these results support the broader vision of continuously evolving frameworks that refine their understanding through sustained interaction with project-specific data, offering a foundational yet adaptable architecture for structure-based drug design.

Nevertheless, important limitations remain. The reliance on public datasets, despite extensive curation, introduces biases from noisy affinity measurements, overrepresented protein families, and uneven chemical diversity. While higher-fidelity datasets partially mitigate these issues, their limited scale constrains model calibration, particularly for underexplored targets. Project-specific adaptation, though powerful, requires carefully chosen objectives and sufficient assay data; otherwise, models risk overfitting to narrow distributions. Additionally, \textsc{Flowr.root} requires the binding pocket to be known and preferably in a holo conformation, leaving the challenge of modeling protein pocket flexibility to future work. Finally, the predictions of \textsc{Flowr.root} remain \textit{in silico} approximations that cannot substitute for experimental validation.

Looking forward, several directions appear promising. Expanding project-level adaptation to include reinforcement learning or active learning frameworks with both physics-based and experimental feedback may enable continuous refinement during discovery campaigns. Further, enhancing synthetic accessibility estimation and coupling to reaction-based generative models would improve downstream feasibility, ensuring that generated molecules are not only potent but also synthetically accessible.

\FloatBarrier
\section*{Methods}
\label{app:methods}

This section outlines the methodology employed in our framework, \textsc{Flowr.root}, for generative structure-aware ligand modeling and binding affinity prediction. We provide a detailed description of the architecture and key components, including framework details about the three-stage training process, the generative modes available for ligand design, and the affinity prediction capabilities. We also discuss the strategies for confidence estimation, inference-time scaling via importance sampling, and domain adaptation as main application in practical drug discovery workflows.

\subsection*{\textsc{Flowr.root}}

\textsc{Flowr.root} is built on the recently proposed \textsc{Flowr} architecture, an $SE(3)$-equivariant flow matching model that learns a mixed continuous/discrete transport map. This map transforms a prior distribution (for example, random noise or fragment anchors for coordinates) to the target ligand distribution within a given protein pocket~\cite{model:flowr}. The model consists of two main components: a pocket encoder and a ligand decoder.

Protein pockets are extracted from holo structures by cutting residues within a defined radius around the binding site. We use a 7{\AA} cutoff radius, which represents a practical compromise between computational tractability and interaction coverage. This cutoff fully encompasses hydrogen bonds directly relevant to ligand binding (which decay rapidly beyond 5.5--6{\AA}), captures vibrational entropy contributions from direct protein-ligand interactions, and includes conformational entropy changes in the binding pocket. While long-range electrostatic effects are system-dependent (primarily affecting charged ligands), the 7{\AA} cutoff provides sufficient coverage for neutral ligands and partial cancellation occurs for charged species. Water-mediated interactions are treated implicitly, consistent with many previous ML approaches. The cutoff is a configurable parameter that can be adjusted at both training and inference time.

These pockets are then encoded using an equivariant self-attention module followed by equivariant feed-forward layers, generating a set of invariant and equivariant protein features~\cite{model:flowr}. The ligand decoder processes noisy ligand coordinates, partial charges, atom types, bond orders, and hybridization states through equivariant self-attention modules, which capture intra-ligand dependencies. A cross-attention layer then integrates contextual information from the pocket features to the hidden ligand features. The pocket encoder processes features on a full-atom basis through $SE(3)$-equivariant self-attention layers using invariant residue embeddings, equivariant coordinate features, and pairwise fully-connected distance information between all pocket atoms. Via fully-connected, equivariant cross-attention, the ligand decoder interacts with the invariant and equivariant pocket encodings for ligand generation. For additional details, we refer to~Cremer et~al.~\cite{model:flowr}.

In \textsc{Flowr.root}, the ligand decoder has three output heads: (1) a structure head that predicts atomic coordinates, atom types, bond orders, charges, and hybridization states; (2) an affinity head comprising four separate potency and binding affinity heads---specifically, for $\mathrm{IC}_{50}$, $\mathrm{p}K_{\mathrm{i}}$, $\mathrm{p}K_{\mathrm{d}}$, and $\mathrm{pEC}_{50}$ prediction---and (3) a confidence head that provides uncertainty estimation based on pLDDT~\cite{folding:alphafold3}.
Importantly, these heads serve distinct yet complementary purposes: the structure head generates geometrically valid ligand conformations within the protein pocket, while the affinity head provides potency and binding affinity estimates that enable (1) in-distribution evaluation of generated structures, (2) inference-time steering toward higher-affinity compounds via importance sampling, and (3) domain adaptation of both structure and affinity prediction through joint finetuning. This joint training paradigm thus facilitates coherent, in-distribution property-guided generation without requiring external scoring functions—avoiding distribution mismatch and additional computational overhead—while supporting effective domain adaptation.

Importantly, \textsc{Flowr.root} builds upon the multi-task training framework of \textsc{Flowr}~\cite{model:flowr}, yet extends it substantially to support four generative modes within a single backbone: (1) \textit{de novo} and, (2) interaction-conditional generation for preserving or enforcing pharmacophoric patterns, (3) scaffold hopping and elaboration, and (4) fragment growing and replacement. Notably, with the latter two \textsc{Flowr.root} supports specifying which atoms to replace while preserving the remainder of the molecule. To enforce locality in these partial modifications, we introduce a flexible prior placement strategy that shifts the generative prior from the zero center-of-mass to the local replacement site. This multi-mode capability enables flexible application across hit identification and hit-to-lead and lead optimization with targeted structural modifications.

The core challenge in protein-ligand modeling, however, lies not just in generation, but in accurately predicting protein-ligand interactions and potency. The sparsity of experimental binding data—where most ligand–protein combinations remain unexplored—poses a substantial obstacle to improving these predictions. While broad-coverage datasets encompass a wide spectrum of protein targets, they are intrinsically sparse, with few ligands tested systematically across different biological contexts. This imbalance leads to biased datasets, which explains why 3D machine and deep learning models—often relying only on protein or ligand descriptors and ignoring interaction patterns—can perform surprisingly well on affinity prediction benchmarks like PDBbind~\cite{volkov2022frustration}. 
Real-world drug discovery requires understanding nuanced SAR across hundreds or even thousands of related molecules within the same binding site—a richness uniquely captured by dense bioactivity landscapes. These landscapes encode redundancy and critical phenomena like activity cliffs, where small structural changes can lead to substantial shifts in bioactivity due to interaction changes.
This motivates the multi-stage training approach of \textsc{Flowr.root}: starting with sparse, lower-fidelity data to learn broad 3D chemistry and binding principles, and progressively refining the model with denser, higher-fidelity and project-specific datasets that capture subtle SAR patterns essential for effective hit expansion to lead optimization. This multi-stage pipeline is intended to closely mirror the practical needs of early drug discovery campaigns, addressing the limitations imposed by data sparsity in current 3D affinity prediction models.
The \textsc{Flowr.root} framework consists of three stages:

We position \textsc{Flowr.root} as a foundation model in the sense formalized by Bommasani et~al.~\cite{bommasani2021foundation}, who define a foundation model as a model that is trained on broad data at scale meant to be adapted (for example, LoRA-fine-tuned) to a range of downstream tasks. Critically, foundation models are described as unfinished, intermediary assets requiring adaptation~\cite{bommasani2021foundation, devlin2019bert, radford2018improving}. In this work, the foundation model property of \textsc{Flowr.root} refers specifically to its broad multi-stage pre-training on diverse chemical and structural data at scale, combined with its adaptability to specific downstream drug discovery tasks via parameter-efficient fine-tuning (LoRA;~\cite{hu2022lora}). We note that zero-shot generalization especially across arbitrary bioactivity landscapes is not implied; rather, we expect the model to provide transferable representations that enable rapid, data-efficient domain adaptation.

\paragraph{Stage 1: Large-scale pre-training.} This stage builds a general-purpose generative prior across chemical and structural distributions, with a broad bioactivity landscape, albeit with varying quality and predictive relevance. The model is trained on a mixture of:
\begin{itemize}
    \item \textbf{Small molecules in vacuum} ($\sim$1.5B datapoints from ZINC3D, PubChem3D, Enamine, OMol25) to ensure broad chemical coverage.
    \item \textbf{Protein–ligand complexes with and without affinity labels} ($\sim$2.5M datapoints from Plinder, BindingMoad, BindingNet, SAIR, KIBA-3D, Kinodata-3D, OMol25), which capture protein-ligand structural diversity, in parts annotated with affinity data. This dataset is split into lower- and higher-fidelity categories. We classify complexes with computational origin (that is, BindingNet, SAIR, KIBA-3D, Kinodata-3D) as lower-fidelity and those stemming from co-crystal data (that is, Plinder, BindingMoad) as higher-fidelity, with the exception of the high-confidence subset of BindingNet, which is treated as higher-fidelity.
\end{itemize}

\paragraph{Stage 2: High-fidelity finetuning.} Pre-trained weights are adapted to thoroughly curated, drug-like, high-quality affinity datasets. In this work, we use \textsc{Spindr} and \textsc{HiQBind} ($\sim$30k complexes) as datasets for multi-target finetuning. However, this stage can be used to finetune \textsc{Flowr.root} on diverse, high-quality in-house data, if available, which teaches the model the fine details of protein-ligand interactions and binding affinity.

\paragraph{Stage 3: Project-specific domain adaptation.} This ensures that \textsc{Flowr.root} can bridge the gap between large-scale, noisy public data and small-scale, high-quality, dense in-assay data, producing distributions that align with specific discovery campaigns. Different domain adaptation strategies are used to finetune and steer the pre-trained \textsc{Flowr.root} model:
\begin{itemize}
    \item \textbf{Inference-time scaling and property steering:} Importance sampling and path reweighting guide the model toward desired properties such as affinity, ADME/T, or synthetic accessibility.
    \item \textbf{Direct preference alignment:} The model is finetuned with project-specific preferences, such as avoiding known liability motifs or penalizing activity cliffs (future work).
    \item \textbf{Finetuning:} Parameter-efficient finetuning, namely Low-Rank Adapation (LoRA)~\cite{hu2022lora}, helps avoid catastrophic forgetting while aligning the model to project-specific structure-activity modes.
\end{itemize}

\subsection*{Training and Inference}

Following the methodology of Cremer et~al.~\cite{model:flowr}, we adopted the training and parameterization scheme from \textsc{FLOWR.multi}. Specifically, we employed a 4-layer pocket encoder with $d_{\text{inv}}^{\text{enc}} = 256$ and a 12-layer ligand decoder with $d_{\text{inv}}^{\text{dec}} = 384$. The equivariant feature dimension was set to $d_{\text{equi}} = 128$ for both the pocket encoder and ligand decoder. For latent attention, we utilized a latent size of 64 with 32 attention heads. Overall, \textsc{Flowr.root} has approximately 33 million trainable parameters. Throughout this work, the model has been trained on one \textsc{Nvidia} H100 node comprising eight H100 GPUs.

To further enhance model stability and generalization, we incorporated residual coordinate and subsequent edge updates using Gaussian radial basis expansions~\cite{model:eqgatdiff}, applied to the layer-wise latent message-passing outputs of the \textsc{Flowr} model backbone~\cite{model:flowr}. Additionally, we included distance and cross-product computations between latent equivariant node features within the message-passing layers, which we found to improve model performance. Specifically, in the pairwise interactions between nodes, given the equivariant queries $q_{\text{equi}} \in \mathbb{R}^{B \times N_q \times 3 \times d_{\text{equi}}}$ and keys $k_{\text{equi}} \in \mathbb{R}^{B \times N_{kv} \times 3 \times d_{\text{equi}}}$, we define the pairwise difference tensor as
\begin{equation}
\mathbf{D} = \mathbf{Q}[:, :, \mathbf{1}, :] - \mathbf{K}[:, \mathbf{1}, :, :],
\end{equation}
and compute the distance features as
\begin{equation}
\mathbf{S} = \left| \mathbf{D} \right|_2 \in \mathbb{R}^{B \times N_q \times N_k \times d}.
\end{equation}
The pairwise cross product tensor is defined as
\begin{equation}
\mathbf{C} = \mathbf{Q}[:, :, \mathbf{1}, :] \times \mathbf{K}[:, \mathbf{1}, :, :],
\end{equation}
where $\times$ denotes the vector cross product over the 3-vector dimension, resulting in
\begin{equation}
\mathbf{C} \in \mathbb{R}^{B \times N_q \times N_k \times 3 \times d}.
\end{equation}
Both $\mathbf{S}$ and $\mathbf{C}$ are stacked to the existing pairwise message tensor as additional features.

The model structure output comprises predictions for atomic coordinates, atom types, charge, hybridization, bond types, bond lengths and bond angles. The overall loss function reads:
\begin{align}
\mathcal{L}_{\mathrm{total}} = &\lambda_{\mathrm{c}} \underbrace{\mathrm{MSE}(\mathbf{X}_{\mathrm{pred}},\ \mathbf{X}_{\mathrm{true}})}_{\text{Coordinate loss}} \ +
\lambda_{\mathrm{t}} \underbrace{\mathrm{CE}(\mathbf{T}_{\mathrm{pred}},\ \mathbf{T}_{\mathrm{true}})}_{\text{Atom type loss}} \ + \lambda_{\mathrm{ch}}\underbrace{\mathrm{CE}(\mathbf{Q}_{\mathrm{pred}},\ \mathbf{Q}_{\mathrm{true}})}_{\text{Charge loss}} \ + \\
&\lambda_{\mathrm{h}}\underbrace{\mathrm{CE}(\mathbf{H}_{\mathrm{pred}},\ \mathbf{H}_{\mathrm{true}})}_{\text{Hybridization loss}} \ +
\lambda_{\mathrm{b}}\underbrace{\mathrm{CE}(\mathbf{B}_{\mathrm{pred}},\ \mathbf{B}_{\mathrm{true}})}_{\text{Bond type loss}}+\lambda_{\mathrm{bl}}\underbrace{\mathcal{L}_{\mathrm{bond\ length}}}_{\text{Bond length loss}}+\lambda_{\mathrm{ba}}\underbrace{\mathcal{L}_{\mathrm{bond\ angle}}}_{\text{Bond angle loss}}
\end{align}
with $\lambda_{i}$ denoting the respective loss weighting.
To encourage geometrically accurate bond lengths, we compute the mean squared error between predicted and ground truth distances for all bonded atom pairs:
\begin{equation}
\mathcal{L}_{\mathrm{bond\ length}} = \frac{1}{|\mathcal{B}|} \sum_{(i,j) \in \mathcal{B}} \left( \|\mathbf{x}_i^{\mathrm{pred}} - \mathbf{x}_j^{\mathrm{pred}}\|_2 - \|\mathbf{x}_i^{\mathrm{true}} - \mathbf{x}_j^{\mathrm{true}}\|_2 \right)^2
\end{equation}
where $\mathcal{B}$ denotes the set of bonded atom pairs extracted from the bond adjacency matrix, considering only the upper triangle to avoid redundancy. To preserve local molecular geometry, we penalize deviations in bond angles for all valid triplets $(i, j, k)$ where atom $j$ is bonded to both atoms $i$ and $k$. For each triplet, the bond angle $\theta_{ijk}$ is computed as:
\begin{equation}
\theta_{ijk} = \arccos\left( \frac{(\mathbf{x}_i - \mathbf{x}_j) \cdot (\mathbf{x}_k - \mathbf{x}_j)}{\|\mathbf{x}_i - \mathbf{x}_j\|_2 \cdot \|\mathbf{x}_k - \mathbf{x}_j\|_2} \right)
\end{equation}
The bond angle loss is then defined using the Huber loss for robustness:
\begin{equation}
\mathcal{L}_{\mathrm{bond\ angle}} = \frac{1}{|\mathcal{A}|} \sum_{(i,j,k) \in \mathcal{A}} \mathrm{Huber}_\delta\left( \theta_{ijk}^{\mathrm{pred}} - \theta_{ijk}^{\mathrm{true}} \right)
\end{equation}
where $\mathcal{A}$ is the set of valid angle triplets, and the Huber loss with threshold $\delta$ is defined as:
\begin{equation}
\mathrm{Huber}_\delta(x) = \begin{cases}
\frac{1}{2}x^2 & \text{if } |x| \leq \delta \\
\delta \left( |x| - \frac{1}{2}\delta \right) & \text{otherwise}
\end{cases}
\end{equation}
We use $\delta = 0.5$ radians to balance sensitivity to small deviations while remaining robust to outliers.

\textsc{Flowr.root} jointly models both continuous (coordinates) and discrete (atom types, bond orders, charges, hybridizations) molecular features. For coordinates, we employ continuous flow matching~\cite{cfm:minibatch-ot}, while discrete flow models~\cite{model:multiflow} are used for categorical variables. The model is trained to recover the original ligand $l_1$ from a noisy ligand $l_t$ by learning the conditional distribution $p^{\theta}_{1|t}(l_1 | l_t, t; \mathcal{P})$, minimizing mean squared error for coordinates and cross-entropy loss for categorical features. Given a pocket structure $\mathcal{P}$, new ligands are generated by iteratively refining an initial noisy ligand $l_0 \sim p_{\text{noise}}$. The generative process follows a learned vector field $v_t^{\theta}$ for continuous features and a discrete integration scheme for categorical attributes~\cite{model:multiflow, model:flowr}.

\paragraph{Anisotropic and reference-ligand-based prior strategies}

\rebuttal{
Standard flow matching employs an isotropic Gaussian $\mathcal{N}(\mathbf{0}, \mathbf{I})$ as the source distribution $p_0$. In partial generation modes---scaffold hopping, scaffold elaboration, linker inpainting, core growing, and fragment growing---a fixed molecular substructure provides geometric context that constrains where new atoms should be placed. An isotropic prior distributes probability mass without directional bias, forcing the learned velocity field $v_t^{\theta}$ to perform long-range, direction-dependent transport to reach the target region. A distinguishing property of flow matching, in contrast to diffusion models where the terminal distribution is determined by the noise schedule, is that $p_0$ need not be a standard Gaussian: any distribution from which efficient sampling is possible suffices~\cite{cfm:minibatch-ot}. The velocity field is trained to transport whatever $p_0$ is supplied toward $p_1$, enabling per-sample prior selection and even mixed isotropic and anisotropic batches within a single training step. \textsc{Flowr.root} exploits this flexibility by constructing task-specific anisotropic Gaussian priors whose covariance encodes the spatial extent and directionality of the generation context.}

\rebuttal{
Two complementary covariance constructions are employed. The shape-based covariance captures the spatial extent of a set of reference atom coordinates. Given a set of $N$ reference atom coordinates $\{\mathbf{r}_i\}_{i=1}^{N}$ (variable atoms or the full molecule, depending on the generation mode), the centered coordinate matrix is
\begin{equation}
  \bar{\mathbf{X}} = \mathbf{X} - \frac{1}{N}\mathbf{1}\mathbf{1}^{\top}\mathbf{X}, \quad \mathbf{X} \in \mathbb{R}^{N \times 3},
\end{equation}
and the sample covariance is
\begin{equation}
  \mathbf{C} = \frac{1}{N-1}\,\bar{\mathbf{X}}^{\top}\bar{\mathbf{X}}.
\end{equation}
Eigendecomposing $\mathbf{C} = \mathbf{V}\,\mathrm{diag}(\lambda_1, \lambda_2, \lambda_3)\,\mathbf{V}^{\top}$ with $\lambda_1 \geq \lambda_2 \geq \lambda_3 \geq 0$, the eigenvalues are first normalized to satisfy $\sum_k \tilde{\lambda}_k = 3$:
\begin{equation}\label{eq:trace_norm}
  \tilde{\lambda}_k = \lambda_k \cdot \frac{3}{\sum_j \lambda_j},
\end{equation}
then clamped to $[\lambda_{\min}, \lambda_{\max}]$ and re-normalized to restore the trace constraint. The shape-based covariance is
\begin{equation}
  \boldsymbol{\Sigma}_{\mathrm{shape}} = \mathbf{V}\,\mathrm{diag}(\lambda_1^{*}, \lambda_2^{*}, \lambda_3^{*})\,\mathbf{V}^{\top}.
\end{equation}
This construction is applied in scaffold hopping and core growing (using the variable-atom coordinates as reference), as well as in scaffold elaboration and linker inpainting (using the full-molecule coordinates to capture the spatial extent of the entire ligand).}

\rebuttal{
The directional covariance encodes a preferred growth direction from a fixed fragment toward a target region. Given the center of mass of the fixed fragment $\mathbf{c}_{\mathrm{src}}$ and a target center $\mathbf{c}_{\mathrm{tgt}}$ (the prior center or the center of mass of the variable atoms), the unit direction is
\begin{equation}
  \mathbf{d} = \frac{\mathbf{c}_{\mathrm{tgt}} - \mathbf{c}_{\mathrm{src}}}{\|\mathbf{c}_{\mathrm{tgt}} - \mathbf{c}_{\mathrm{src}}\|_2}.
\end{equation}
An orthonormal basis $\{\mathbf{v}_2, \mathbf{v}_3, \mathbf{d}\}$ is obtained via Gram--Schmidt orthogonalization, yielding a rotation matrix $\mathbf{R}_{\mathrm{dir}} = [\mathbf{v}_2 \mid \mathbf{v}_3 \mid \mathbf{d}] \in SO(3)$. With an elongation factor $\alpha > 1$ (default $\alpha = 2$) and the trace normalization constraint $\mathrm{tr}(\boldsymbol{\Sigma}) = 3$, the eigenvalues become
\begin{equation}
  \lambda_{\perp} = \frac{3}{2 + \alpha}, \quad \lambda_{\parallel} = \frac{3\alpha}{2 + \alpha},
\end{equation}
and the directional covariance reads
\begin{equation}
  \boldsymbol{\Sigma}_{\mathrm{dir}} = \mathbf{R}_{\mathrm{dir}}\,\mathrm{diag}(\lambda_{\perp}, \lambda_{\perp}, \lambda_{\parallel})\,\mathbf{R}_{\mathrm{dir}}^{\top}.
\end{equation}
This construction is used for fragment growing, where the direction points from the fixed fragment toward the prior center, and composes with the reference-ligand center-of-mass placement described below.}

\rebuttal{
Both covariance types satisfy the trace invariant $\mathrm{tr}(\boldsymbol{\Sigma}) = 3$, ensuring $\mathbb{E}[\|\mathbf{x}_0\|^2] = 3$ for $\mathbf{x}_0 \sim \mathcal{N}(\mathbf{0}, \boldsymbol{\Sigma})$, the same second moment as the isotropic prior. This guarantees consistent velocity field magnitudes across prior types and enables stable mixed-batch training in which isotropic and anisotropic samples coexist. Since both $\boldsymbol{\Sigma}_{\mathrm{shape}}$ and $\boldsymbol{\Sigma}_{\mathrm{dir}}$ are derived from reference atom coordinates, they transform equivariantly under rigid-body transformations of the protein--ligand complex, preserving the $SE(3)$-equivariance of the model. Sampling from $\mathcal{N}(\mathbf{0}, \boldsymbol{\Sigma})$ is performed via the Cholesky decomposition $\boldsymbol{\Sigma} = \mathbf{L}\mathbf{L}^{\top}$, yielding $\mathbf{x}_0 = \mathbf{z}\,\mathbf{L}^{\top}$ with $\mathbf{z} \sim \mathcal{N}(\mathbf{0}, \mathbf{I})$.}

\rebuttal{
Orthogonal to the covariance construction, the prior point cloud can be spatially relocated to improve locality. For global inpainting modes, the prior cloud is shifted to the center of mass of the to-be-generated atoms in the reference ligand. During training, Gaussian noise with standard deviation $\sigma_{\mathrm{noise}}$ is added to this placement for robustness. Covariance controls the shape of the prior while placement controls its location; the two compose independently. For substructure and fragment inpainting, the prior cloud is shifted to the center of mass of the to-be-replaced fragment in the reference ligand, and an isotropic $\mathcal{N}(\mathbf{0}, \mathbf{I})$ prior suffices since the spatial placement itself provides sufficient locality.}

\paragraph{Potency and binding affinity prediction}
In drug discovery, a variety of parameters, including $\mathrm{IC}_{50}$, $K_{\mathrm{i}}$, $K_{\mathrm{d}}$, and $\mathrm{EC}_{50}$, are reported as potency and binding affinity measures. These parameters differ in their definitions and experimental setups, which complicates their direct comparison. For example, $K_{\mathrm{d}}$ is an equilibrium constant that directly measures the strength with which a ligand binds to its target protein. On the other hand, a $K_{\mathrm{i}}$ is an equilibrium constant indicating how well a given inhibitor inhibits the binding of a natural substrate. In other words, while $K_{\mathrm{d}}$ measures the stickiness of a molecule to a target, $K_{\mathrm{i}}$ measures how well the inhibitor blocks the enzyme from its natural substrate. Though these values often correlate, they are not always interchangeable. Likewise, $\mathrm{EC}_{50}$ and $\mathrm{IC}_{50}$ are used for biological assays. Though they correlate with $K_{\mathrm{d}}$ and $K_{\mathrm{i}}$, their interpretation depends on assay conditions. For better readability, for the remainder of this work we refer to potency and binding affinity prediction as affinity prediction.

Therefore, in contrast to prior work~\cite{model:aev-plig, boltz2_preprint}, we predict each potency measure separately, aligning better with the complexity of drug-target data. Specifically, \textsc{Flowr.root} employs separate prediction heads for each affinity type ($\mathrm{pIC}_{50}$, $\mathrm{p}K_{\mathrm{i}}$, $\mathrm{p}K_{\mathrm{d}}$, $\mathrm{pEC}_{50}$), with each head trained separately on its corresponding data type. This design explicitly avoids treating heterogeneous affinity labels as interchangeable.
Given invariant and equivariant ligand features $\mathbf{h}_i^{\mathrm{inv}}, \mathbf{h}_i^{\mathrm{equi}}$, pocket features $\mathbf{p}_j^{\mathrm{inv}}, \mathbf{p}_j^{\mathrm{equi}}$, and ligand-pocket interaction features $\mathbf{e}_{ij}$ extracted from the ligand decoder's latent message-passing module, the affinity head computes:

\begin{align}
\mathbf{f}_i^{\mathrm{lig}} &= \mathrm{Gate}(\mathbf{h}_i^{\mathrm{inv}}, \mathbf{h}_i^{\mathrm{equi}}) \\
\mathbf{z}^{\mathrm{lig}} &= \mathrm{MLP}_{\mathrm{lig}}\left(\left[\frac{1}{C}\sum_i \mathbf{f}_i^{\mathrm{lig}},\; \frac{1}{|\mathcal{L}|}\sum_i \mathbf{f}_i^{\mathrm{lig}}\right]\right) \\
\mathbf{f}_j^{\mathrm{pocket}} &= \mathrm{Gate}(\mathbf{p}_j^{\mathrm{inv}}, \mathbf{p}_j^{\mathrm{equi}}) \\
\mathbf{z}^{\mathrm{pocket}} &= \mathrm{MLP}_{\mathrm{pocket}}\left(\frac{1}{|\mathcal{P}|}\sum_j \mathbf{f}_j^{\mathrm{pocket}}\right) \\
\mathbf{z}^{\mathrm{int}} &= \mathrm{MLP}_{\mathrm{int}}\left(\frac{\sum_{i,j} \mathbf{e}_{ij} m_i m_j}{\sum_{i,j} m_i m_j}\right)
\end{align}
where $\mathrm{Gate}$ is a Gated equivariant block~\cite{model:painn} to combine equivariant and invariant feature tensors, and $\mathrm{MLP}$ is a multi-layer perceptron with two linear layers and SiLU activation. The input to $\mathrm{MLP}_{\mathrm{lig}}$ is the concatenation of normalized invariant features, where $C=100$, and $|\mathcal{L}|$ denotes the number of atoms in the ligand. The concatenated feature tensor $\mathbf{z} = [\mathbf{z}^{\mathrm{lig}}, \mathbf{z}^{\mathrm{pocket}}, \mathbf{z}^{\mathrm{int}}]$ is passed to task-specific heads defined as

\begin{equation}
y_{\mathrm{affinity}} = \mathrm{ReLU}(\mathrm{MLP}_{\mathrm{affinity}}(\mathbf{z})).
\end{equation}
The affinity prediction loss is computed using the mean Huber loss between predicted and true affinity values in log units ($\mathrm{pIC}_{50}$, $\mathrm{p}K_{\mathrm{d}}$, $\mathrm{p}K_{\mathrm{i}}$, $\mathrm{pEC}_{50}$) through
\begin{equation}
\mathcal{L}_{\text{affinity}} = \frac{1}{N} \sum_{i=1}^N \mathrm{Huber}\left(\hat{y}_i, y_i\right),
\end{equation}
where $\hat{y}_i$ and $y_i$ are the predicted and true affinity values for valid samples $i$, and $N$ is the number of valid samples.

To convert affinity values to experimental free energy $\Delta G$ values for comparison with physics-based models, we use the following formula:
\begin{equation}
\Delta G = -RT \ln 10 \cdot pK,
\end{equation}
where $R = 1.987 \times 10^{-3}$ kcal K$^{-1}$ mol$^{-1}$ is the universal gas constant and $T = 298$ K is the absolute temperature and pK being the respective affinity type.

\paragraph{Role of joint structure--affinity training.}
\rebuttal{We note that the affinity head is architecturally lightweight, comprising four small MLPs that operate on pooled ligand, pocket, and interaction features, totalling $\sim$743K parameters ($\sim$2.3\% of the 33M total). The affinity Huber loss constitutes one of eight loss terms; the remaining seven---coordinate MSE, atom-type CE, bond-type CE, charge CE, hybridization CE, bond-length MSE, and bond-angle Huber---collectively dominate the gradient signal to the shared backbone by a substantial margin. While the affinity loss gradients do propagate through the shared representations during training, their relative contribution is minor, consistent with the well-established finding that multi-task learning from related auxiliary objectives provides implicit regularization rather than degradation of the primary task~\cite{ruder2017mtl, crawshaw2020mtl}. Multiple lines of empirical evidence underline this: (i) the comparison between \textsc{Flowr} (no affinity head) and \textsc{Flowr.root}$^{\text{base}}$ (with affinity head), trained from scratch on the same data and protocol, shows that all structural quality improvements (for example, PB-validity, strain energy) are attributable to explicit architectural enhancements (improved equivariant message-passing, residual coordinate updates, bond-length and bond-angle losses) rather than multi-task training; (ii) the unconditional model \textsc{Flowr.root}$^{\text{base}}_{\text{uncond.}}$, which contains no pocket encoder or affinity head, independently achieves leading structural quality on GEOM-Drugs; and (iii) under affinity-guided importance sampling (see Methods), where the affinity head actively steers generation, structural quality metrics remain largely preserved (PB-validity: 0.93$\rightarrow$0.91; strain energy: 48.46$\rightarrow$54.42\,kcal/mol), demonstrating robustness even under maximal affinity-head influence. The purpose of joint training is therefore not to improve structural generation, but to enable a unified workflow supporting in-distribution property-guided generation, joint structure--affinity domain adaptation via LoRA finetuning, multi-target selectivity campaigns, and affinity-annotated fragment elaboration---capabilities that would be lost entirely without the affinity head.}

\paragraph{Confidence prediction: Predicted local distance difference test (pLDDT)}

We also predict pLDDT confidence scores to assess the reliability of generated ligand structures. Given predicted ligand samples $\hat{l}_1 = (\hat{X}, \hat{T}, \hat{Q}, \hat{H}, \hat{B})$ and ground-truth ligand-protein coordinates $X_l \in \mathbb{R}^{N_l \times 3}$ and $X_p \in \mathbb{R}^{N_p \times 3}$, we compute distance matrices $D, \hat{D} \in \mathbb{R}_{+}^{N_l \times N_p}$ for true and predicted ligand-protein distances, respectively. We consider distances below $12.0${\AA} with $b_i = \text{clamp}(\sum_{j=1}^{N_p} (D_{i,j} < 12.0), \min=1)$ neighbors for each ligand atom.

Using distance thresholds $\tau=\{\frac{1}{2}, 1, 2, 4\}$ and L1 distance errors $L=|D-\hat{D}|$, the LDDT score for each ligand atom is obtained through
\begin{equation}
\text{LDDT}_{i} = \frac{1}{b_i} \sum_{j=1}^{N_p} \left[ \frac{1}{|\tau|} \sum_{c\in \tau } (L_{ij} < c) \right] \in (0, 1).
\end{equation}
Each atom's LDDT score is binned into $k=50$ categories for multi-class classification. The confidence head, $f_\phi$, shares the ligand decoder backbone with reduced depth ($l=8$ layers), taking the generated structure $\hat{l}_1$ as input and outputting invariant logits $\text{pLDDT}_i \in \mathbb{R}^{50}$ for cross-entropy loss minimization.

\paragraph{Training and sampling overview}

We summarize the training and sampling procedures in Algorithms~\ref{alg:training} and~\ref{alg:sampling}. The complete ligand representation comprises coordinates~$\mathbf{X}$, atom types~$\mathbf{A}$, formal charges~$\mathbf{Q}$, hybridization states~$\mathbf{H}$, and bond types~$\mathbf{B}$. For brevity, Algorithms~\ref{alg:training}--\ref{alg:sampling} detail only coordinates (continuous) and atom types (discrete, $F$ categories); all other categorical features follow the identical discrete flow procedure.

\begin{algorithm}[ht]
\caption{Training (one step)}
\label{alg:training}
\begin{algorithmic}[1]
\Require Dataset $\mathcal{D} = \{(\mathcal{P}_i, l_{1,i})\}$, model $f_\theta$
\State Sample $(\mathcal{P}, l_1)$ from $\mathcal{D}$ with $l_1 = (\mathbf{X}_1, \mathbf{A}_1, \dots)$
\State Sample generation mode $m$ (de novo, scaffold hopping, scaffold elaboration, linker inpainting, core/fragment growing, interaction-conditional)
\State Extract fragment mask $\mathbf{m} \in \{0,1\}^N$\quad($m_i{=}1$: fixed; $m_i{=}0$: variable; $\mathbf{m}{=}\mathbf{0}$ for de novo)
\State Sample priors: $\mathbf{X}_0 \sim \mathcal{N}(\mathbf{0}, \boldsymbol{\Sigma})$,\quad $\mathbf{A}_0 \sim \mathrm{Uniform}(\{1,\dots,F\})^N$ (one-hot) \Comment{$\boldsymbol{\Sigma}$: anisotropic or $\mathbf{I}_3$}
\State Overwrite fixed atoms: $\mathbf{X}_{0,i} \leftarrow \mathbf{X}_{1,i},\; \mathbf{A}_{0,i} \leftarrow \mathbf{A}_{1,i}\;\forall\, i$ where $m_i{=}1$
\State Optionally shift variable-atom prior to reference-ligand center of mass
\State Sample $t \sim \mathcal{U}(0,1)$; set $t_i \leftarrow 1$ for all fixed atoms ($m_i{=}1$)
\State Interpolate: $\mathbf{X}_t = (1{-}t)\mathbf{X}_0 + t\,\mathbf{X}_1 + \sigma(t)\,\boldsymbol{\epsilon}$,\quad $\boldsymbol{\epsilon} \sim \mathcal{N}(\mathbf{0},\mathbf{I})$
\State Unmask atom types: $A_{t,i} = A_{1,i}$ w.p.\ $t$,\; $A_{t,i} = A_{0,i}$ w.p.\ $1{-}t$ \Comment{Independently per atom}
\State Forward: $(\hat{\mathbf{X}}_1, \hat{\mathbf{A}}_1) = f_\theta(l_t, t;\, \mathcal{P})$ \Comment{Model predicts target $l_1$}
\State $\mathcal{L} = \lambda_{\mathrm{c}}\,\mathrm{MSE}(\hat{\mathbf{X}}_1, \mathbf{X}_1) + \lambda_{\mathrm{t}}\,\mathrm{CE}(\hat{\mathbf{A}}_1, \mathbf{A}_1) + \mathcal{L}_{\mathrm{aux}}$ \Comment{See $\mathcal{L}_{\mathrm{total}}$ above}
\State Update $\theta \leftarrow \theta - \eta\,\nabla_\theta \mathcal{L}$
\end{algorithmic}
\end{algorithm}

\begin{algorithm}[ht]
\caption{Sampling / Inference}
\label{alg:sampling}
\begin{algorithmic}[1]
\Require Pocket $\mathcal{P}$, optional fragment mask $\mathbf{m}$, num.\ atoms $N$, steps $K$, $\Delta t = 1/K$, model $f_\theta$, coord.\ noise $\sigma_c$, inpainting frequency $N_{\mathrm{inp}} \leq K$ (default $K$)
\State Construct $l^{(0)}$: $\mathbf{X}^{(0)} \sim \mathcal{N}(\mathbf{0}, \boldsymbol{\Sigma})$,\; $\mathbf{A}^{(0)} \sim \mathrm{Uniform}(\{1,\dots,F\})^N$ (one-hot); initialize $t \leftarrow 0$
\State If $\mathbf{m} \neq \mathbf{0}$: fix coordinates/types for atoms with $m_i{=}1$; set $t_i{=}1$
\For{$k = 0, \dots, K{-}1$}
  \State $(\hat{\mathbf{X}}_1, \hat{\mathbf{A}}_1) = f_\theta(l^{(k)}, t;\, \mathcal{P})$ \Comment{$\hat{\mathbf{A}}_1 = \mathrm{softmax}(\cdot)$}
  \State $v_t^{\theta} = (\hat{\mathbf{X}}_1 - \mathbf{X}^{(k)})\,/\,(1 - t)$;\quad $g_t = (t{+}\epsilon)^{-1}\,\mathbf{1}[t < t_c]$ \Comment{Velocity; inv.\ noise sched.}
  \State $\mathbf{X}^{(k+1)} = \mathbf{X}^{(k)} + \Delta t \cdot v_t^{\theta}$ \Comment{Euler step (ODE, default)}
  \State \quad or stochastic: $s_t {=} \tfrac{t\,v_t^{\theta} - \mathbf{X}^{(k)}}{1-t}$;\; $\mathbf{X}^{(k+1)} {=} \mathbf{X}^{(k)} {+} \Delta t\,(v_t^{\theta} {+} g_t\, s_t) {+} \Delta t\sqrt{2g_t\sigma_c}\,\boldsymbol{\epsilon}$, $\boldsymbol{\epsilon}{\sim}\mathcal{N}(\mathbf{0},\mathbf{I})$ \Comment{SDE}
  \State $P(\mathbf{A}^{(k+1)}{=}s \mid \mathbf{A}^{(k)}{=}j) = \begin{cases} \frac{\Delta t}{1-t}\,\hat{\mathbf{A}}_{1,s} & s \neq j \\ 1 - \sum_{s' \neq j} \frac{\Delta t}{1-t}\,\hat{\mathbf{A}}_{1,s'} & s = j \end{cases}$
  \State $\mathbf{A}^{(k+1)} \sim \mathrm{Cat}(P)$ \Comment{Discrete flow step}
  \State If $\mathbf{m} \neq \mathbf{0}$ \textbf{and} $k \bmod \lfloor K / N_{\mathrm{inp}} \rfloor = 0$: restore fixed atoms from fragment \Comment{Periodic inpainting}
  \State $t \leftarrow t + \Delta t$
\EndFor
\State Final pass: $(\hat{\mathbf{X}}_1, \hat{\mathbf{A}}_1) = f_\theta(l^{(K)}, t{\approx}1;\, \mathcal{P})$ \Comment{Corrector step}
\State \Return $\hat{l}_1 = (\hat{\mathbf{X}}_1,\; \arg\max\, \hat{\mathbf{A}}_1)$
\end{algorithmic}
\end{algorithm}

\subsection*{Inference-time scaling}
Inference-time scaling is a powerful approach that allocates additional computation during generation, rather than solely relying on larger architectures~\cite{snell2024scalingllmtesttimecompute, setlur2024rewardingprogressscalingautomated}. Originally developed for large language models, this approach has been extended to diffusion and flow matching models. Diffusion models leverage stochasticity through particle sampling~\cite{wu2023practical, model:pilot} and Feynman-Kac steering~\cite{singhal2025a}, where multiple trajectories are generated and resampled based on reward functions. Flow matching models, despite being deterministic, achieve inference-time scaling through SDE-based generation, interpolant conversion, and Rollover Budget Forcing (RBF) for adaptive resource allocation~\cite{kim2025inferencetimescalingflowmodels}, improving upon diffusion-based advances~\cite{ma2025inferencetimescalingdiffusionmodels}.

In our work, we steer generative trajectories to sample from the conditional (target) distribution
\begin{equation}
p_\phi(l|y, P) \propto r(y|l, \mathcal{P}) p_\theta(l|\mathcal{P}),
\end{equation}
where $y$ is a vector of properties such as binding affinity, and $r$ represents a reward function that encodes user preferences.

\paragraph{Steering via Importance Sampling} The target distribution is usually intractable, especially when applying neural generative models that transport samples over time, such as flow or diffusion models. One way to sample from the target distribution is to generate $B$ particles and resample them based on importance weights derived from scores $\exp(\lambda r(y, l, \mathcal{P}))$, following the principle of importance sampling.

The sequential Monte Carlo (SMC) method~\cite{smc_del_moral_doucet} extends this concept to a time-sequential setting by maintaining $B$ particles and updating their importance weights over time~\cite{wu2023practical}. A straightforward weighting scheme involves evaluating the current batch of particles $\{l_{i, \tau } \}_{i=1}^B$ at time $\tau$, computing their scores $\bigl \{ \hat{y}_i=r(l_{i,\tau}, \mathcal{P}) \bigr \}_{i=1}^B$, and obtaining resampling weights through softmax normalization: $w_{i} = \frac{\exp(\hat{y}_i)}{\sum_j \exp(\hat{y}_j)}$. Particles are then resampled from the discrete distribution $\{ l_i \}_{i=1}^B \sim \text{Multinomial}(B, w=(w_1, w_2, \dots, w_B))$ with probability proportional to their weights, as done in \textsc{Pilot}~\cite{model:pilot}.

\subsection*{Quantum Mechanical Calculations on Protein-Ligand Complexes}
To benchmark \textsc{Flowr.root}'s affinity prediction head, we perform quantum mechanical (QM) calculations on protein-ligand complexes and correlate the results with predicted affinities. Calculations were conducted using ULYSSES~\cite{Menezes:ULYSSES}, a general-purpose semi-empirical library offering multiple molecular Hamiltonians. We employ GFN2-xTB~\cite{xtb_gfn2} with ALPB aqueous solvation~\cite{alpb}. GFN2-xTB is a simplified density functional method that parametrizes interparticle interaction integrals against reference data, substantially accelerating calculations~\cite{xtb_gfn2}—a critical advantage for large atomic systems such as protein-ligand complexes. Additionally, GFN2-xTB provides a balanced treatment of electrostatic interactions and incorporates state-of-the-art dispersion corrections, enabling accurate descriptions of non-bonded complexes.
In this work, the QM validation of \textsc{Flowr.root} used full protein domains, containing over 5000 atoms. Such a system size, required for meaningful physics-based evaluations of the granular details distinguishing different ligands, necessitates using semi-empirical Hamiltonians. Further, we only consider non-covalent binders, and, for such cases, intermolecular interactions are treated at a level similar to full Density Functional Theory (DFT). Finally, we found in previous work that GFN2-xTB provides accurate evaluations of protein-ligand complexes, in agreement with experimental SAR data \cite{menezes:silmitasertib,NOWACKI2025117979,doi:10.1021/acschemneuro.4c00365,GRYGIER2025148124}.

Regarding the choice of GFN2-xTB for non-covalent protein-ligand binding energetics: (1) The SQM2.20 method demonstrated strong correlations between semi-empirical QM-derived binding energies and experimental data for multiple targets including BACE1, which we also investigate---notably, such correlations were not observed with force-field approaches or contemporary ML scoring functions~\cite{pecina2024sqm220}. (2) GFN2-xTB was explicitly parameterized using non-covalent interaction datasets, employing the D4 dispersion correction and an updated electrostatics model~\cite{xtb_gfn2}, making it particularly suited for non-bonded interactions that dominate protein-ligand binding. (3) Large QM regions are essential for reliably capturing binding energetics and selectivity effects; for example, DYRK1A and DYRK1B are highly homologous kinases (over 85\% sequence identity), and treating only small binding-site regions would be insufficient to distinguish relative affinities. We emphasize that the QM calculations serve as an independent validation of predicted affinity trends rather than absolute binding energies.

Benchmark calculations on protein complexes extracted from the Schr{\"o}dinger FEP dataset already contained protonated protein structures. For these cases, \textsc{Flowr.root}-generated ligands had to be protonated using open babel~\cite{obabelpaper}. Calculations were run on the full complexes, water molecules were excluded and solvation was treated only implicitly, except when otherwise stated. In the case of the kinase selectivity test, structures were extracted directly from the PDB and processed with open babel. This includes protonation of the protein pockets and of all ligands, along with their predicted total charges.

\section*{Data Availability}

The following public datasets were used in this study:
\begin{itemize}
  \item Plinder [\url{https://www.plinder.sh}]~\cite{dataset:plinder}.
  \item Binding MOAD [\url{https://github.com/Mugsie1969/BindingMOAD}].
  \item SPINDR [\url{https://doi.org/10.5281/zenodo.15991056}]~\cite{model:flowr}.
  \item HiQBind [\url{https://doi.org/10.6084/m9.figshare.27430305}]~\cite{dataset:hiqbind}.
  \item SAIR [\url{https://huggingface.co/datasets/SandboxAQ/SAIR}]~\cite{dataset:sair}.
  \item BindingNet [\url{http://bindingnet.huanglab.org.cn}].
  \item BindingNetv2 [\url{http://bindingnetv2.huanglab.org.cn}]~\cite{dataset:bindingnetv2}.
  \item CrossDocked2020 [\url{http://bits.csb.pitt.edu/files/crossdock2020/}]~\cite{dataset:crossdocked}.
  \item Kinodata-3D [\url{https://doi.org/10.5281/zenodo.10852507}]~\cite{dataset:kinodata}.
  \item KIBA [\url{https://tdcommons.ai/multi_pred_tasks/dti/\#kiba}]~\cite{dataset:kiba}.
  \item Davis [\url{https://tdcommons.ai/multi_pred_tasks/dti/\#davis}]~\cite{dataset:davis}.
  \item ZINC20 [\url{https://zinc20.docking.org}]~\cite{dataset:zinc}.
  \item PubChem3D [\url{https://ftp.ncbi.nlm.nih.gov/pubchem/Compound_3D/}]~\cite{dataset:pubchem}.
  \item Enamine REAL [\url{https://enamine.net/compound-collections/real-compounds}].
  \item OMol25 [\url{https://huggingface.co/facebook/OMol25}]~\cite{dataset:omol25}.
  \item GEOM-Drugs [\url{https://doi.org/10.7910/DVN/JNGTDF}]~\cite{dataset:geom}.
  \item OpenFE IndustryBenchmarks2024 [\url{https://doi.org/10.5281/zenodo.17245549}]~\cite{openfe,openfe_results}.
  \item PDE10A benchmark [\url{https://doi.org/10.1007/s10822-022-00478-x}]~\cite{tosstorff2022pde10a}.
\end{itemize}

The following co-crystal structures from the RCSB Protein Data Bank were used: 3PE1 [\url{http://doi.org/10.2210/pdb3PE1/pdb}], 6KHF [\url{http://doi.org/10.2210/pdb6KHF/pdb}], 4GIH [\url{http://doi.org/10.2210/pdb4GIH/pdb}], 2Q70 [\url{http://doi.org/10.2210/pdb2Q70/pdb}], 4ZSP [\url{http://doi.org/10.2210/pdb4ZSP/pdb}], 3FT8 [\url{http://doi.org/10.2210/pdb3FT8/pdb}], and 5SF4 [\url{http://doi.org/10.2210/pdb5SF4/pdb}].

Curated training data, model checkpoints and generated ligand sets are deposited on Zenodo [\url{https://doi.org/10.5281/zenodo.20069588}] and on Google Drive [\url{https://drive.google.com/drive/folders/1NWpzTY-BG_9C4zXZndWlKwdu7UJNCYj8?usp=sharing}].

The four proprietary in-house structure--activity datasets used for the LoRA finetuning experiments in Fig.~\ref{fig:in_house} are not publicly available, as they comprise commercially sensitive medicinal-chemistry data of Pfizer Inc. covering active drug-discovery projects.

Additional data are provided in the Supplementary Information and the accompanying Source Data file.

Source data are provided with this paper.

\section*{Code Availability}

The source code for \textsc{Flowr.root} is publicly available on GitHub at \url{https://github.com/jule-c/flowr_root} under the MIT licence. The exact version of the code used to produce the results in this paper is archived on Zenodo at \url{https://doi.org/10.5281/zenodo.20068203}~\cite{flowrroot_zenodo}. Pretrained model checkpoints are mirrored on Zenodo (see Data Availability).

\FloatBarrier

\newpage

\section*{Acknowledgements}
This work was supported by Pfizer Worldwide Research and Development (J.C., T.L., M.M.G., D.-A.C.). F.M.\ acknowledges funding from Helmholtz Munich.

\section*{Author Contributions}
J.C. conceived the study; developed the methodology; performed experiments, validation, formal analysis; implemented the software; curated the data; created the visualizations; and wrote the original draft of the manuscript. T.L. assisted with experiments. M.M.G. contributed to study design. E.S. assisted with data preparation. F.M. performed validation, formal analysis; created visualizations and wrote the original draft of the manuscript. D.-A.C. provided supervision. All authors reviewed and edited the manuscript. 

\section*{Competing Interests}
The authors declare no competing interests.


\clearpage

\begingroup
\setcounter{figure}{0}
\setcounter{table}{0}
\setcounter{equation}{0}
\renewcommand{\thefigure}{S\arabic{figure}}
\renewcommand{\thetable}{S\arabic{table}}
\renewcommand{\theequation}{S\arabic{equation}}
\captionsetup[figure]{labelformat=empty}
\captionsetup[table]{labelformat=empty}
\renewcommand{\refname}{Supplementary References}

\section*{Supplementary Information}
\addcontentsline{toc}{section}{Supplementary Information}
\FloatBarrier

\FloatBarrier

\section*{Methods}
\label{S:app:methods}

\subsection*{\textsc{Flowr.ui}: A graphical user interface for interactive ligand design}
\label{S:app:flowr_ui}

\begin{figure}[htb!]
    \centering
    \includegraphics[width=1.0\textwidth]{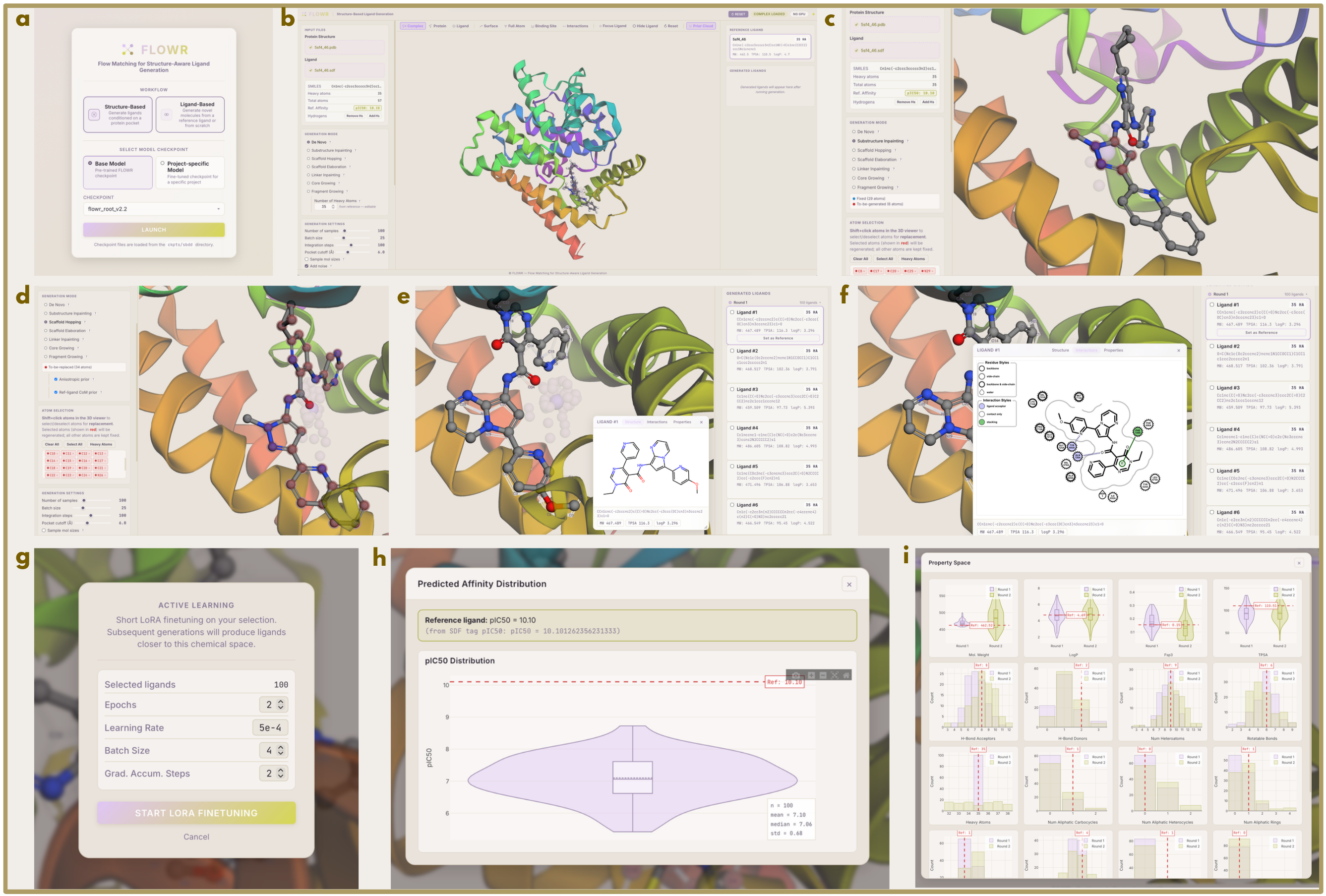}
    \caption{\textbf{Supplementary Fig.~1.} \rebuttal{\textbf{Overview of the \textsc{Flowr.ui} graphical user interface for interactive ligand design.} \textbf{a}~Login and model selection page, where users choose between structure-based (SBDD) and ligand-based (LBDD) workflows and select a base model checkpoint or a project-specific finetuned model. \textbf{b}~Main interface layout showing the interactive 3D protein-ligand viewer alongside generation mode selection, atom selection controls, and sampling parameter configuration. \textbf{c}~Substructure inpainting example: user-selected atoms (highlighted) define the replacement region, with the corresponding anisotropic Gaussian prior cloud shifted to the selected substructure. \textbf{d}~Scaffold hopping mode with optional user-guided atom selection for Murcko scaffold replacement while preserving R-groups. \textbf{e}~Results panel displaying generated ligands with 2D structure depictions, interaction diagrams, and computed molecular properties (TPSA, LogP, molecular weight, among others). \textbf{f}~Protein-ligand interaction diagram for a selected generated ligand, showing hydrogen bonds and salt bridges with distance annotations. \textbf{g}~Active learning interface for LoRA finetuning of the loaded \textsc{Flowr.root} checkpoint, with configurable training parameters (number of epochs, batch size, learning rate). \textbf{h}~Predicted potency distribution of generated ligands across generation rounds. \textbf{i}~Molecular property distributions of generated ligands compared to the reference ligand.}}
    \label{S:fig:flowr_ui}
\end{figure}

\rebuttal{\textsc{Flowr.ui} is a browser-based graphical user interface for interactive, structure-aware ligand design using the \textsc{Flowr.root} generative model (Supplementary Fig.~\ref{S:fig:flowr_ui}). The application implements a two-tier client-server architecture: a CPU-bound FastAPI~\cite{S:tool:fastapi} frontend server handles molecular file management, input validation, property computation, interaction detection, and chemical space analysis, while a separate GPU-bound FastAPI worker service loads the \textsc{Flowr.root} model checkpoint and executes ligand generation. This separation enables deployment on HPC clusters where the frontend operates on login or CPU nodes and GPU resources are allocated on demand via SLURM, with automatic release after a configurable idle timeout.}

\rebuttal{\textsc{Flowr.ui} supports both structure-based (SBDD, conditioned on a protein pocket) and ligand-based (LBDD, molecule-only) design workflows. Seven generation modes are available: (1)~\textit{de novo} generation, optionally conditioned on a reference ligand; (2)~substructure inpainting, replacing user-selected atoms while retaining fixed substructures; (3)~scaffold hopping, replacing the Murcko scaffold while preserving R-groups; (4)~scaffold elaboration, decorating a retained ring scaffold with new functional groups; (5)~linker inpainting, regenerating atoms connecting ring systems; (6)~core growing, retaining a ring system core and regenerating the remainder; and (7)~fragment growing, extending from a placed fragment with a configurable anisotropic Gaussian prior center. Users select atoms for retention or replacement directly in the 3D viewer; inpainting masks are computed in real time using pure RDKit-based~\cite{S:tool:rdkit} algorithms without GPU access.}

\rebuttal{The frontend renders protein structures as interactive 3D cartoon representations with optional molecular surfaces and binding-site residue highlighting (3Dmol.js~\cite{S:tool:3dmoljs}), 2D molecular depictions (RDKit.js~\cite{S:tool:rdkit}), and analytical plots (Plotly.js~\cite{S:tool:plotly}). Gaussian prior clouds representing the spatial sampling distribution are visualized as 3D point clouds with interactive repositioning. Post-generation analysis includes chemical space projections via PCA, $t$-SNE~\cite{S:tool:tsne}, or UMAP~\cite{S:tool:umap} of Morgan fingerprints~\cite{S:tool:morgan} (radius~2, 2048~bits), property distribution plots for 20~molecular descriptors (including molecular weight, LogP, TPSA, hydrogen bond donors and acceptors, rotatable bonds, ring counts, fraction of $\mathrm{sp}^{3}$ carbons, and structural alerts), affinity distribution tracking, and protein-ligand interaction visualization (hydrogen bonds and salt bridges) rendered as 3D distance-labeled overlays and 2D interaction diagrams.}

\rebuttal{Generated ligands undergo configurable post-processing: validity and uniqueness filtering, Tanimoto-based diversity filtering across generation rounds, RDKit~\cite{S:tool:rdkit} or GFN2-xTB~\cite{S:tool:gfn2xtb} geometry optimization, property-based filtering, and optional ADMET model filtering. Results are accumulated over multiple rounds with cross-round diversity enforcement. Binding affinities (IC$_{50}$, K$_i$, K$_d$, EC$_{50}$) are automatically parsed from uploaded SDF property tags with unit detection and p-value conversion, enabling direct comparison between reference and generated compounds. \textsc{Flowr.ui} further implements an active learning loop: users select promising ligands, trigger LoRA~\cite{S:tool:lora} finetuning of the loaded \textsc{Flowr.root} checkpoint on the GPU worker, and regenerate with the adapted model, enabling iterative structure-activity relationship refinement within a single session. All generated ligands can be ranked, selected, and exported as SDF files.}

\rebuttal{For HPC deployment, the server manages the full SLURM lifecycle---job submission with configurable resource parameters, compute node discovery via \texttt{squeue}, worker health probing, and automatic GPU deallocation. Users access the interface via SSH port forwarding. File transfers between frontend and worker are authenticated using HMAC-based tokens, and the application enforces content security policies, rate limiting, and input sanitization.}

\subsection*{Datasets}
\label{S:app:datasets}

\begin{figure}[h]
    \centering
    \includegraphics[width=1.0\textwidth]{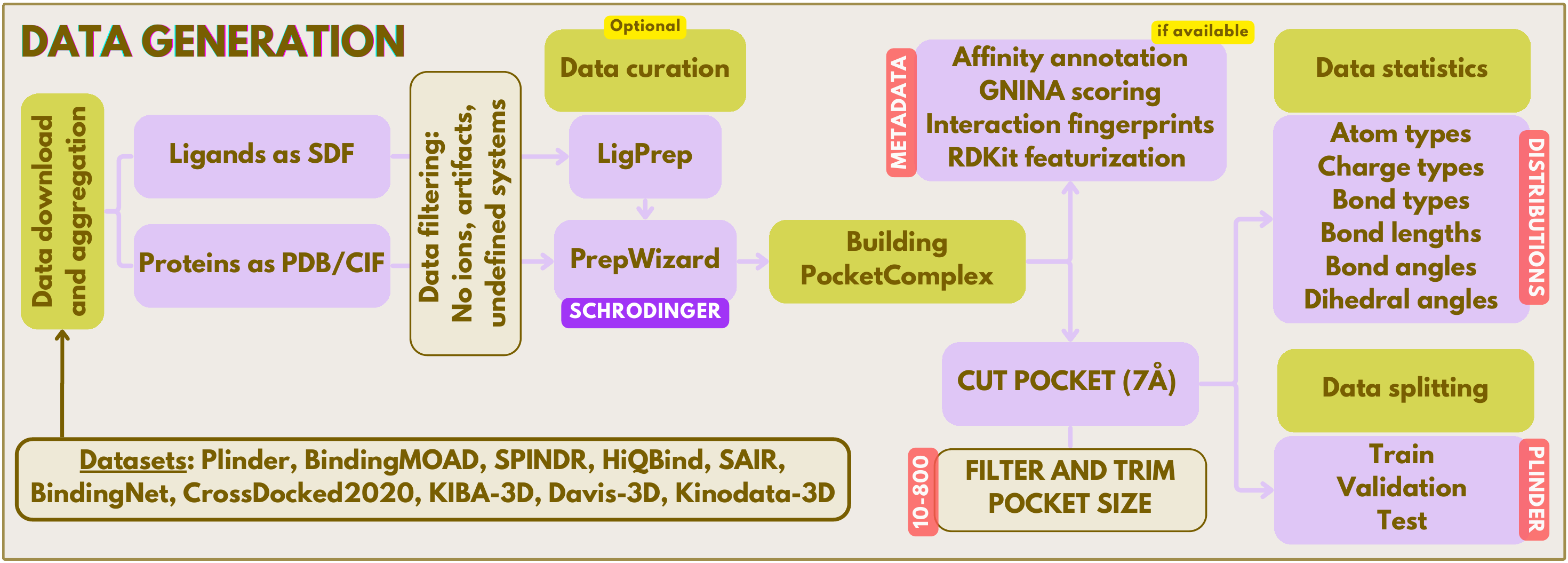}
    \caption{\textbf{Supplementary Fig.~2.} \textbf{Overview of the dataset generation pipeline used in this work.} Dataset generation workflow comprising data filtering, curation via Schrodinger's LigPrep and PrepWizard, building of metadata-annotated internal representation, and calculation of molecule statistics.}
    \label{S:fig:dataset_generation}
\end{figure}

To comprehensively train and evaluate \textsc{Flowr.root} for structure-aware ligand design, we leverage a diverse collection of public datasets spanning both small molecules and biomolecular complexes. Our dataset selection encompasses three primary categories: (1) large-scale small molecule databases for training conformational and chemical diversity, (2) large-scale computationally generated datasets that bridge the gap between available experimental data and the scale and augmentation required for deep learning applications, and (3) small-scale experimental protein-ligand complex datasets for higher-fidelity structure-based modeling. This multi-faceted approach ensures robust model training across diverse chemical, conformational, and biological/bio-activity spaces while maintaining high structural quality standards.

For our small molecule database, we utilized Zinc3D, PubChem3D, Enamine REAL, and OMol25.
For our protein-ligand database, we assembled a comprehensive collection of protein–ligand complex datasets by aggregating and standardizing data from multiple sources, including Plinder, BindingMOAD, SPINDR, HiQBind, SAIR, BindingNet, CrossDocked2020, KIBA-3D, Davis-3D, and Kinodata-3D. As visualized in Supplementary Fig.~\ref{S:fig:dataset_generation}, after rigorous preprocessing, filtering, and preparation of both ligands and proteins, each complex was converted into a unified internal representation and annotated with extensive metadata, such as affinity values, if available, and molecular descriptors. This harmonized dataset enables systematic analysis of chemical composition, structural diversity, and affinity distributions across all included sources, providing a robust foundation for downstream modeling and benchmarking.

While SPINDR and HiQBind provide preprocessed and well-curated co-crystal data resources, we applied additional comprehensive curation to selected datasets, namely Plinder, BindingMoad, SAIR, KIBA-3D and Davis-3D using Schrödinger's LigPrep and PrepWizard tools.
We used LigPrep to generate multiple protonated molecular conformations considering among other things different tautomeric states at physiological pH (7.4 $\pm$ 2.0), utilizing the OPLS4 force field and Epik for accurate p$K_a$ prediction aligned with the reference ligand via the maximum common substructure (MCS). PrepWizard handled protein preparation through side chain completion, protonation state determination using Epik and PROPKA, termini capping, water molecule sampling within 10.0 \AA, constrained hydrogen and overall restrained minimization (0.3 \AA\ RMSD tolerance) using the S-OPLS force field.
Unless otherwise stated, protein pockets were extracted using a 7{\AA} cutoff radius around the respective reference ligands, with constraints of a minimum of 10 and maximum of 800 pocket atoms per complex to ensure computational tractability while preserving essential binding site information.

Importantly, throughout all protein-ligand datasets we kept a consistent dataset split following the provided Plinder~\cite{S:dataset:plinder} train, validation and test set splits to avoid data leakage as best as possible enabling a stringent downstream evaluation.
For more details about the datasets we refer to Supplementary Section~\ref{S:app:datasets}, where we provide an overview of the aforementioned dataset generation, curation, and preprocessing pipeline, and a set of different dataset statistics.

\paragraph{Zinc3D}
Zinc3D~\cite{S:dataset:zinc} is a subset of the ZINC20 database containing pre-computed low-energy conformations for commercially available compounds. By providing ready-to-use 3D structures, ZINC3D eliminates the computational overhead of on-the-fly conformer generation, thereby accelerating virtual screening campaigns and structure-based drug discovery workflows. We utilized 646,663,126 molecules with their provided conformations.

\paragraph{PubChem3D}
PubChem3D~\cite{S:dataset:pubchem} extends the widely-used PubChem database by providing up to 500 computed 3D conformations for millions of bioactive compounds. This resource offers extensive conformational diversity, crucial for training robust generative models. We employed 10 conformations per molecule from approximately 93 million compounds, resulting in 928,649,525 total conformations.

\paragraph{Enamine REAL}
The Enamine REAL database comprises commercially available, synthetically accessible compounds widely employed in virtual screening and drug design. Its well-curated chemical space supports generative design strategies focused on drug-likeness and synthetic feasibility. Following~Cremer et~al.~\cite{S:model:pilot}, we used a diversity subset of the Enamine REAL database. We employed OpenEye's Omega software with default parameters to generate up to five conformers per molecule, yielding 111,389,149 conformations.

\paragraph{OMol25}
OMol25~\cite{S:dataset:omol25} is a comprehensive dataset containing over 100 million density functional theory (DFT) calculations at the $\omega$B97M-V/def2-TZVPD level of theory. The dataset covers systems up to 350 atoms with exceptional chemical and structural diversity across 83 elements. We utilized two subsets: the small molecules collection (21,352,259 structures) comprising recomputed versions of widely-used datasets (ANI-2X, Orbnet Denali, SPICE2, Solvated Protein Fragments, and ~30\% of GEOM) upgraded to consistent high-level DFT theory, and the biomolecules subset (5,180,233 structures), encompassing protein-ligand interactions derived from BioLiP2 and additional complexes generated through docking drug-like molecules from GEOM, ChEMBL, and ZINC20. The biomolecule dataset features protein pocket environments processed through molecular dynamics simulation with appropriate capping and protonation state sampling.

\paragraph{KIBA-3D}
We introduce KIBA-3D, a kinase-focused dataset derived from the KIBA bioactivity dataset~\cite{S:dataset:kiba}. Using Schrödinger's Glide, we performed exhaustive cross-docking of the KIBA ligand space across 172 kinase targets, generating 333,670 protein–ligand complexes spanning 2,038 unique ligands. This dataset provides a dense bioactivity landscape ideally suited for affinity guided, pocket-conditioned generative modeling and kinase-specific benchmarking applications.

\paragraph{Davis-3D}
DAVIS-3D is a structurally augmented version of the original DAVIS kinase bioactivity dataset~\cite{S:dataset:davis}, created as part of the Folding-Docking-Affinity (FDA) framework~\cite{S:dataset:davis-3d}. The dataset was generated by computationally folding protein structures using ColabFold and determining protein-ligand binding conformations through DiffDock, a deep learning-based docking model. This process transforms the original sequence-based DAVIS dataset into a collection of three-dimensional protein-ligand binding structures, of which we utilized 12,982 complexes with binding affinities.

\paragraph{Kinodata-3D}
Kinodata-3D~\cite{S:dataset:kinodata} is a curated collection of kinase complexes processed using cross-docking methodologies. We utilized 94,211 complexes from the combined mid- to higher-confidence subsets, which include binding affinity annotations and are preprocessed with a 5{\AA} protein-ligand pocket cutoff.

\paragraph{BindingNet}
BindingNet~\cite{S:dataset:bindingnetv1,S:dataset:bindingnetv2} comprises 689,796 modeled protein-ligand binding complexes across 1,794 protein targets. The dataset was constructed using an enhanced template-based modeling workflow that incorporates pharmacophore and molecular shape similarities alongside chemical similarity. Structures are categorized by template matching quality into high-confidence (232,030), moderate-confidence (164,912), and low-confidence (292,813) subsets, all of which we employed with their respective binding affinity annotations.

\paragraph{Plinder}
Plinder~\cite{S:dataset:plinder} represents the largest and most comprehensively annotated protein-ligand interaction (PLI) dataset, containing 449,383 PLI systems with over 500 annotations per complex. The dataset encompasses diverse interaction types including multi-ligand systems, oligonucleotides, peptides, and saccharides. Plinder introduced an approach for generating training and evaluation splits that minimizes task-specific leakage while maximizing test set quality.
After removing complexes containing ligand artifacts, ions, or undefined ligands, and processing using Schrödinger's LigPrep and PrepWizard, we utilized 250,633 protein-ligand systems.

\paragraph{SAIR}
The Structurally Augmented IC50 Repository (SAIR)~\cite{S:dataset:sair} is the largest publicly available dataset of protein-ligand 3D structures with binding affinity annotations, addressing the scarcity of high-quality experimental structures for deep learning applications. The original dataset contains 5,244,285 computationally generated structures across 1,048,857 unique protein-ligand systems from ChEMBL and BindingDB, with structures folded using the Boltz-1x model~\cite{S:model:boltz-1}.
We applied stringent filtering criteria, retaining only PoseBusters-valid~\cite{S:benchmarking:posebusters} complexes with negative AutoDock-Vina scores~\cite{S:docking:autodock-vina}, confidence scores $\geq$0.8, interaction PTM $\geq$0.6, and IPTM $\geq$0.8, yielding 1,781,634 complexes. Further processing using Schrödinger's LigPrep and PrepWizard resulted in 1,564,677 curated complexes with IC50 annotations.

\paragraph{BindingMOAD}
BindingMOAD~\cite{S:dataset:bindingmoad} is a comprehensive database developed over two decades (2001-2025), containing 41,409 protein-ligand complexes with affinity coverage for 15,223 complexes (37\%) and 20,387 unique ligands. After preprocessing with Schrödinger's LigPrep and PrepWizard, we obtained 33,286 complexes with diverse binding affinity annotations.

\paragraph{SPINDR}
SPINDR~\cite{S:model:flowr} is a higher-fidelity dataset of protein–ligand complexes curated for interaction-aware modeling. The dataset emphasizes accurate binding site geometries and ligand poses, supporting physically consistent generative modeling. Complexes are thoroughly cleaned to focus on drug-like, non-covalent interactions using Schrödinger's PrepWizard, with protein-ligand interactions annotated using ProLIF~\cite{S:bouysset_prolif_2021}. We employed all 35,627 provided complexes with partial binding affinity annotations.

\paragraph{HiQBind}
HiQBind~\cite{S:dataset:hiqbind} is a curated dataset addressing structural artifacts in widely-used datasets, like PDBbind. Containing over 18,000 unique PDB entries and 30,000 protein-ligand complex structures, it matches binding free energies from BioLiP, Binding MOAD, and BindingDB with co-crystallized PDB complexes. The dataset employs strictly open-source curation tools with multiple quality control modules for steric clash detection, ligand structure fixing, protein completion, and hydrogen addition protocols. We utilized all 31,571 complexes with binding affinity annotations.

\paragraph{Schrodinger FEP+ dataset}
The Schrodinger FEP+ dataset~\cite{S:ross2023maximal} comprises a large and diverse collection of protein-ligand complexes, each featuring congeneric series of small molecules with experimentally measured binding affinities (K$_d$, K$_i$, or IC$_{50}$). Designed as a benchmark for assessing the accuracy and reproducibility of free energy perturbation (FEP) methods, the dataset emphasizes high-quality structural data, including X-ray structures and carefully curated protein and ligand preparations. It covers a broad chemical space, including challenging cases such as macrocycles, charge-changing transformations, and buried water displacement. This dataset is intended to support the development, validation, and comparison of computational methods for predicting relative binding affinities, providing a robust foundation for downstream applications in drug discovery and molecular design.

\paragraph{Data heterogeneity and quality control.}

The use of heterogeneous, multi-fidelity data---including computationally generated protein--ligand complexes---is a deliberate and principled design choice supported by both foundational machine learning theory and recent domain-specific evidence. Transfer learning research has consistently shown that pretraining on diverse, even noisy, data improves downstream task performance: Yosinski et~al.~\cite{S:yosinski2014transferable} demonstrated that transferred features outperform random initialization even across distant tasks, while Mahajan et~al.~\cite{S:mahajan2018weakly} showed that pretraining on 3.5 billion weakly labeled images transfers remarkably well to clean benchmarks. In physics, McCabe et~al.~\cite{S:mccabe2023mpp} showed that a single model pretrained on heterogeneous physics simulations outperforms task-specific baselines when finetuned on unseen systems.

Within drug discovery, the value of computationally generated structural data has been independently validated. Valsson et~al.~\cite{S:model:aev-plig} demonstrated that augmenting training data with BindingNet complexes improved FEP benchmark performance (PCC $0.41 \rightarrow 0.59$, Kendall $\tau$ $0.26 \rightarrow 0.42$). Most directly relevant, Hsu et~al.~\cite{S:hsu2025aipredicted} systematically showed that augmentation benefits depend critically on structural quality: high-confidence BindingNet structures substantially improve binding affinity prediction ($\tau = 0.80$ correlation between PCC and training set size), while moderate- and low-confidence structures provide negligible or negative returns. Co-folded structures (Boltz-1x) were shown to substitute effectively for experimental structures when filtered by confidence.

Our pipeline implements quality control measures that align with these findings: (1)~SAIR complexes are filtered from $\sim$~5M to $\sim$~1.5M using stringent multi-criteria thresholds (PoseBusters-validity, Vina scores, Boltz-1x confidence); (2)~BindingNet complexes are stratified by confidence, with only the high-confidence subset treated as higher-fidelity data; and (3)~the progressive training paradigm ensures that Stage~2 finetuning on curated experimental data (SPINDR, HiQBind) overrides residual noise from lower-fidelity Stage~1 sources.

\paragraph{Dataset Statistics}
\begin{figure}[htb!]
    \centering
    \includegraphics[width=1.0\textwidth]{figures/all_datasets_overview.pdf}
    \caption{\textbf{Supplementary Fig.~3.} \textbf{Dataset chemistry statistics.} Distributions of ligand atom types, protein atom types, and protein residue types across multiple datasets. We show normalized frequencies of each type aggregated over all datasets, and per-dataset distributions visualized as heatmaps (logarithmic scale for atom types), highlighting differences in composition between datasets. Only the 20 standard amino acids are shown for residue types; hydrogens are excluded from atom type analyses. Per-dataset sample sizes ($n$, number of ligands or complexes) are listed in the corresponding paragraphs of Supplementary Section~\ref{S:app:datasets}; distributions are descriptive (no statistical test applied).}
    \label{S:fig:datasets_overview}
\end{figure}

\begin{figure}[htb!]
    \centering
    \includegraphics[width=1.0\textwidth]{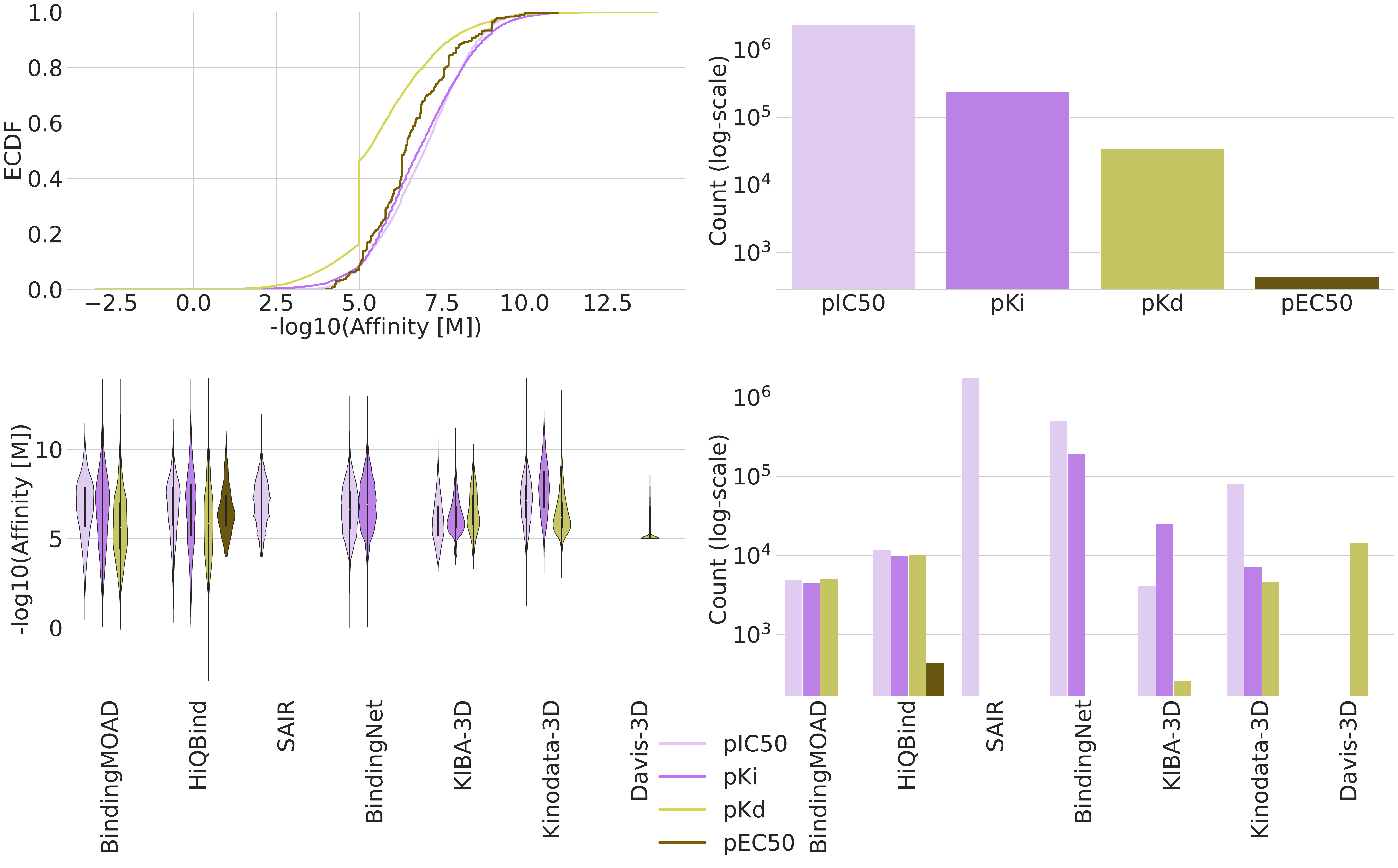}
    \caption{\textbf{Supplementary Fig.~4.} \textbf{Dataset affinity statistics.} Overview of affinity data distributions and dataset coverage. We show empirical cumulative distribution functions (ECDFs) of affinity values for each affinity type across all datasets; overall counts of each affinity type (log scale); and distribution of affinity values per dataset and affinity type (horizontal violin plots), as well as affinity type counts per dataset (horizontal bar plots, log scale). Legend indicates affinity type color coding used throughout all plots. Per-dataset sample sizes ($n$, number of affinity-annotated complexes) are listed in the corresponding paragraphs of Supplementary Section~\ref{S:app:datasets}; distributions are descriptive (no statistical test applied).}\label{S:fig:affinity_overview}
\end{figure}

\begin{figure}[htb!]
    \centering
    \includegraphics[width=1.0\textwidth]{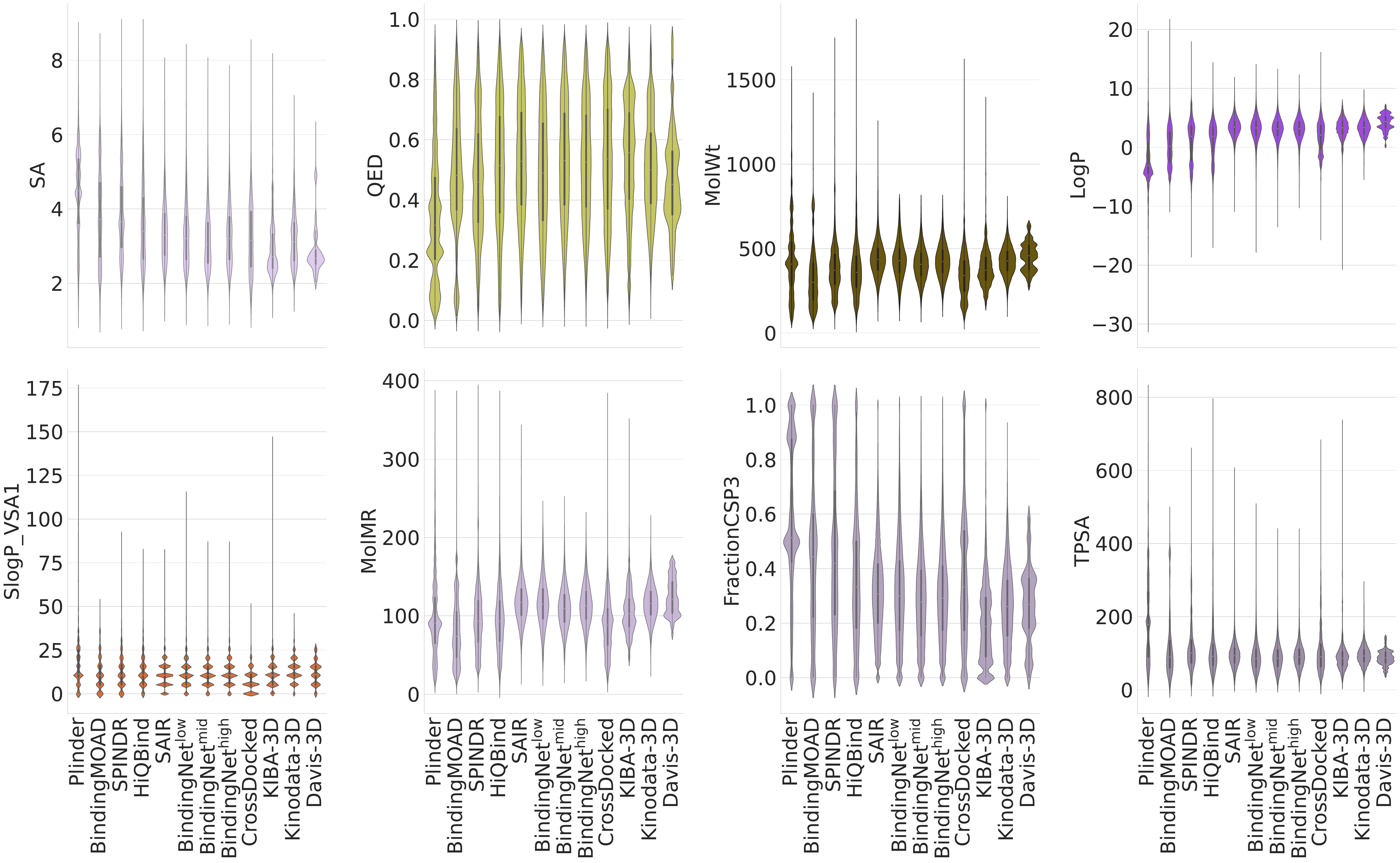}
    \caption{\textbf{Supplementary Fig.~5.} \textbf{Dataset property statistics.} Comparison of key continuous molecular properties of ligands—including molecular weight, logP, molar refractivity, and topological polar surface area—across all datasets. The distributions reveal substantial variability in ligand size, polarity, and hydrophobicity, reflecting the chemical diversity and distinct selection criteria of each dataset. Per-dataset sample sizes ($n$, number of ligands) are listed in the corresponding paragraphs of Supplementary Section~\ref{S:app:datasets}; distributions are descriptive (no statistical test applied).}\label{S:fig:cont_features}
\end{figure}

\begin{figure}[htb!]
    \centering
    \includegraphics[width=1.0\textwidth]{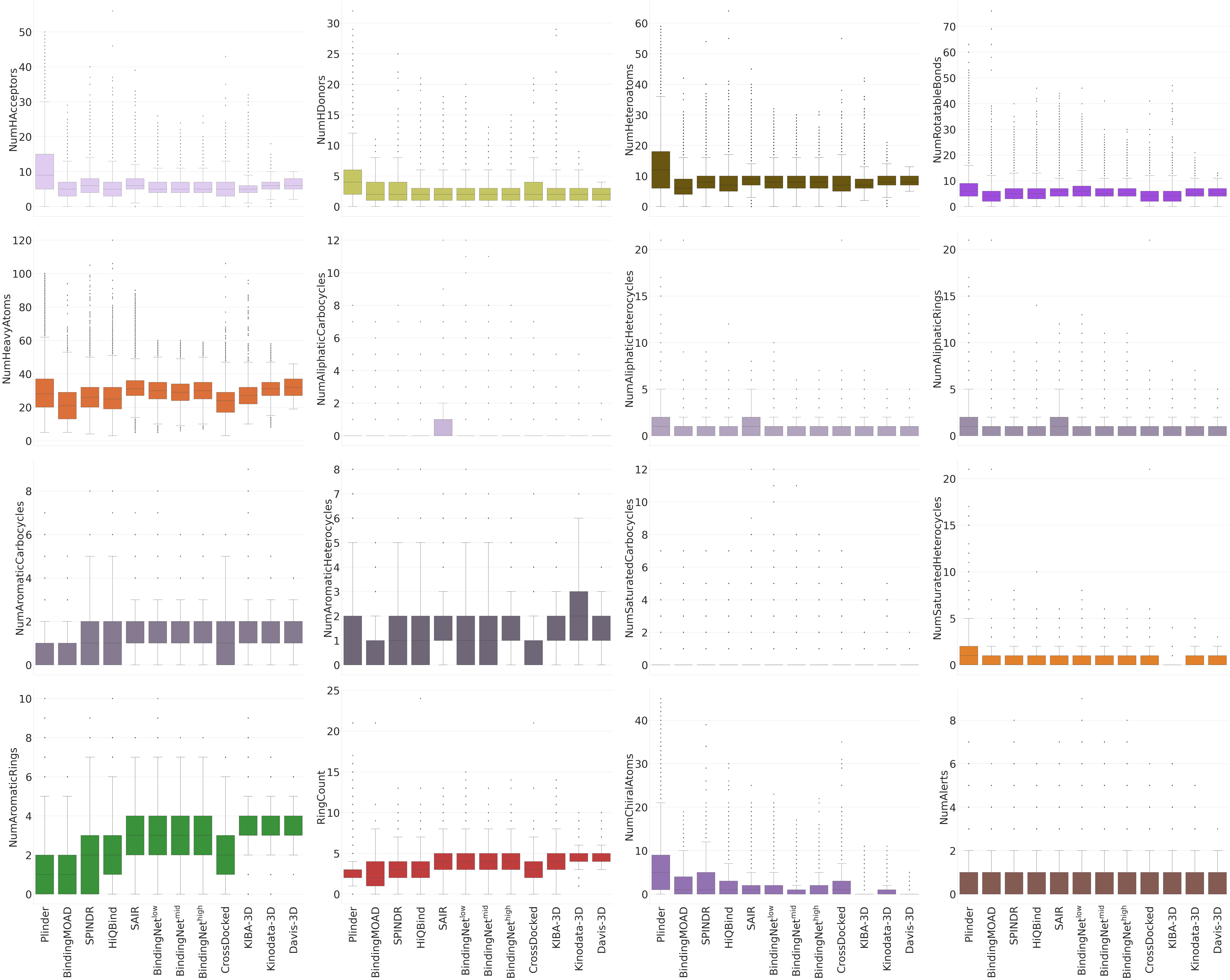}
    \caption{\textbf{Supplementary Fig.~6.} \textbf{Dataset property statistics.} Comparison of the distributions of discrete ligand features such as hydrogen bond acceptors and donors, rotatable bonds, ring systems, and chiral centers for each dataset. The observed differences highlight variations in molecular complexity, flexibility, and functional group composition among the datasets, providing insight into their respective chemical spaces. Per-dataset sample sizes ($n$, number of ligands) are listed in the corresponding paragraphs of Supplementary Section~\ref{S:app:datasets}; distributions are descriptive (no statistical test applied).}\label{S:fig:disc_features}
\end{figure}

In Supplementary Section~\ref{S:app:datasets} in Supplementary Fig.~\ref{S:fig:dataset_generation} we visualize the data generation framework that we established for this work. As outlined in Supplementary Section~\ref{S:app:datasets} we used as protein-ligand complex data Plinder, BindingMOAD, SPINDR, HiQBind, SAIR, BindingNet, CrossDocked2020, KIBA-3D, Davis-3D, and Kinodata-3D. After downloading and aggregating the different data sources, we split proteins and ligands, if not already done, into separate files. We save ligands as SDF and proteins as PDB or CIF files. Afterwards, we filter the data and remove systems that classify the ligands either as ions, artifacts or not definable. Then, we use Schrodinger's LigPrep to prepare the ligands and Schrodinger's PrepWizard to prepare the protein-ligand complex (see Supplementary Section~\ref{S:app:datasets} for more details). Successfully prepared systems are converted into our internal pocket-ligand representation encapsulated as PocketComplex instances for easy handling and data management. We retrieve pockets by cutting around the reference ligands with a cutoff of 7A and, if necessary, trim the pocket to the max size of 800 atoms. Every PocketComplex instance is annotated with comprehensive metadata spanning affinity values (if available), GNINA scores, interaction fingerprints, and selected RDKit features, like molecular weight, logP, TPSA, number of hydrogen acceptors and donors, number of rotatable bonds, number of aliphatic rings and many more. Finally, for all datasets we calculate distribution statistics comprising atom, charge, and bond type distributions as well as bond angles, bond lengths and dihedral angles distributions. Importantly, all datasets are splitted following the Plinder train, validation and test splits.

In Supplementary Fig.~\ref{S:fig:datasets_overview}, we compared the chemical composition of all protein–ligand datasets used in this work by analyzing the distributions of ligand atom types, protein atom types, and protein residue types. The aggregated distributions (top row) reveal the overall prevalence of each type, while the per-dataset heatmaps (bottom row) illustrate how these distributions vary between datasets. For atom types, a logarithmic scale was used to emphasize less frequent elements. Only the 20 standard amino acids were considered for residue type comparisons, and hydrogens were excluded from atom type analyses to improve interpretability.

To provide a comprehensive overview of the affinity data, Supplementary Fig.~\ref{S:fig:affinity_overview} summarizes both the distribution and coverage of affinity types across all datasets. The ECDFs display the range and distribution of affinity values for each affinity type, highlighting differences in spread and modality. The overall affinity type counts, presented on a logarithmic scale, reveal a pronounced imbalance, with IC50 being the most abundant. Dataset-specific distributions are visualized using horizontal violin plots, while horizontal bar plots summarize the number of measurements for each affinity type within each dataset. The majority of affinity values originate from the SAIR and BindingNet datasets. Notably, BindingNet also provides Ki measurements, while HiQBind and BindingMOAD offer a balanced representation of IC50, Ki, and Kd values. HiQBind is the only dataset containing EC50 measurements. In contrast, Davis-3D is exclusively annotated with Kd values. This overview facilitates the identification of data biases and gaps that may influence downstream analyses.

To systematically assess the chemical diversity of ligands across datasets, we analyzed the distributions of several continuous molecular descriptors in Supplementary Fig.~\ref{S:fig:cont_features}, including molecular weight, logP, molar refractivity, and topological polar surface area (TPSA). These features capture key aspects of ligand size, hydrophobicity, polarizability, and polarity, which are relevant for molecular recognition and drug-likeness. The results reveal pronounced differences between datasets: for example, BindingNet and SAIR contain ligands with a broader range of molecular weights and higher TPSA values, while datasets such as HiQBind and BindingMOAD exhibit more constrained distributions. These trends reflect the varying selection criteria and source domains of the datasets, and highlight the importance of considering chemical diversity in benchmarking and model development.

We further characterized the datasets by comparing the distributions of discrete ligand features in Supplementary Fig.~\ref{S:fig:disc_features}, including the number of hydrogen bond acceptors and donors, rotatable bonds, ring systems, and chiral centers. These properties provide insight into molecular complexity, flexibility, and the presence of functional groups relevant for binding interactions. The analysis demonstrates that datasets such as BindingNet and SAIR encompass ligands with higher numbers of rotatable bonds and greater ring system diversity, whereas datasets like Davis-3D and KIBA-3D are more restricted in these aspects. Notably, the number of chiral centers and structural alerts also varies substantially, underscoring differences in stereochemical complexity and potential reactivity. Together, these comparisons elucidate the distinct chemical spaces sampled by each dataset and inform the interpretation of downstream modeling results.

\paragraph{KIBA-3D}
The KIBA dataset was originally introduced by Tang \textit{et al.}~\cite{S:tang2014} as a large-scale benchmark for kinase–ligand binding affinity prediction, integrating heterogeneous bioactivity readouts into a standardized KIBA score. The original dataset contained $52,498$ chemical compounds, $467$ kinase targets and over $240,000$ interaction records. While comprehensive, the original KIBA dataset was highly scattered, due to an imbalance between the number of ligand-kinase pairs and associated measurements. To improve the density and ensure more reliable evaluation, subsequent studies discarded the ligands and targets with less than 10 interaction records. This procedure reduced the dataset to $2,094$ ligands and $229$ kinase targets.
For this work, we prepared the 3D extension of KIBA (KIBA-3D) based on kinase targets with experimentally solved structures deposited in the PDB. Supplementary Fig.~\ref{S:fig:kiba_data} shows examples of our docking results, illustrating that Glide could reproduce near-native binding poses for selected kinase inhibitors with conserved bidentate hydrogen bond pattern in the hinge region (for example, PDBID:2XB7-CHEMBL461876, glide gscore $-11,4$, PDBID:6LVM-CHEMBL1968590, glide gscore $-11,9$). The KIBA-3D dataset serves as a large-scale, structure-consistent benchmark for evaluating generative and predictive models.

\begin{figure}[htb!]
    \centering
    \includegraphics[width=1.0\textwidth]{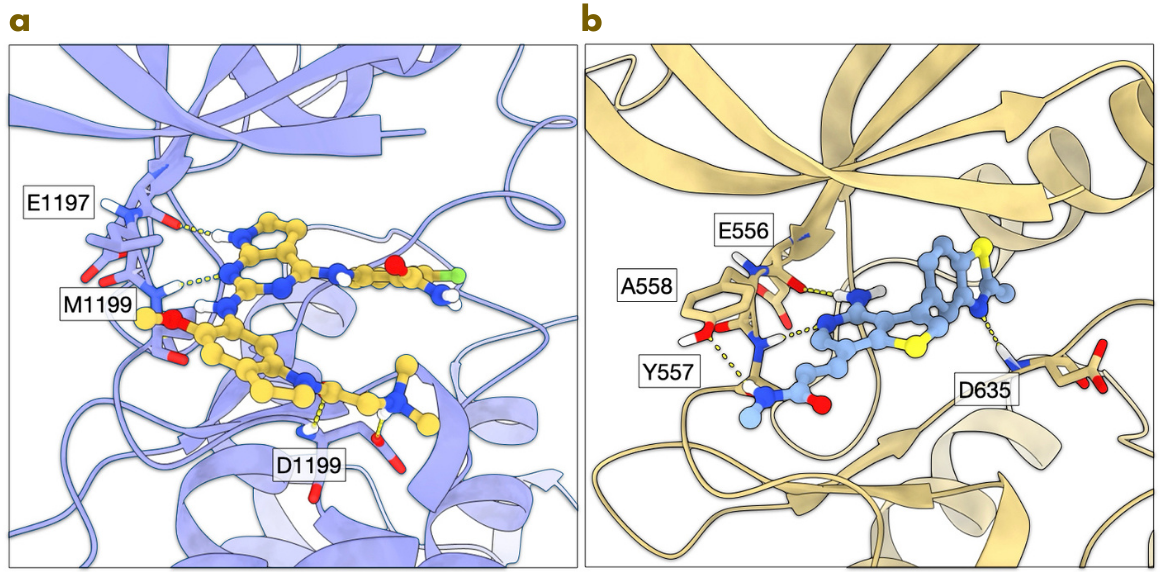}
    \caption{\textbf{Supplementary Fig.~7.} \textbf{Example visualizations of protein-ligand complexes in KIBA-3D.} \textbf{a} The structure of anaplastic lymphoma kinase (ALK, PDB ID: 2XB7, blue) with docked CHEMBL461876. The pyrrolopyrimidine moiety is positioned in the adenine region of the catalytic pocket, forming a bidentate hydrogen bond (yellow dashed lines) with the main chains of hinge residues Glu1197 and Met1199. \textbf{b} the structure of the tyrosine kinase fibroblast growth factor receptor 3 (FGFR3, PDB ID: 6LVM, yellow) accommodates an aminopyridine derivative (CHEMBL1968590) in a similar manner, with a conserved bidentate H-bond pattern in the hinge region (Glu556 and Ala558), complemented by an additional H-bond to the Tyr557 side chain. Furthermore, the nitrogen of the thiazole ring forms a hydrogen bond with Asp635 from the functional DFG motif. Both poses were favourable scored, with Glide docking gscore below $-11$ ($-11.4$ for the 2XB7--CHEMBL461876 complex and $-11.9$ for the 6LVM--CHEMBL1968590 complex). Representative panels: $n=2$ docked complexes shown out of $333{,}670$ Glide-docked KIBA-3D complexes spanning $2{,}094$ unique ligands and $174$ kinase targets.}\label{S:fig:kiba_data}
\end{figure}

\paragraph{Target protein structure selection for KIBA-3D}

Kinase targets from the KIBA dataset were mapped to their UniProt identifiers. For each UniProt accession, X-ray crystal structures were selected according to a set of heuristics to ensure biological relevance of the binding site. Only entries covering the kinase domain were considered. Among the available structures, the highest resolution one was chosen, with preference for holo over apo forms. Complexes with ATP, ADP, or staurosporine were deprioritized, as these ligands tend to stabilize non-representative conformations of the binding pocket; in particular, staurosporine, aside from being a highly promiscuous kinase inhibitor has a flat shape that inflate the catalytic site and biases the pocket toward accommodating larger ligands. Structures with fewer missing residues were favored, and cases where the bound small molecule occupied an allosteric or noncanonical pocket were excluded. The resulting curated set provides $174$ kinase structures suitable for docking of the KIBA ligand set.

\paragraph{Docking}
We performed docking with the Maestro suite (Schrödinger Release 2023-2: Maestro, Schrödinger, LLC, New York, NY, 2023) to generate consistent protein–ligand complexes across the $174$ kinase targets. Small molecules were first optimized with LigPrep. All the $2094$ ligands were processed. The dataset covers a broad chemical space, with molecular weights ranging from $180.15$ to $1381.64$ Da (median $376.49$ Da). Protein structures were processed using the Protein Preparation Workflow, including protonation with Epik and propka~\cite{S:johnston2023epik} and restrained minimization with the OPLS4 force field~\cite{S:sastry2013protein}. Missing residues were built with Prime. Docking grids were centered on the kinase domain's hinge region, covering the canonical ATP-binding pocket used by first-type kinase inhibitors. Docking was performed with the Glide SP protocol~\cite{S:yang2021efficient}, with no constraints applied, generating one pose per ligand.

\subsection*{Additional results}
\label{S:app:additional_results}

\paragraph{GEOM-Drugs and CrossDocked2020 Benchmarks}
Complete benchmark results on the unconditional GEOM-Drugs dataset and pocket-conditional CrossDocked2020 dataset are provided below, demonstrating that \textsc{Flowr.root} achieves leading performance across chemical validity, geometric accuracy, and energetic stability metrics.

\begin{table}[h]
\caption{\textbf{Supplementary Table~1.} \textbf{Evaluation and comparison of~\textsc{Flowr.root} unconditional base model on~\textsc{Geom-Drugs}.} Benchmark comparison of the non-pretrained~\textsc{Flowr.root} ligand-only base model against~\textsc{EQGAT-diff},~\textsc{ADiT},~\textsc{SemlaFlow},~\textsc{Megalodon},~\textsc{FlowMol3} on the~\textsc{Geom-Drugs} dataset. We follow the conventions in this field and sample 10,000 molecules with molecule sizes randomly sampled from the test set. We evaluate the performance of~\textsc{Flowr.root} on RDKit- and PoseBusters-validity, the change in potential energy resulting from GFN2-xTB minimization ($\Delta E_{\mathrm{relax}}$) and the all-atom RMSD between the predicted and GFN2-xTB minimized conformations (Relax RMSD). For each model, $n=10{,}000$ molecules were sampled per replicate; values are the sample mean with the 95\% confidence interval over five replicate runs with different random seeds. For $\Delta E_{\mathrm{relax}}$ and Relax RMSD we report the median over the same $n=10{,}000$ molecules per replicate.}
\label{S:tab:results_geom_app}
\centering
\begin{adjustbox}{width=1.0\textwidth,center}
\begin{sc}
\begin{tabular}{l|c|ccc|cc|cc}
\toprule
Model & RDKit-valid(\%)$\uparrow$ & PB-valid(\%)$\uparrow$ & $\Delta E_{\mathrm{relax}}$$\downarrow$ & \text{Relax RMSD}$\downarrow$ & Params (M) \\
\toprule
\textsc{\textsc{EQGAT-diff}} & 86.0 {\tiny$\pm$ 0.9} & 77.6 {\tiny$\pm$ 0.8} & 6.51 {\tiny$\pm$ 0.16} & 0.60 {\tiny$\pm$ 0.01} & 12 \\
\textsc{\textsc{ADiT}} & 99.9 {\tiny$\pm$ 0.0} & 82.7 {\tiny$\pm$ 0.8} & 79.32 {\tiny$\pm$ 1.00} & 1.30 {\tiny$\pm$ 0.02} & 150 \\
\textsc{\textsc{SemlaFlow}} & 95.5 {\tiny$\pm$ 0.5} & 88.5 {\tiny$\pm$ 1.3} & 31.9 {\tiny$\pm$ 2.30} & 0.24 {\tiny$\pm$ 0.03} & 40 \\
\textsc{\textsc{Megalodon}} & 94.8 {\tiny$\pm$ 0.3} & 86.6 {\tiny$\pm$ 0.7} & 3.17 {\tiny$\pm$ 0.11} & 0.41 {\tiny$\pm$ 0.01} & 60 \\
\textsc{FlowMol3} & 99.9 {\tiny$\pm$ 0.1} & 91.9 {\tiny$\pm$ 0.7} & 3.83 {\tiny$\pm$ 0.08} & 0.39 {\tiny$\pm$ 0.01} & 6\\
\midrule
\textsc{Flowr.root}$^{\text{base}}_{\text{uncond.}}$ & 98.5 {\tiny$\pm$ 0.2} & 94.0 {\tiny$\pm$ 0.2} & 3.65 {\tiny$\pm$ 0.07} & 0.07 {\tiny$\pm$ 0.02} & 34 \\
\toprule
Train set & 100.0 {\tiny$\pm$ 0.0} & 93.2 {\tiny$\pm$ 0.1} & - & -\\
\bottomrule
\end{tabular}
\end{sc}
\end{adjustbox}
\end{table}

On the \textsc{Geom-Drugs} dataset, \textsc{Flowr.root} achieves a PoseBusters-validity of 94.0\%, surpassing all other models including \textsc{FlowMol3} (91.9\%). The model demonstrates exceptional geometric precision with a median relaxation RMSD of only 0.07~\AA, substantially lower than all baselines. The relaxation energy of 3.65 kcal/mol indicates that generated conformations are close to local energy minima.

\begin{table}[h]
\caption{\textbf{Supplementary Table~2.} \textbf{Evaluation and comparison of~\textsc{Flowr.root} base model on~\textsc{CrossDocked2020}.} Benchmark comparison of the non-pretrained~\textsc{Flowr.root} base model against~\textsc{Pocket2Mol},~\textsc{TargetDiff},~\textsc{DiffSBDD},~\textsc{Pilot},~\textsc{DrugFlow} and~\textsc{Flowr} on the~\textsc{CrossDocked2020} test dataset. We follow the conventions in this field and sample 100 ligands per test target, of which there are 100. We evaluate the most expressive metrics, namely PoseBusters-validity, GenBench3D strain energy, AutoDock-Vina scores and the Wasserstein distance of the generated ligands' bond angles (BondA.W1) and bond lengths (BondL.W1) distributions relative to the test set. Sample sizes: $n=100$ generated ligands per target across $n=100$ \textsc{CrossDocked2020} test-set targets. Values are mean $\pm$ standard deviation;.}
\label{S:tab:results_crossdocked_app}
\centering
\begin{adjustbox}{width=1.0\textwidth,center}
\begin{sc}
\begin{tabular}{l|c|ccc|cc|cc}
\toprule
Model & PB-valid$\uparrow$ & Strain$\downarrow$ & Vina score$\downarrow$ & Vina score$^{\text{min}}$$\downarrow$ & BondA.W1$\downarrow$ & BondL.W1 [$10^{-2}$]$\downarrow$ & Size & Time (s)$\downarrow$ \\
\toprule
\textsc{\textsc{Pocket2Mol}} & 0.76 {\tiny$\pm$ 0.39} & 147.22 {\tiny$\pm$ 61.41} & -4.72 {\tiny$\pm$ 1.47} & -5.80 {\tiny$\pm$ 1.26} & 2.04 & 0.66 & 17.04 {\tiny$\pm$ 4.11} & 2320 {\tiny$\pm$ 45}\\
\textsc{\textsc{DiffSBDD}} & 0.38 {\tiny$\pm$ 0.46} & 519.03 {\tiny$\pm$ 251.32} & -2.97 {\tiny$\pm$ 5.21} & -4.71 {\tiny$\pm$ 3.30} & 7.00 & 0.51 & 24.85 {\tiny$\pm$ 8.94} & 160.31 {\tiny$\pm$ 73.30}\\
\textsc{\textsc{TargetDiff}} & 0.57 {\tiny$\pm$ 0.46} & 294.89 {\tiny$\pm$ 136.32} & -5.20 {\tiny$\pm$ 1.79} & -5.82 {\tiny$\pm$ 1.60} & 7.76 & 0.42& 22.79 {\tiny$\pm$ 9.46} & 3228 {\tiny$\pm$ 121} \\
\textsc{\textsc{DrugFlow}} & 0.75 {\tiny$\pm$ 0.39} & 120.21 {\tiny$\pm$ 73.28} & -5.66 {\tiny$\pm$ 1.78} & -6.10 {\tiny$\pm$ 1.62} & 2.11 & 0.38 & 21.14 {\tiny$\pm$ 6.81} & - \\
\textsc{Pilot} & 0.83 {\tiny$\pm$ 0.33} & 110.48 {\tiny$\pm$ 87.47} & -5.73 {\tiny$\pm$ 1.72} & -6.21 {\tiny$\pm$ 1.65} & 1.75 & 0.33 & 22.58 {\tiny$\pm$ 9.77} & 295.42 {\tiny$\pm$ 117.35}\\
\midrule
\textsc{Flowr} & 0.92 {\tiny$\pm$ 0.22} & 87.83 {\tiny$\pm$ 74.30} & -6.29 {\tiny$\pm$ 1.56} & -6.48 {\tiny$\pm$ 1.45} & 0.96 & 0.27 & 22.28 {\tiny$\pm$ 9.78} & 12.05 {\tiny$\pm$ 8.01} \\
\textsc{Flowr.root}$^{\text{base}}$ & 0.97 {\tiny$\pm$ 0.22} & 67.13 {\tiny$\pm$ 53.05} & -7.76 {\tiny$\pm$ 0.55} & -7.93 {\tiny$\pm$ 0.42} & 0.91 & 0.22 & 22.41 {\tiny$\pm$ 8.95} & 15.43 {\tiny$\pm$ 6.22} \\
\toprule
Test set & 0.95 {\tiny$\pm$ 0.21} & 75.62 {\tiny$\pm$ 57.29} & -6.44 {\tiny$\pm$ 2.74} & -6.46 {\tiny$\pm$ 2.61} & - & - & 22.75 {\tiny$\pm$ 9.90} & -\\
\bottomrule
\end{tabular}
\end{sc}
\end{adjustbox}
\end{table}

On \textsc{CrossDocked2020}, \textsc{Flowr.root} achieves a PoseBusters-validity of 0.97, surpassing all models including the test set reference (0.95). The strain energy of 67.13 kcal/mol is substantially lower than FLOWR (87.83) and PILOT (110.48), indicating energetically favorable binding poses. \textsc{Flowr.root} also achieves the best AutoDock-Vina score of -7.76 kcal/mol while maintaining efficient inference times (15.43s per ligand).

\paragraph{Pocket-conditional ligand generation: \textsc{Spindr}}

We evaluate the non-pretrained \textsc{Flowr.root}$^{base}$ model across three conditional generation modes and compare against the corresponding \textsc{Flowr.multi}~\cite{S:model:flowr} baselines. For interaction-conditional generation, where the model is guided by predefined protein-ligand interactions, results are shown in Supplementary Table~\ref{S:tab:flowr_root_interactions}). We also assess scaffold inpainting, where functional groups are provided as context and the mod~el generates the core scaffold (Supplementary Table~\ref{S:tab:flowr_root_scaffold}). Conversely, for functional group inpainting, where the scaffold is given and the model decorates it with functional groups
(Supplementary Table~\ref{S:tab:flowr_root_func_group}).

\begin{table*}[t!]
\caption{\textbf{Supplementary Table~3.} \textbf{Benchmark of the (non-pretrained) \textsc{Flowr.root}$^{\text{base}}$ model against \textsc{Flowr} on the \textsc{Spindr} test set}. \textsc{Flowr.root}$^{\text{base}}_{\text{large}}$ denotes a base model with $\sim$1.5x the number of parameters to see the effect of model scaling on the \textsc{Spindr} dataset. We report RDKit- and PoseBusters-validity of generated ligands, the GenBench3D strain energy and the AutoDock-Vina score. We also state the Wasserstein distance of generated ligands for the bond angles and bond lengths distribution to the \textsc{Spindr} test set. Novelty, uniqueness and diversity measure the capability of the model to explore the chemical space both in 2D and 3D with the latter evaluating uniqueness and diversity of conformers of the same molecule (Note: if the list of generated ligands for a target contains duplicated molecules, otherwise they become zero). RDKit's QED evaluation, SAScore, the molecular weight as well as the logP values evaluate drug-likeness of generated ligands. All presented values are mean values taken for 100 sampled ligands per test set target. The test dataset comprises 225 test set targets. Note, both RDKit- and PoseBusters-validity are evaluated on the raw generated set of 100 ligands per target. All other metrics are calculated on the subset of RDKit-valid ligands. Values are mean $\pm$ standard deviation across n = 225 \textsc{Spindr} test-set targets. Reported values are descriptive.}
\label{S:tab:full_results_root}
\centering
\begin{adjustbox}{width=1.0\textwidth,center}
\begin{sc}
\begin{tabular}{l|c|c|c|c}
\toprule
Metric & Test set &~\textsc{Flowr} &~\textsc{Flowr.root}$^{\text{base}}$ &~\textsc{Flowr.root}$^{\text{base}}_{\text{large}}$ \\
\midrule
RDKit-validity & 1.00 {\tiny$\pm$ 0.00} & 0.94 {\tiny$\pm$ 0.24} & 0.98 {\tiny$\pm$ 0.13} & 0.98 {\tiny$\pm$ 0.14} \\
PB-validity & 0.99 {\tiny$\pm$ 0.02} & 0.88 {\tiny$\pm$ 0.21} & 0.97 {\tiny$\pm$ 0.10} & 0.98 {\tiny$\pm$ 0.09} \\
\bottomrule
Strain energy & 43.27 {\tiny$\pm$ 41.85} & 90.05 {\tiny$\pm$ 52.18} & 50.36 {\tiny$\pm$ 34.59} & 47.67 {\tiny$\pm$ 34.59} \\
Vina score & -7.69 {\tiny$\pm$ 2.00} & -6.93 {\tiny$\pm$ 0.92} & -7.52 {\tiny$\pm$ 0.84} & -7.56 {\tiny$\pm$ 0.85} \\
Vina score (minimized) & -7.88 {\tiny$\pm$ 2.00} & -7.22 {\tiny$\pm$ 0.92} & -7.71 {\tiny$\pm$ 0.85} & -7.74 {\tiny$\pm$ 0.85} \\
\bottomrule
BondAnglesW1 & - & 1.08 & 0.60 & 0.68 \\
BondLengthsW1 [10$^{-2}$] & - & 0.35 & 0.43 & 0.40 \\
\bottomrule
Novelty & 1.00 {\tiny$\pm$ 0.00} & 0.94 {\tiny$\pm$ 0.23} & 1.00 {\tiny$\pm$ 0.00} & 1.00 {\tiny$\pm$ 0.00} \\
Uniqueness2D & 0.92 {\tiny$\pm$ 0.10} & 0.94 {\tiny$\pm$ 0.13} & 0.89 {\tiny$\pm$ 0.18} & 0.89 {\tiny$\pm$ 0.18} \\
Uniqueness3D & - & 0.50 {\tiny$\pm$ 0.20} & 0.45 {\tiny$\pm$ 0.22} & 0.43 {\tiny$\pm$ 0.18} \\
Diversity2D & 0.92 {\tiny$\pm$ 0.04} & 0.86 {\tiny$\pm$ 0.05} & 0.83 {\tiny$\pm$ 0.10} & 0.82 {\tiny$\pm$ 0.09} \\
Diversity3D & - & 0.21 {\tiny$\pm$ 0.12} & 0.13 {\tiny$\pm$ 0.11} & 0.12 {\tiny$\pm$ 0.11} \\
\bottomrule
SA & 0.66 {\tiny$\pm$ 0.12} & 0.67 {\tiny$\pm$ 0.13} & 0.67 {\tiny$\pm$ 0.14} & 0.66 {\tiny$\pm$ 0.14} \\
QED & 0.49 {\tiny$\pm$ 0.22} & 0.52 {\tiny$\pm$ 0.21} & 0.49 {\tiny$\pm$ 0.20} & 0.50 {\tiny$\pm$ 0.20} \\
Rings & 2.98 {\tiny$\pm$ 1.42} & 2.68 {\tiny$\pm$ 1.35} & 3.33 {\tiny$\pm$ 1.52} & 3.41 {\tiny$\pm$ 1.56} \\
Aromatic Rings & 1.84 {\tiny$\pm$ 1.31} & 1.52 {\tiny$\pm$ 1.16} & 2.07 {\tiny$\pm$ 1.39} & 2.13 {\tiny$\pm$ 1.35} \\
HAcceptors & 7.30 {\tiny$\pm$ 4.49} & 6.67 {\tiny$\pm$ 4.23} & 7.36 {\tiny$\pm$ 4.78} & 7.39 {\tiny$\pm$ 4.79} \\
HDonors & 2.62 {\tiny$\pm$ 1.68} & 2.52 {\tiny$\pm$ 1.68} & 2.49 {\tiny$\pm$ 1.67} & 2.50 {\tiny$\pm$ 1.65} \\
LogP & 0.29 {\tiny$\pm$ 3.48} & 0.29 {\tiny$\pm$ 3.31} & 0.59 {\tiny$\pm$ 3.52} & 0.54 {\tiny$\pm$ 3.45} \\
MolWt & 390.43 {\tiny$\pm$ 119.82} & 350.10 {\tiny$\pm$ 114.00} & 384.26 {\tiny$\pm$ 121.50} & 384.45 {\tiny$\pm$ 121.68} \\
Lipinski & 4.00 {\tiny$\pm$ 1.34} & 4.35 {\tiny$\pm$ 1.05} & 4.25 {\tiny$\pm$ 1.11} & 4.27 {\tiny$\pm$ 1.10} \\
\bottomrule
\end{tabular}
\end{sc}
\end{adjustbox}
\end{table*}

\begin{table}[t!]
\caption{\textbf{Supplementary Table~4.} \textbf{Benchmark of the (non-pretrained) \textsc{Flowr.root}$^{\text{base}}$ model against \textsc{Flowr.multi} running interaction-/pharmacophore-conditional generation on the \textsc{Spindr} test set}. Given ligand atoms forming interactions with the protein pocket as context, the model generates the remaining structure. We report RDKit- and PoseBusters-validity of generated ligands, the GenBench3D strain energy and the AutoDock-Vina score. We also state the Wasserstein distance of generated ligands for the bond angles and bond lengths distribution to the \textsc{Spindr} test set. Novelty, uniqueness and diversity measure the capability of the model to explore the chemical space both in 2D and 3D with the latter evaluating uniqueness and diversity of conformers of the same molecule (Note: if the list of generated ligands for a target contains duplicated molecules, otherwise they become zero). RDKit's QED evaluation, SAScore, the molecular weight as well as the logP values evaluate drug-likeness of generated ligands. All presented values are mean values taken for 100 sampled ligands per test set target. The test dataset comprises 225 complexes. Note, both RDKit- and PoseBusters-validity are evaluated on the raw generated set of 100 ligands per target. All other metrics are calculated on the subset of RDKit-valid ligands. Values are mean $\pm$ standard deviation across n = 225 \textsc{Spindr} test-set targets. Reported values are descriptive.}
\label{S:tab:flowr_root_interactions}
\centering
\begin{adjustbox}{width=1.0\textwidth,center}
\begin{sc}
\begin{tabular}{l|c|c|c}
\toprule
Metric & Test set & \textsc{Flowr.multi}$^{\mathrm{interact.-cond.}}$ & \textsc{Flowr.root}$^{\text{base}}_{\mathrm{interact.-cond.}}$ \\
\midrule
RDKit-validity & 1.00\,$_{\scriptscriptstyle\pm0.00}$ & 0.93\,$_{\scriptscriptstyle\pm0.25}$ & 0.98\,$_{\scriptscriptstyle\pm0.15}$ \\
PB-validity & 0.99\,$_{\scriptscriptstyle\pm0.02}$ & 0.86\,$_{\scriptscriptstyle\pm0.19}$ & 0.95\,$_{\scriptscriptstyle\pm0.11}$ \\
\midrule
Strain energy & 43.27\,$_{\scriptscriptstyle\pm41.85}$ & 107.60\,$_{\scriptscriptstyle\pm93.07}$ & 56.60\,$_{\scriptscriptstyle\pm43.19}$ \\
Vina score & -7.69\,$_{\scriptscriptstyle\pm2.00}$ & -7.18\,$_{\scriptscriptstyle\pm0.83}$ & -7.57\,$_{\scriptscriptstyle\pm0.67}$ \\
Vina score (minimized) & -7.88\,$_{\scriptscriptstyle\pm2.00}$ & -7.48\,$_{\scriptscriptstyle\pm0.80}$ & -7.82\,$_{\scriptscriptstyle\pm0.68}$ \\
\midrule
BondAnglesW1 & - & 1.17 & 0.61 \\
BondLengthsW1 [10$^{-2}$] & - & 0.43 & 0.37 \\
\midrule
Novelty & 1.00\,$_{\scriptscriptstyle\pm0.00}$ & 0.93\,$_{\scriptscriptstyle\pm0.26}$ & 1.00\,$_{\scriptscriptstyle\pm0.00}$ \\
Uniqueness2D & 0.92\,$_{\scriptscriptstyle\pm0.10}$ & 0.83\,$_{\scriptscriptstyle\pm0.26}$ & 0.74\,$_{\scriptscriptstyle\pm0.32}$ \\
Uniqueness3D & - & 0.40\,$_{\scriptscriptstyle\pm0.21}$ & 0.36\,$_{\scriptscriptstyle\pm0.21}$ \\
Diversity2D & 0.92\,$_{\scriptscriptstyle\pm0.04}$ & 0.82\,$_{\scriptscriptstyle\pm0.08}$ & 0.79\,$_{\scriptscriptstyle\pm0.10}$ \\
Diversity3D & - & 0.06\,$_{\scriptscriptstyle\pm0.07}$ & 0.09\,$_{\scriptscriptstyle\pm0.08}$ \\
\midrule
SA & 0.66\,$_{\scriptscriptstyle\pm0.12}$ & 0.67\,$_{\scriptscriptstyle\pm0.13}$ & 0.66\,$_{\scriptscriptstyle\pm0.13}$ \\
QED & 0.49\,$_{\scriptscriptstyle\pm0.22}$ & 0.50\,$_{\scriptscriptstyle\pm0.21}$ & 0.49\,$_{\scriptscriptstyle\pm0.20}$ \\
Rings & 2.98\,$_{\scriptscriptstyle\pm1.42}$ & 2.98\,$_{\scriptscriptstyle\pm1.38}$ & 3.33\,$_{\scriptscriptstyle\pm1.52}$ \\
Aromatic Rings & 1.84\,$_{\scriptscriptstyle\pm1.31}$ & 1.79\,$_{\scriptscriptstyle\pm1.22}$ & 2.05\,$_{\scriptscriptstyle\pm1.34}$ \\
HAcceptors & 7.30\,$_{\scriptscriptstyle\pm4.49}$ & 7.23\,$_{\scriptscriptstyle\pm4.44}$ & 7.25\,$_{\scriptscriptstyle\pm4.60}$ \\
HDonors & 2.62\,$_{\scriptscriptstyle\pm1.68}$ & 2.75\,$_{\scriptscriptstyle\pm1.58}$ & 2.72\,$_{\scriptscriptstyle\pm1.56}$ \\
LogP & 0.29\,$_{\scriptscriptstyle\pm3.48}$ & 0.41\,$_{\scriptscriptstyle\pm3.43}$ & 0.35\,$_{\scriptscriptstyle\pm3.44}$ \\
MolWt & 390.43\,$_{\scriptscriptstyle\pm119.82}$ & 379.85\,$_{\scriptscriptstyle\pm115.96}$ & 382.45\,$_{\scriptscriptstyle\pm117.86}$ \\
Lipinski & 4.00\,$_{\scriptscriptstyle\pm1.34}$ & 4.29\,$_{\scriptscriptstyle\pm1.11}$ & 4.31\,$_{\scriptscriptstyle\pm1.08}$ \\
\bottomrule
\end{tabular}
\end{sc}
\end{adjustbox}
\end{table}

\begin{table}[t!]
\caption{\textbf{Supplementary Table~5.} \textbf{Benchmark of the (non-pretrained) \textsc{Flowr.root}$^{\text{base}}$ model against \textsc{Flowr.multi} on the \textsc{Spindr} test set for scaffold hopping}. Given functional groups as context, the model generates the remaining scaffold structure. We report RDKit- and PoseBusters-validity of generated ligands, the GenBench3D strain energy and the AutoDock-Vina score. We also state the Wasserstein distance of generated ligands for the bond angles and bond lengths distribution to the \textsc{Spindr} test set. Novelty, uniqueness and diversity measure the capability of the model to explore the chemical space both in 2D and 3D with the latter evaluating uniqueness and diversity of conformers of the same molecule (Note: if the list of generated ligands for a target contains duplicated molecules, otherwise they become zero). RDKit's QED evaluation, SAScore, the molecular weight as well as the logP values evaluate drug-likeness of generated ligands. All presented values are mean values taken for 100 sampled ligands per test set target. The test dataset comprises 225 test set targets. Note, both RDKit- and PoseBusters-validity are evaluated on the raw generated set of 100 ligands per target. All other metrics are calculated on the subset of RDKit-valid ligands. Values are mean $\pm$ standard deviation across n = 225 \textsc{Spindr} test-set targets. Reported values are descriptive.}
\label{S:tab:flowr_root_scaffold}
\centering
\begin{adjustbox}{width=1.0\textwidth,center}
\begin{sc}
\begin{tabular}{l|c|c|c}
\toprule
Metric & Test set & \textsc{Flowr.multi}$^{\text{scaffold-inpaint}}$ & \textsc{Flowr.root}$^{\text{base}}_{\text{scaffold-inpaint}}$ \\
\midrule
RDKit-validity & 1.00\,$_{\scriptscriptstyle\pm0.00}$ & 0.92\,$_{\scriptscriptstyle\pm0.26}$ & 0.98\,$_{\scriptscriptstyle\pm0.14}$ \\
PB-validity & 0.99\,$_{\scriptscriptstyle\pm0.02}$ & 0.86\,$_{\scriptscriptstyle\pm0.17}$ & 0.95\,$_{\scriptscriptstyle\pm0.12}$ \\
\midrule
Strain energy & 43.27\,$_{\scriptscriptstyle\pm41.85}$ & 105.32\,$_{\scriptscriptstyle\pm95.47}$ & 56.40\,$_{\scriptscriptstyle\pm44.10}$ \\
Vina score & -7.69\,$_{\scriptscriptstyle\pm2.00}$ & -7.10\,$_{\scriptscriptstyle\pm0.71}$ & -7.27\,$_{\scriptscriptstyle\pm0.63}$ \\
Vina score (minimized) & -7.88\,$_{\scriptscriptstyle\pm2.00}$ & -7.34\,$_{\scriptscriptstyle\pm0.72}$ & -7.51\,$_{\scriptscriptstyle\pm0.64}$ \\
\midrule
BondAnglesW1 & - & 1.14 & 0.46 \\
BondLengthsW1 [10$^{-2}$] & - & 0.58 & 0.33 \\
\midrule
Novelty & 1.00\,$_{\scriptscriptstyle\pm0.00}$ & 0.87\,$_{\scriptscriptstyle\pm0.33}$ & 1.00\,$_{\scriptscriptstyle\pm0.00}$ \\
Uniqueness2D & 0.92\,$_{\scriptscriptstyle\pm0.10}$ & 0.70\,$_{\scriptscriptstyle\pm0.33}$ & 0.71\,$_{\scriptscriptstyle\pm0.29}$ \\
Uniqueness3D & - & 0.31\,$_{\scriptscriptstyle\pm0.20}$ & 0.35\,$_{\scriptscriptstyle\pm0.15}$ \\
Diversity2D & 0.92\,$_{\scriptscriptstyle\pm0.04}$ & 0.78\,$_{\scriptscriptstyle\pm0.08}$ & 0.75\,$_{\scriptscriptstyle\pm0.10}$ \\
Diversity3D & - & 0.07\,$_{\scriptscriptstyle\pm0.12}$ & 0.06\,$_{\scriptscriptstyle\pm0.06}$ \\
\midrule
SA & 0.66\,$_{\scriptscriptstyle\pm0.12}$ & 0.65\,$_{\scriptscriptstyle\pm0.13}$ & 0.66\,$_{\scriptscriptstyle\pm0.13}$ \\
QED & 0.49\,$_{\scriptscriptstyle\pm0.22}$ & 0.49\,$_{\scriptscriptstyle\pm0.22}$ & 0.48\,$_{\scriptscriptstyle\pm0.21}$ \\
Rings & 2.98\,$_{\scriptscriptstyle\pm1.42}$ & 2.93\,$_{\scriptscriptstyle\pm1.36}$ & 2.93\,$_{\scriptscriptstyle\pm1.37}$ \\
Aromatic Rings & 1.84\,$_{\scriptscriptstyle\pm1.31}$ & 1.60\,$_{\scriptscriptstyle\pm1.20}$ & 1.82\,$_{\scriptscriptstyle\pm1.30}$ \\
HAcceptors & 7.30\,$_{\scriptscriptstyle\pm4.49}$ & 7.54\,$_{\scriptscriptstyle\pm4.40}$ & 7.43\,$_{\scriptscriptstyle\pm4.49}$ \\
HDonors & 2.62\,$_{\scriptscriptstyle\pm1.68}$ & 2.80\,$_{\scriptscriptstyle\pm1.67}$ & 2.69\,$_{\scriptscriptstyle\pm1.81}$ \\
LogP & 0.29\,$_{\scriptscriptstyle\pm3.48}$ & 0.07\,$_{\scriptscriptstyle\pm3.46}$ & 0.30\,$_{\scriptscriptstyle\pm3.59}$ \\
MolWt & 390.43\,$_{\scriptscriptstyle\pm119.82}$ & 382.87\,$_{\scriptscriptstyle\pm117.73}$ & 388.79\,$_{\scriptscriptstyle\pm120.44}$ \\
Lipinski & 4.00\,$_{\scriptscriptstyle\pm1.34}$ & 4.20\,$_{\scriptscriptstyle\pm1.18}$ & 4.20\,$_{\scriptscriptstyle\pm1.17}$ \\
\bottomrule
\end{tabular}
\end{sc}
\end{adjustbox}
\end{table}

\begin{table}[t!]
\caption{\textbf{Supplementary Table~6.} \textbf{Benchmark of the (non-pretrained) \textsc{Flowr.root}$^{\text{base}}$ model for functional group inpainting against \textsc{Flowr.multi} on the \textsc{Spindr} test set}. Given a scaffold as context, the model generates the remaining functional groups. We report RDKit- and PoseBusters-validity of generated ligands, the GenBench3D strain energy and the AutoDock-Vina score. We also state the Wasserstein distance of generated ligands for the bond angles and bond lengths distribution to the\textsc{Spindr} test set. Novelty, uniqueness and diversity measure the capability of the model to explore the chemical space both in 2D and 3D with the latter evaluating uniqueness and diversity of conformers of the same molecule (Note: if the list of generated ligands for a target contains duplicated molecules, otherwise they become zero). RDKit's QED evaluation, SAScore, the molecular weight as well as the logP values evaluate drug-likeness of generated ligands. All presented values are mean values taken for 100 sampled ligands per test set target. The test dataset comprises 225 test set targets. Note, both RDKit- and PoseBusters-validity are evaluated on the raw generated set of 100 ligands per target. All other metrics are calculated on the subset of RDKit-valid ligands. Values are mean $\pm$ standard deviation across n = 225 \textsc{Spindr} test-set targets. Reported values are descriptive.}
\label{S:tab:flowr_root_func_group}
\centering
\begin{adjustbox}{width=1.0\textwidth,center}
\begin{sc}
\begin{tabular}{l|c|c|c}
\toprule
Metric & Test set & \textsc{Flowr.multi}$^{\mathrm{func.-group.-inpaint}}$ & \textsc{Flowr.root}$^{\text{base}}_{\mathrm{func.-group.-inpaint}}$ \\
\midrule
RDKit-validity & 1.00\,$_{\scriptscriptstyle\pm0.00}$ & 0.93\,$_{\scriptscriptstyle\pm0.25}$ & 0.98\,$_{\scriptscriptstyle\pm0.15}$ \\
PB-validity & 0.99\,$_{\scriptscriptstyle\pm0.02}$ & 0.88\,$_{\scriptscriptstyle\pm0.13}$ & 0.98\,$_{\scriptscriptstyle\pm0.10}$ \\
\midrule
Strain energy & 43.27\,$_{\scriptscriptstyle\pm41.85}$ & 86.26\,$_{\scriptscriptstyle\pm78.31}$ & 52.52\,$_{\scriptscriptstyle\pm36.64}$ \\
Vina score & -7.69\,$_{\scriptscriptstyle\pm2.00}$ & -7.41\,$_{\scriptscriptstyle\pm0.67}$ & -7.35\,$_{\scriptscriptstyle\pm0.58}$ \\
Vina score (minimized) & -7.88\,$_{\scriptscriptstyle\pm2.00}$ & -7.72\,$_{\scriptscriptstyle\pm0.59}$ & -7.62\,$_{\scriptscriptstyle\pm0.59}$ \\
\midrule
BondAnglesW1 & - & 0.84 & 0.58 \\
BondLengthsW1 [10$^{-2}$] & - & 0.52 & 0.27 \\
\midrule
Novelty & 1.00\,$_{\scriptscriptstyle\pm0.00}$ & 0.94\,$_{\scriptscriptstyle\pm0.23}$ & 1.00\,$_{\scriptscriptstyle\pm0.00}$ \\
Uniqueness2D & 0.92\,$_{\scriptscriptstyle\pm0.10}$ & 0.74\,$_{\scriptscriptstyle\pm0.28}$ & 0.75\,$_{\scriptscriptstyle\pm0.24}$ \\
Uniqueness3D & - & 0.35\,$_{\scriptscriptstyle\pm0.12}$ & 0.37\,$_{\scriptscriptstyle\pm0.15}$ \\
Diversity2D & 0.92\,$_{\scriptscriptstyle\pm0.04}$ & 0.77\,$_{\scriptscriptstyle\pm0.07}$ & 0.75\,$_{\scriptscriptstyle\pm0.09}$ \\
Diversity3D & - & 0.02\,$_{\scriptscriptstyle\pm0.01}$ & 0.06\,$_{\scriptscriptstyle\pm0.07}$ \\
\midrule
SA & 0.66\,$_{\scriptscriptstyle\pm0.12}$ & 0.67\,$_{\scriptscriptstyle\pm0.13}$ & 0.66\,$_{\scriptscriptstyle\pm0.13}$ \\
QED & 0.49\,$_{\scriptscriptstyle\pm0.22}$ & 0.51\,$_{\scriptscriptstyle\pm0.21}$ & 0.51\,$_{\scriptscriptstyle\pm0.21}$ \\
Rings & 2.98\,$_{\scriptscriptstyle\pm1.42}$ & 3.29\,$_{\scriptscriptstyle\pm1.46}$ & 3.27\,$_{\scriptscriptstyle\pm1.50}$ \\
Aromatic Rings & 1.84\,$_{\scriptscriptstyle\pm1.31}$ & 1.91\,$_{\scriptscriptstyle\pm1.33}$ & 1.84\,$_{\scriptscriptstyle\pm1.30}$ \\
HAcceptors & 7.30\,$_{\scriptscriptstyle\pm4.49}$ & 6.90\,$_{\scriptscriptstyle\pm4.21}$ & 7.00\,$_{\scriptscriptstyle\pm4.38}$ \\
HDonors & 2.62\,$_{\scriptscriptstyle\pm1.68}$ & 2.61\,$_{\scriptscriptstyle\pm1.62}$ & 2.53\,$_{\scriptscriptstyle\pm1.73}$ \\
LogP & 0.29\,$_{\scriptscriptstyle\pm3.48}$ & 0.83\,$_{\scriptscriptstyle\pm3.33}$ & 0.69\,$_{\scriptscriptstyle\pm3.29}$ \\
MolWt & 390.43\,$_{\scriptscriptstyle\pm119.82}$ & 380.10\,$_{\scriptscriptstyle\pm115.69}$ & 381.96\,$_{\scriptscriptstyle\pm118.04}$ \\
Lipinski & 4.00\,$_{\scriptscriptstyle\pm1.34}$ & 4.40\,$_{\scriptscriptstyle\pm1.02}$ & 4.34\,$_{\scriptscriptstyle\pm1.07}$ \\
\bottomrule
\end{tabular}
\end{sc}
\end{adjustbox}
\end{table}

\FloatBarrier
\paragraph{Domain Adaptation via Finetuning: Project-1}
\begin{figure}[htb!]
    \centering
    \includegraphics[width=1.0\columnwidth]{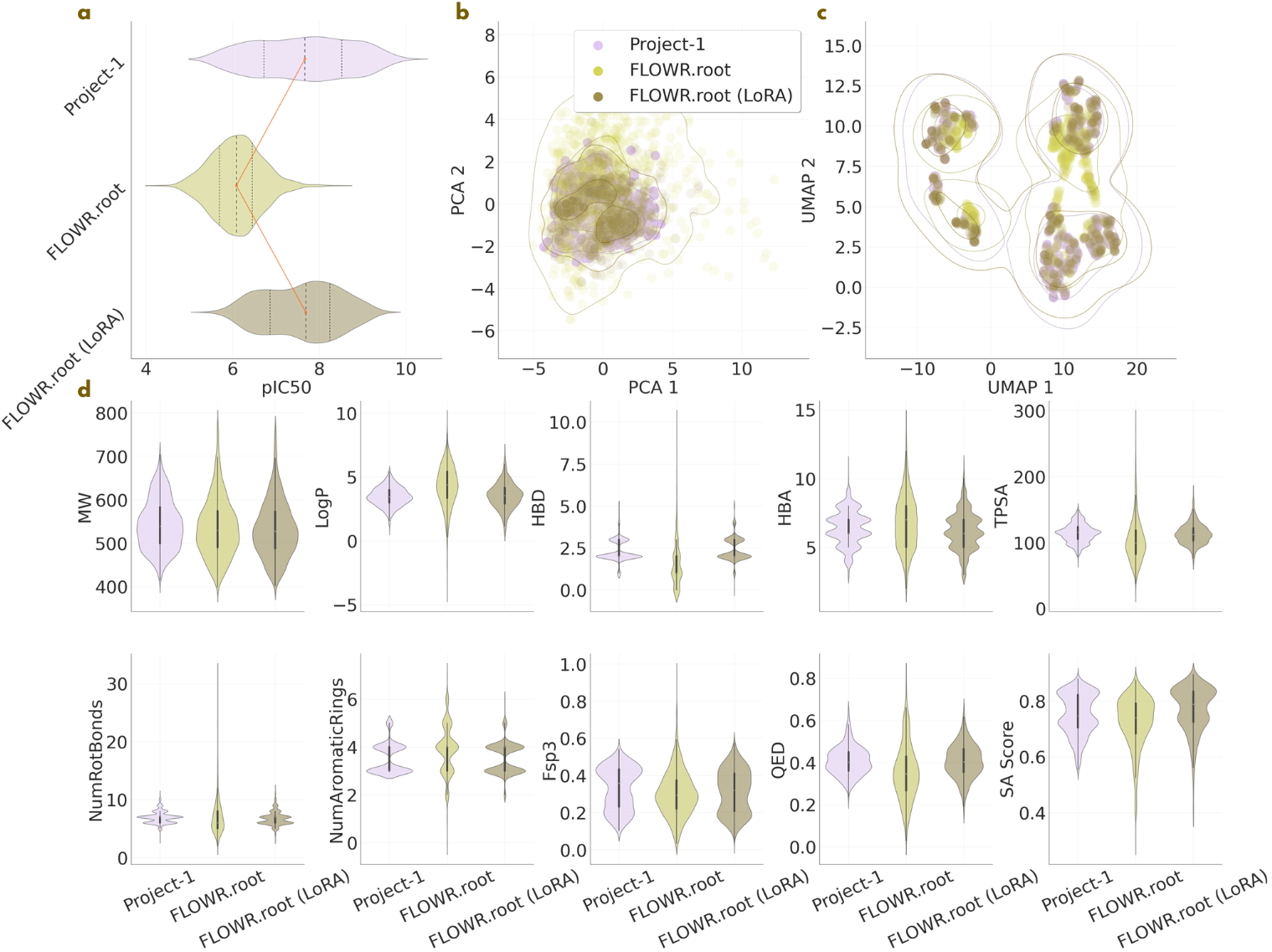}
    \caption{\textbf{Supplementary Fig.~8.} \textbf{Evaluation of \textsc{Flowr.root}-generated samples on the in-house Project-1 test dataset.} \textbf{a} Comparing the distribution of pIC$_{50}$ values of generated ligands across test set complexes between \textsc{Flowr.root} and LoRA-finetuned \textsc{Flowr.root}. The pIC$_{50}$ values for the test set are experimental, and otherwise predicted. \textbf{b} Depiction of chemical space comparison via PCA analysis showing the first two principal components. \textbf{c} Depiction of chemical space comparison via 2D UMAP analysis. \textbf{d} Distribution comparison regarding different chemical properties, namely molecular weight (MW), logP, number of hydrogen donors (HBD) and acceptors (HBA), topological surface area (TPSA), number of rotatable bonds (NumRotBonds) and aromatic rings (NumAromaticRings), fraction of $\mathrm{sp}^{3}$ carbons (Fsp3), druglikeness (QED) and synthesizability (SA Score). Panels \textbf{a} and \textbf{d} show kernel density estimates over $n=$1000 generated ligands per test-set complex across $n=$1000 Project-1 test complexes;.}
    \label{S:fig:project_1}
\end{figure}

In Supplementary Fig.~\ref{S:fig:project_1}, we visualize the structure–activity landscape of the test data from one of our in-house project datasets, comparing it to the distribution of ligands generated by both \textsc{Flowr.root} and its LoRA-finetuned variant. Following finetuning, we observe substantial adaptation to the SAR characteristics of the in-house distribution across predicted potency as well as a broad range of chemical properties, including the number of hydrogen bond donors and acceptors, rotatable bonds, and TPSA. These results underscore the effectiveness of the proposed domain adaptation approach. As expected, \textsc{Flowr.root} performs substantially worse when comparing chemical feature distributions; however, structural fidelity remains high. The mean PoseBusters validity is 0.90$\pm{0.28}$, and the mean strain energy is 48.04$\pm{19.19}$ (compared to a mean reference strain energy 45.16$\pm{11.43}$). Thus, in contrast to the zero-shot affinity prediction capabilities of \textsc{Flowr.root}, zero-shot structure prediction yields physically realistic structures.

\paragraph{Domain Adaptation via Finetuning: PDE10A}

\begin{figure}[htb!]
    \centering
    \includegraphics[width=1.0\columnwidth]{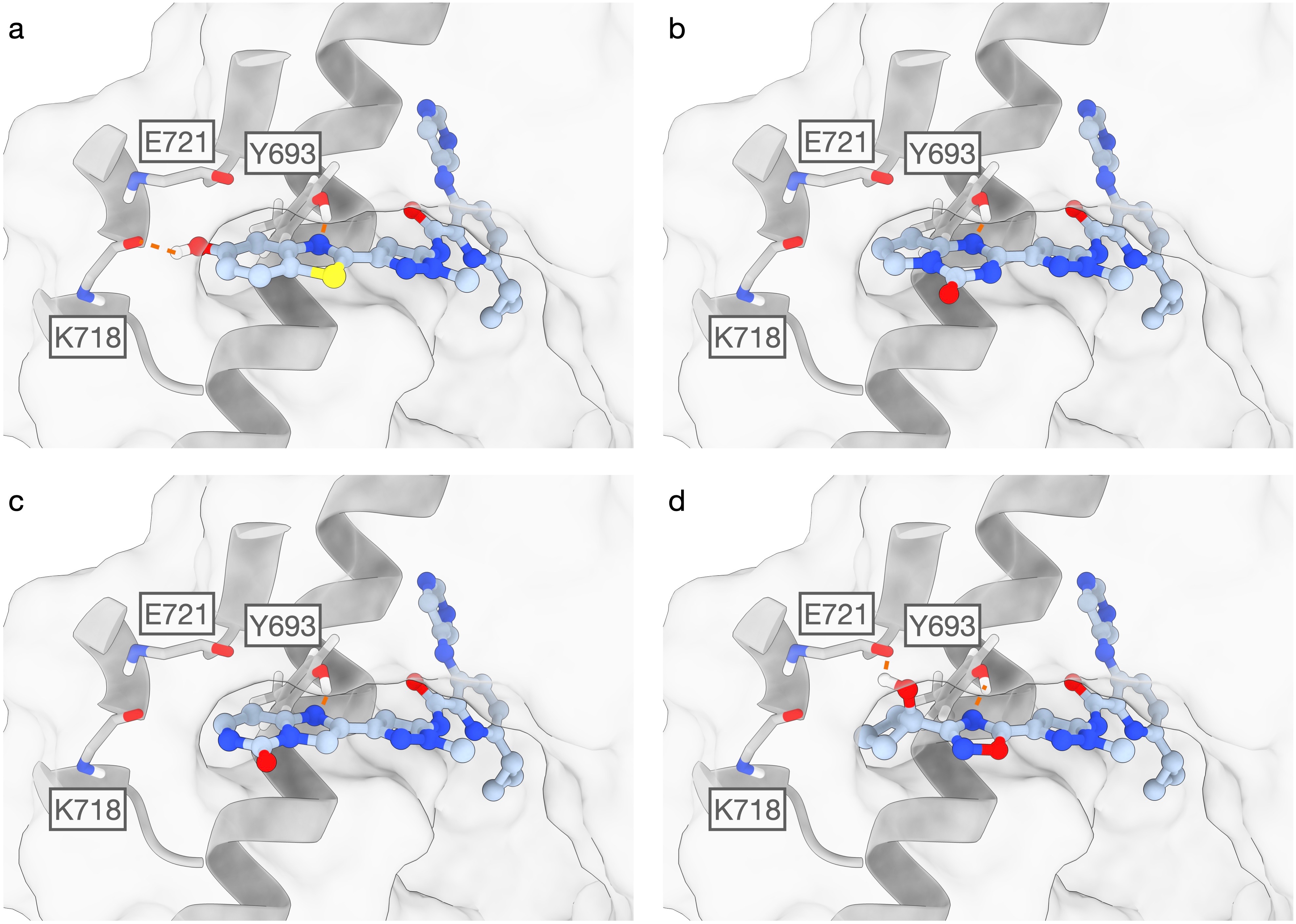}
    \caption{\textbf{Supplementary Fig.~9.} \textbf{Visualization of \textsc{Flowr.root}-predicted ligands running fragment replacement on PDE10A (PDB ID: 5SF4)}. Several ligand chemotypes with different interaction patterns around the protein pocket. \textbf{a} capturing K718's main chain with a hydrogen bond. \textbf{b} and \textbf{c} similar chemotypes capable of avoiding interactions with the residues E721 and K718. \textbf{d} A ligand capable of capturing E721 with a hydrogen bond.}
    \label{S:fig:pde10a}
\end{figure}

\begin{figure}[htb!]
    \centering
    \includegraphics[width=1.0\columnwidth]{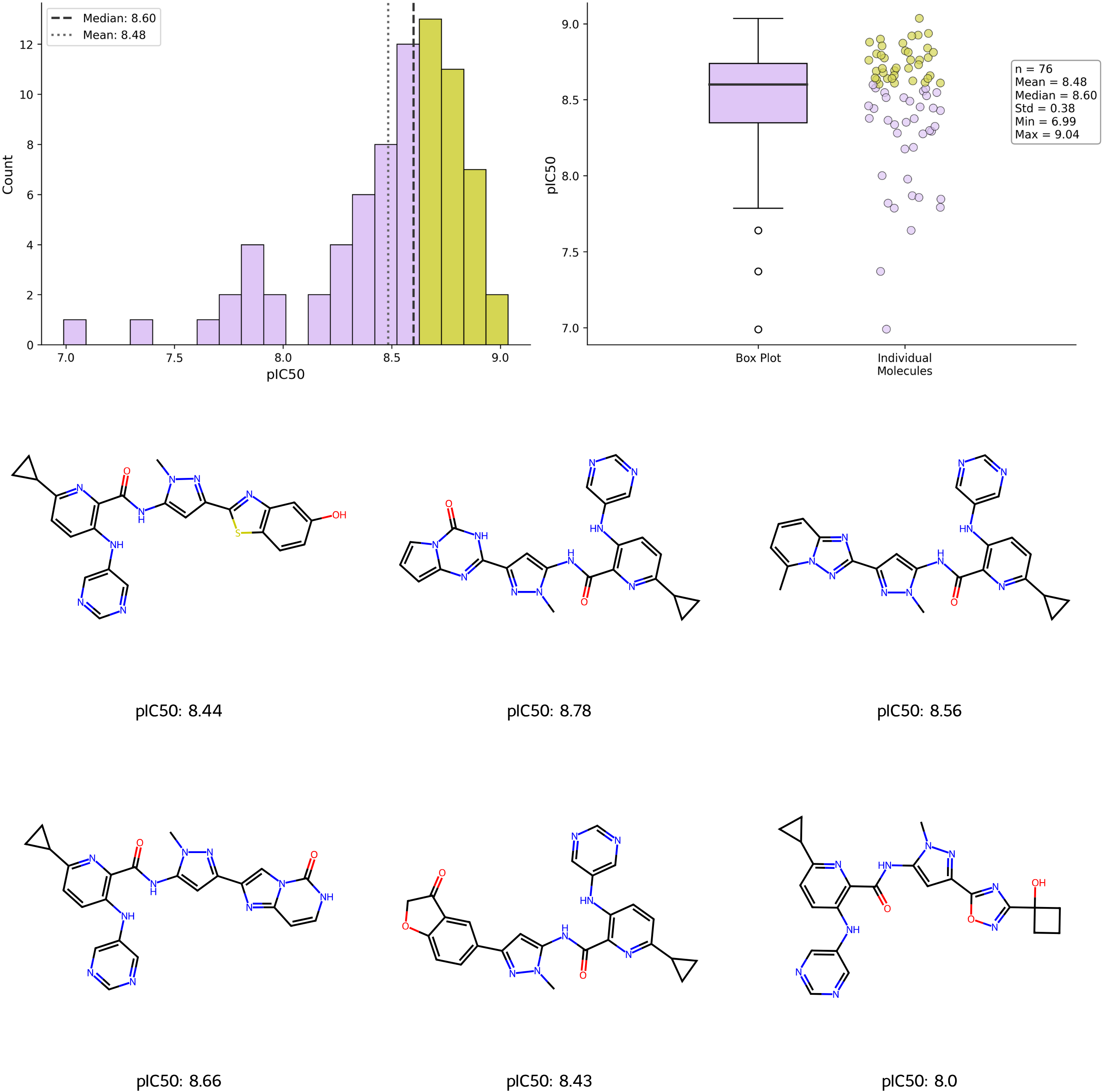}
    \caption{\textbf{Supplementary Fig.~10.} \textbf{Analysis of \textsc{Flowr.root}-predicted ligands after running fragment replacement on PDE10A (PDB ID: 5SF4).} Distribution over predicted pIC$_{50}$ values of generated ligands within the 5SF4 protein pocket and a selection over promising candidates after interaction-analysis. Histogram over $n=$76 generated ligands.}
    \label{S:fig:pde10a_mols}
\end{figure}

In Supplementary Fig.~\ref{S:fig:pde10a}, we visualize \textsc{Flowr.root}-generated ligands using fragment replacement and growing, allowing the model to extend the quinoline ring of compound 5SF4\_46 from the PDE10A dataset within the 5SF4 protein pocket. The generated ligands effectively explore hydrogen bonding opportunities with the hydroxyl group of Y693 while extending in alternative directions, such as towards residues E721 and K718. In most cases, the generated ligands tend to avoid interactions with these residues. However, in two instances (Supplementary Fig.~\ref{S:fig:pde10a}\textbf{a} and Supplementary Fig.~\ref{S:fig:pde10a}\textbf{d}), the ligand developed functional groups capable of forming hydrogen bonds with the main chain atoms of E721 and K718. The generated structures maintain high physical plausibility, with a mean PoseBusters-validity of $0.89\pm{0.31}$ and a mean strain energy of $93.92\pm{15.48}$~kcal/mol (compared to a reference ligand strain energy of $83.50$~kcal/mol).

\FloatBarrier
\subsection*{Case studies}
\paragraph{Hsp90: Heat shock protein 90}
We also performed QM validation of \textsc{Flowr.root}'s affinity head prediction on the heat shock protein 90 (Hsp90), Supplementary Fig.~\ref{S:fig:hsp90}. Unlike the cases in the main text, the dynamic range of predicted affinities is substantially smaller, a consequence of the narrower pocket filled with water molecules. We performed two sets of benchmarks, on different sets: test1 includes more ligands with higher variation in functional groups (Supplementary Fig.~\ref{S:fig:hsp90}\textbf{a}-\textbf{c}); test2 includes explicit water in the QM calculations (Supplementary Fig.~\ref{S:fig:hsp90}\textbf{d}-\textbf{f}). In all cases, critical interactions, like the hydrogen bond to Asp93, are retained over the ligand space. In test1, \textsc{Flowr.root} tried to explore interactions in a lipophilic subpocket of Hsp90 by replacing a terminal phenyl ring with a pyridyl, to realize the latter is less likely to be stabilized in the lipophilic environment. The model also replaced the pyrimidine ring with a quinazoline, in order to reduce the ligand's degrees of freedom, leading to less penalizing changes in entropy upon binding. This is seen when comparing the worst and best binders of the series. Finally, \textsc{Flowr.root} also tried to explore substitutions on the pyrimidine/quinazoline rings to further grow the ligand. As this is a solvent-exposed domain, $\mathrm{NH}_{2}$ groups are favored and lead to better interaction patterns. In the case of test2, we explicitly included water molecules in the QM calculations, on top of implicit aqueous environment. This resulted in substantially better correlation between \textsc{Flowr.root} and the QM binding energies, although some of the poorer ligands had poorer QM scores due to soft clashes with the water molecules. This indicates that, although \textsc{Flowr.root} implicitly learns the composition and orientations of solvation layers, future work might involve giving the model the ability to capture these dynamically.

\begin{figure}[htp!]
    \centering
    \includegraphics[width=1.0\columnwidth]{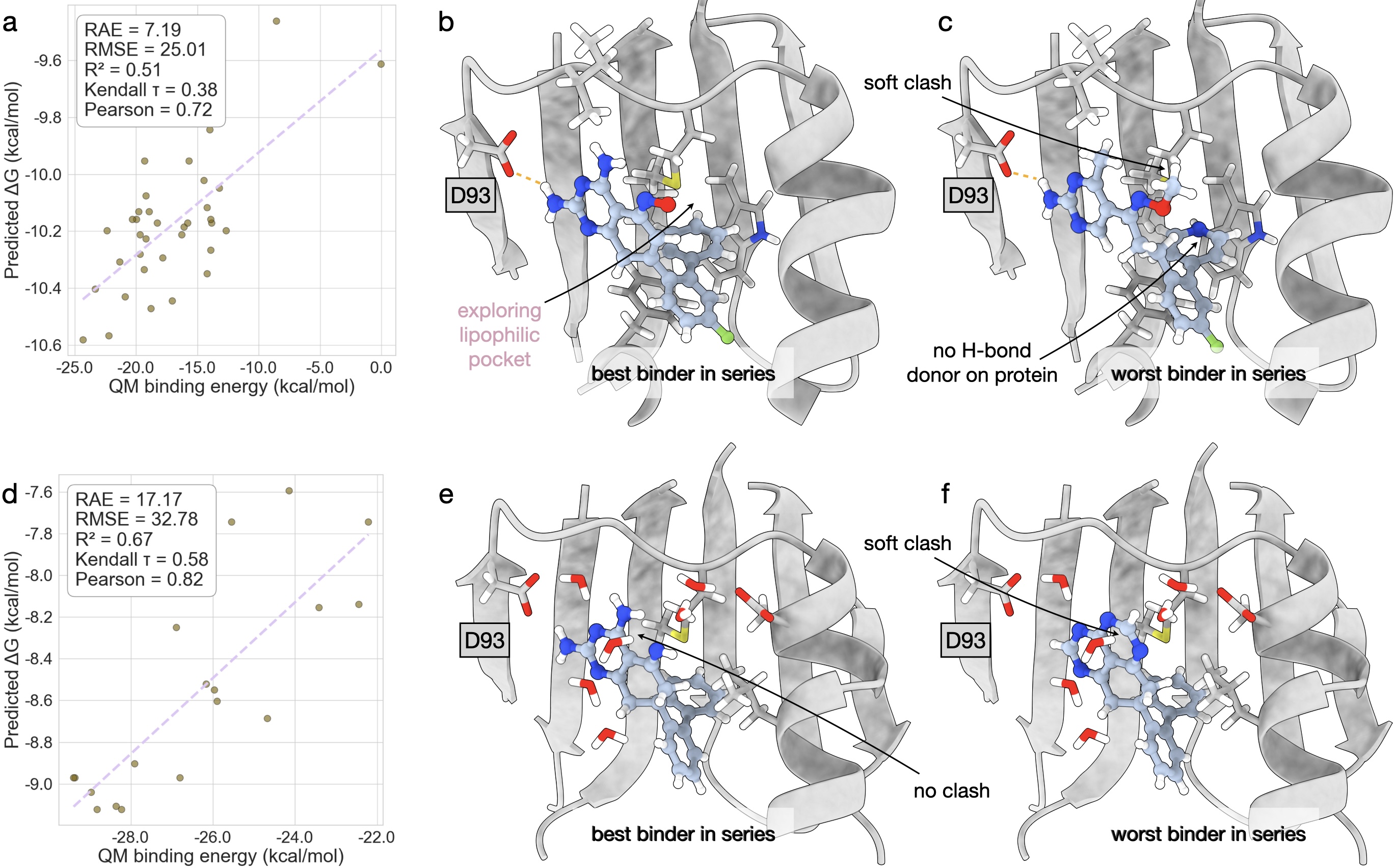}
    \caption{\textbf{Supplementary Fig.~11.} \textbf{\textsc{Flowr.root} quantum mechanical validation on Hsp90}. Test-1 validation, with more ligands but without explicit waters. \textbf{a} Correlation between \textsc{Flowr.root} affinity metrics and QM binding energies. \textbf{b} Schematic representation of the best binder in the series. \textbf{c} Schematic representation of the worst binder in the series. Test-2 validation, with explicit waters considered in the QM calculations. \textbf{d} Correlation between \textsc{Flowr.root} affinity metrics and QM binding energies. \textbf{e} Schematic representation of the best binder in the series. \textbf{f} Schematic representation of the worst binder in the series. In \textbf{a} and \textbf{d}, points are individual ligands ($n=$33 ligands for test-1; $n=$17 ligands for test-2). Reported correlation coefficients are Pearson's $r$ from a two-sided test; exact $p$-values: \textbf{[p=??]}. HSP90 reference complex: PDB \texttt{3FT8}.}
    \label{S:fig:hsp90}
\end{figure}

\subsection*{Related Works}
\label{S:app:related_works}

Diffusion- and flow-based generative models have emerged as leading frameworks in machine learning~\cite{S:ho2020ddpm,S:song2020sde,S:rombach2022ldm}. Denoising Diffusion Probabilistic Models (DDPMs) learn a reverse process to transform noise into data samples~\cite{S:ho2020ddpm}, while score-based models formulate generation via reverse-time stochastic differential equations (SDEs) or their probability-flow ordinary differential equation (ODE) counterparts~\cite{S:song2020sde}. Flow matching, in turn, directly regresses the velocity field along a prescribed probability path to train continuous normalizing flows~\cite{S:lipman2022flowmatching,S:liu2022rectified}, with stochastic-interpolant theory offering a unified perspective on diffusion and flow models~\cite{S:albergo2022building,S:albergo2023stochinterp}.
Beyond applications in natural language processing and computer vision, these models are increasingly applied to biology and chemistry, fueling advances in generative chemistry for drug discovery~\cite{S:model:edm,S:model:midi,S:model:eqgatdiff,S:model:diffsbdd,S:model:targetdiff,S:model:pilot,S:model:multiflow,S:model:semlaflow,S:model:flowr}. In drug discovery, designing small molecules that selectively bind to specific protein targets remains a critical challenge. Diffusion and flow models have proven effective at capturing complex molecular and structural distributions, advancing methods from 3D molecular generation in structure-based drug design (SBDD) to protein–ligand (co-)folding~\cite{S:folding:alphafold3,S:boltz2_preprint}.


One important subfield of AI-driven SBDD is pocket-conditional, structure-aware ligand design. Models like DiffSBDD~\cite{S:model:diffsbdd} and TargetDiff~\cite{S:model:targetdiff} have demonstrated that $SE(3)$-equivariant diffusion models can generate ligands directly within protein pockets. Follow-up studies have focused on improving chemical validity and pose accuracy~\cite{S:model:eqgatdiff,S:model:molcraft} while coupling pocket conditioning with large-scale pre-training and multi-objective importance sampling to guide generation toward potency and synthesizability~\cite{S:model:pilot}. Additionally, AI-driven models have increasingly integrated fragment priors into generative pipelines~\cite{S:imrie2020delinker,S:voloboev2024fragment2d,S:zhang2024fraggen,S:lee2025fragfm,S:guo2023linkinvent,S:model:difflinker}. In fragment-based drug discovery (FBDD), which focuses on hit-to-lead optimization, models typically specialize either in \textit{de novo} design or fragment-based strategies. However, practical lead optimization requires flexible navigation between these regimes, such as fragment growing or scaffold hopping under interaction constraints, which most current frameworks do not adequately address. Furthermore, many models do not incorporate the geometric constraints imposed by protein pockets. Recent work proposes a unified flow matching approach that integrates \textit{de novo}, interaction-constrained, and fragment-based generation under pocket conditioning with efficient ODE sampling~\cite{S:model:flowr}.

However, the effectiveness of generated ligands ultimately depends on their potency and binding affinity, which requires reliable affinity prediction. While experimental validation remains indispensable, computational prioritization during the design phase relies on accurate affinity predictions to effectively navigate target-relevant regions of chemical space. Classical scoring functions, such as AutoDock Vina~\cite{S:trott2010vina}, Glide~\cite{S:friesner2004glide}, and GOLD~\cite{S:jones1997gold}, offer computational efficiency but often lack the necessary accuracy for reliable prioritization. Physics-based methods like free energy perturbation (FEP) and absolute binding free energy (ABFE) calculations provide higher precision~\cite{S:wang2015fepplus,S:mey2020bestpractices,S:mobley2012alchemical,S:alibay2022abfe,S:feng2022abfe,S:Ries2024_abfe}, yet their computational cost prohibits application to large-scale generative campaigns, restricting their use to small subsets of candidates where absolute affinity data is critical. Machine learning–based scoring functions improve throughput but suffer from dataset bias and limited generalization~\cite{S:jimenez2018kdeep}. Recent approaches using data augmentation have shown modest improvements~\cite{S:model:aev-plig}, though they lack explicit structural information and remain insufficient for robust affinity prediction. The Boltz-2 ligand–protein co-folding model achieves near-FEP accuracy on selected targets with substantial speed advantages~\cite{S:boltz2_preprint}. However, its decoupled architecture—where the affinity head is trained independently from the structural module—creates dependence on structure prediction quality, requires co-folding prior to each affinity prediction, and may limit co-adaptation between protein pocket and ligand geometry, particularly during project-specific finetuning.

These limitations motivate a framework that jointly learns structure and affinity while supporting fast, controllable, structure-aware generation. A model that learns the joint probability distribution over ligand geometry, binding pose, and affinity within the protein pocket context would enable efficient, potency-guided ligand generation with on-the-fly ranking capabilities. Crucially, joint training facilitates simple project-specific adaptation through finetuning—an essential capability given the distinct constraints of drug discovery campaigns, ranging from ADME/T requirements to R-group and scaffold novelty constraints, diverse assay readouts, and medicinal chemistry heuristics.
A fundamental challenge in structure-based generative modeling lies in the inherent disconnect between public-domain training data and project-specific structure-activity relationships (SARs). While models can achieve broad coverage of chemical space through ligand generation, generalizing across unseen bioactivity landscapes represents a fundamentally more complex problem. We posit that expecting universal generalization without adaptation is unrealistic; instead, models should function as dynamic companions that continuously refine their understanding of project-specific SARs through sustained interaction with incoming data. This paradigm shift necessitates moving from static models to efficient iterative refinement processes where model utility grows through continuous adaptation. Additionally, techniques such as multi-objective guidance via inference-time importance sampling~\cite{S:model:pilot} can help steering generation toward pre-specified, desired properties, while deeper distribution mismatches or activity cliffs can be addressed through direct preference alignment~\cite{S:drugflow} when suitable data is available.

While large ligand-only resources such as ZINC and PubChem offer abundant chemical diversity~\cite{S:dataset:zinc,S:dataset:pubchem}, high-quality protein–ligand complexes with reliable affinity annotations remain scarce and noisy. Databases like PDBbind provide valuable training data~\cite{S:dataset:pdbbind}, yet suffer from quality and coverage limitations; curated subsets such as HiQBind address these issues and even increase sample sizes~\cite{S:dataset:hiqbind}, yet remain limited in scale. The Plinder dataset~\cite{S:dataset:plinder} expands scale by broadly curating the PDB, but most entries lack affinity annotations. More recent resources—BindingNet~\cite{S:dataset:bindingnetv2}, Kinodata-3D~\cite{S:dataset:kinodata}, and especially SAIR~\cite{S:dataset:sair}—improve chemical space coverage and include affinity data, though they do so with reduced accuracy.
This heterogeneous data landscape, however, motivates a multi-stage training paradigm: by systematically combining available resources ranked by fidelity, we can exploit their complementary strengths. Large-scale, lower-fidelity data establishes broad chemical space coverage and foundational structural understanding, while subsequent refinement on higher-quality, curated datasets sharpens affinity prediction and structural accuracy. Critically, this approach yields foundation models capable of efficient adaptation to project-specific objectives.

In the following, we discuss related works in more details for 3D molecular generation, pocket-aware ligand design, fragment-based drug discovery, and binding affinity prediction methods.

\paragraph{Molecule generation} Early neural approaches to 3D molecular generation explored autoregressive models that sequentially add coordinates and atom types while maintaining geometric consistency. Symmetry-aware models such as G-SchNet~\cite{S:model:gschnet} and its conditional inverse-design follow-up~\cite{S:model:cgschnet} demonstrated that enforcing $E(3)$ symmetries and conditioning signals can substantially improve the validity and controllability of generated 3D structures. Subsequent work proposed explicit autoregressive flows~\cite{S:model:autoregressive_flow}. In parallel, the diffusion modeling paradigm matured from its non-equilibrium thermodynamics roots~\cite{S:diffusion:original} through score-based~\cite{S:song2020sde} and variational formulations~\cite{S:diffusion:ddpm,S:kingma2021vdm}. These ideas rapidly translated to molecular geometry: Xu et~al.~\cite{S:xu2022geodiff} and Jing et~al.~\cite{S:model:torsional_diffusion} showed that $E(3)$-equivariant denoisers enable high-quality conformation generation by diffusing in Cartesian and torsional spaces, respectively.

The first $E(3)$-equivariant diffusion model for joint continuous coordinate and atom type generation was EDM~\cite{S:model:edm}. Follow-ons pushed the design space along several axes: discrete and continuous combination with explicit bond order learning~\cite{S:model:midi}, and enhanced denoising learning objectives~\cite{S:model:eqgatdiff}, establishing strong baselines.

Thereafter, diffusion-based methods pioneered pocket-aware \textit{de novo} design, also utilizing $E(3)$-equivariant networks. Notably, DiffSBDD~\cite{S:model:diffsbdd} and TargetDiff~\cite{S:model:targetdiff} demonstrated that conditional diffusion models can generate diverse, target-specific ligands within the protein pocket, adhering to symmetry constraints and enabling task versatility through sampling controls. More recently, PILOT~\cite{S:model:pilot} combined large-scale pre-training, pocket conditioning, and property guidance, highlighting the importance of multi-objective steering (for example, drug-likeness, synthesizability) under structure constraints. Together, these works established that pocket-aware 3D diffusion can simultaneously respect symmetry, improve pose realism, and support versatile constraints via conditioning and guided sampling.

Meanwhile, flow matching~\cite{S:lipman2022flowmatching} (FM) was proposed refining continuous normalizing flows by directly regressing a time-dependent velocity field that pushes a simple prior to the data distribution, offering faster sampling and flexible priors. \textsc{SemlaFlow}~\cite{S:model:semlaflow} introduced a scalable $SE(3)$-equivariant architecture (Semla) trained via flow matching, achieving state-of-the-art unconditional 3D molecule generation with substantial speed-ups. Pushing FM into SBDD, FLOWR~\cite{S:model:flowr} extended the Semla-style backbones with a dedicated pocket encoder and mixed continuous/categorical FM, supporting multi-mode \textit{de novo}, interaction-guided, and fragment-based generation in a single model. FLOWR reports large speedups over pocket-diffusion baselines.

\paragraph{Fragment-based drug discovery} Fragment-based drug discovery (FBDD) motivates models that initiate from fragments and perform growth, linking, or merging under pocket constraints. Early deep generative linker design incorporated 3D information into graph models (DeLinker) \cite{S:imrie2020delinker}, while SyntaLinker and AutoLinker utilized conditional transformers to synthesize linkers directly in the SMILES space, given fragment pairs and constraints \cite{S:yang2020syntalink,S:feng2022syntalinkhybrid}. Recent advances have introduced $E(3)$-equivariant models: DiffLinker formulates linker generation as an $E(3)$-equivariant conditional diffusion, explicitly learning 3D geometry between fragment anchors \cite{S:model:difflinker}. For broader medicinal-chemistry workflows, Link-INVENT extends REINVENT with reinforcement learning to optimize linkers for multiple objectives, demonstrated on fragment linking, scaffold hopping, and PROTAC design \cite{S:guo2023linkinvent}. STRIFE extracts target-specific pharmacophoric features to steer elaboration in 3D \cite{S:hadfield2022strife}, while AutoFragDiff integrates fragment-wise, autoregressive diffusion with pocket conditioning to improve local 3D geometry during growth \cite{S:ghorbani2023autofragdiff}. For scaffold hopping, DiffHopp employs an $E(3)$-equivariant graph diffusion model tailored for scaffold replacement conditioned on a protein–ligand complex \cite{S:torge2023diffhopp}, and TurboHopp accelerates pocket-conditioned 3D scaffold hopping with consistency models and reinforcement learning-based preference optimization \cite{S:yoo2024turbohopp}.

\paragraph{Binding Affinity Prediction}
Estimating the change in free energy upon binding (\(\Delta G_{\mathrm{bind}}\), or affinity) accurately remains a cornerstone of structure-enabled small-molecule discovery. Binding affinity is relevant for all early stages of drug discovery, starting from hit identification, where the goal is to find tight and selective binders, through hit-to-lead and lead optimization, where potency must be balanced with absorption, distribution, metabolism, excretion, safety, toxicity, and efficacy considerations. Given the astronomical size of chemical space, computer-aided drug design (CADD) is indispensable to select and prioritize candidates \textit{in silico} before spending scarce experimental resources~\cite{S:reymond2010chemicalspace,S:jorgensen2004manyroles}.

Classical structure-based approaches to affinity prediction span knowledge-based scoring and physics-based models grounded in molecular mechanics~\cite{S:liu2015classification,S:gilson2007calculation}. Heuristic docking scores offer speed at the expense of physical rigor. Empirical scoring functions such as AutoDock Vina, Glide, or GOLD remain widely used due to speed, but show inconsistent results between targets \cite{S:trott2010vina,S:friesner2004glide,S:jones1997gold}. Semi-empirical and QM / MM scoring have closed part of the gap at an intermediate cost, for example, SQM2.20 achieves DFT-quality affinity estimates in minutes, but only on selected targets~\cite{S:pecina2024sqm220,S:molani2024qmmm}.

End-point methods such as MM-PBSA and MM-GBSA combine molecular mechanics with continuum solvation to approximate the $\Delta G_{\mathrm{bind}}$ from MD snapshots at a relatively low cost and remain widely used when throughput is critical~\cite{S:kollman2000accresearch,S:homeyer2012mmpbsa,S:still1990gb,S:gohlke2004rasraf}. Alchemical binding free energy methods, absolute (ABFE) and relative (RBFE), trade throughput for accuracy ~\cite{S:aldeghi2016_abfe,S:boresch2003abfe,S:gilson1997statmech,S:Fu2022_abfe,S:feng2022abfe,S:cournia2017rbfe,S:Ross2023_rbfe}. Modern workflows based on the free energy perturbation theory (FEP)~\cite{S:zwanzig1954perturbation} have achieved impressive accuracy on suitable congeneric series~\cite{S:abel2017enhanced,S:wang2015fepplus,S:ross2023maximal}, but remain sensitive to force fields and system preparation.

Machine learning (ML) offers a complementary path to rapid affinity estimation by learning structure–activity relationships directly from data. The early ML scoring functions used interaction fingerprints and hand-made descriptors~\cite{S:ballester2010rfscore,S:wojcikowski2019plec,S:kundu2018fundamental,S:boyles2020learningligand}. Sequence-based CPI / DTA models (for example, DeepDTA) encode proteins and ligands from 1D inputs to predict binding affinity~\cite{S:ozturk2018deepdta}, while more recent deep architectures, such as 3D convolutional neural networks and graph neural networks, operate more holistically on complex geometry and interaction graphs~\cite{S:jimenez2018kdeep,S:jiang2021interactiongraphnet,S:karlov2020graphdelta,S:nguyen2021graphdta,S:li2021sign,S:meli2022deepreview,S:model:aev-plig}.

ML models are typically trained and evaluated on community benchmarks (for example, CASF)~\cite{S:li2018casf2013,S:su2019casf2016}. However, strong in-benchmark performance does not guarantee generalization. Multiple analyses show that models can overfit ligand biases, struggle on out-of-distribution (OOD) targets, or even partially fit to noise~\cite{S:volkov2022frustration,S:scantlebury2023generalizability,S:yang2020predictingorpretending,S:crusius2025fittingnoise}. This limits their reliability in prospective campaigns and underscores the need for approaches that encode biophysical constraints, reduce dataset shortcuts, and validate on OOD benchmarks.

Compounding these challenges is data scarcity: structure-based learning ideally requires reliable affinity measurements paired with high-resolution 3D protein–ligand complexes. Although data augmentation is a mainstay in computer vision and NLP~\cite{S:paulin2023syntheticcv,S:pellicer2023nlpda}, generating meaningful molecular data that respect stereochemistry, conformational physics, and pocket geometry remains non-trivial. However, combining ChEMBL and PDBbind through comparative complex structure and enhanced template-based modeling resulted in the BindingNet resource, comprising \textit{ca.} 690k complexes. This substantially densifies the bioactivity landscape compared to the PDBbind alone~\cite{S:dataset:bindingnetv1,S:dataset:bindingnetv2}.

A recent advancement is Boltz-2, a co-folding foundation model that predicts complex structures and based on that binding affinity, approaching FEP-level accuracy on certain targets while running orders of magnitude faster, makes large-scale affinity ranking feasible~\cite{S:boltz2_preprint}. Boltz-2's affinity module couples structural inference with potency prediction, providing a stronger supervisory signal than \textit{post hoc} scoring and highlighting the value of unified structure–affinity modeling in end-to-end pipelines~\cite{S:boltz2_preprint}.

\endgroup

\end{document}